\documentclass[a4paper,russian,12pt]{report}
\usepackage{latexsym}
\usepackage[cp1251]{inputenc}
\usepackage[russian]{babel}
\usepackage[unicode]{hyperref}

\def\bc{\begin{center}
}
\def\ec{
\end{center}}

\def\pdashright#1{
  \begin{picture}(20,8)
  \multiput (-5,3)(5,0){5}{\line(1,0){3}}
  \put (20,3){\vector(1,0){5}}
  \put (9,7){\makebox(1,1){$\scriptstyle #1$}}
  \end{picture} }

\def\pdownl#1{
  \begin{picture}(8,20)
  \put (4,20){\vector(0,-1){28}}
  \put (1,6){\makebox(1,1)[r]{$\scriptstyle #1$}}
  \end{picture} }
  
\def\pdownr#1{
  \begin{picture}(8,20)
  \put (4,20){\vector(0,-1){28}}
  \put (7,6){\makebox(1,1)[l]{$\scriptstyle #1$}}
  \end{picture} }

\def\pright#1{
  \begin{picture}(20,8)
  \put (-5,3){\vector(1,0){30}}
  \put (9,8){\makebox(1,1){$\scriptstyle #1$}}
  \end{picture} }

\def\diagrw#1{{
  \def\normalbaselines{\baselineskip20pt \lineskip3pt \lineskiplimit3pt }
  \matrix{#1}}}

\def\be#1{\begin{equation}\label{#1}}
\def\ee{\end{equation}}
\def\re#1{(\ref{#1})}

\def\bn{\begin{enumerate}}
\def\en{\end{enumerate}}
\def\bi{\begin{itemize}}
\def\ei{\end{itemize}}
\def\i{\item}
\def\dequiv#1{\;\mathop{\sim}\limits_{#1}\;}
\def\dapprox#1{\;\mathop{\approx}\limits_{#1}\;}

\def\sieq#1{\;\mathop{\sim}\limits_{#1}\;}

\def\vo{\;\mathop{\to}\limits_{r}\;}
\def\blackbox{\vrule height 7pt width 7pt depth 0pt}

\def\eam{\mathbin{{\mathop{=}\limits^{\mbox{\scriptsize def}}}}}

\def\by{\begin{array}{llllllllllllll}}
\def\ey{\end{array}}

\def\bullettri{\bigtriangleup}

\def\bi{\begin{itemize}}

\def\ei{\end{itemize}}
\def\bn{\begin{enumerate}}
\def\en{\end{enumerate}}
\def\i{\item}

\def\bi{\begin{itemize}}

\def\ei{\end{itemize}}
\def\bn{\begin{enumerate}}
\def\en{\end{enumerate}}
\def\i{\item}

\def\bc{\begin{center}\begin{tabular}{l}}
\def\bcr{\begin{center}\begin{tabular}{rll}}
\def\ec{\end{tabular}\end{center}}

\def\bi{\begin{itemize}}

\def\ei{\end{itemize}}
\def\bn{\begin{enumerate}}
\def\en{\end{enumerate}}
\def\i{\item}

\newcounter{theorem}
\newcounter{lemma}

\title{Теория вероятностных автоматов\\
Часть 1}

\author{А.М.Миронов}
\date{ }
\begin{document}

\maketitle
\tableofcontents


\chapter*{Введение}


Понятие {\bf вероятностного автомата (ВА)} впервые
было сформулировано в 1963 г. в основополагающей работе М. Рабина \cite{rabin}. Данное понятие 
возникло 
как синтез понятий конечного детерминированного 
автомата \cite{hmu}
и цепи Маркова \cite{kskcm} и было
предназначено для 
построения математических моделей динамических систем, 
в которых присутствует неопределённость,
описываемая статистическими закономерностями.
Эта неопределённость 
связана: 
\bi\i с неточностью знаний 
о состояниях, в которых моделируемые 
системы находятся в процессе своего функционирования, и \i  с недетерминированностью правил изменения этих состояний. \ei
Неопределённость в ВА
может быть вызвана различными причинами, 
которые подразделяются на два класса.
\bn\i Причины из первого класса
связаны с природой системы,
моделируемой вероятностным автоматом.
К ним относятся:
\bi\i влияние
случайных факторов 
на функционирование системы, например: 
случайные сбои компонентов системы или отказы в их работе,  случайное 
изменение  условий функционирования анализируемой системы,
случайность потока заявок в системе массового обслуживания
и т. п.;
\i несовершенство (или невозможность)  точного измерения состояний этой системы.\ei
\i Второй класс причин 
связан с преднамеренным внесением неточности и неопределённости
в математические модели анализируемых систем.
Это делается
в тех случаях, когда точные модели анализируемых систем
имеют неприемлемо высокую сложность и проведение анализа
поведения таких систем возможно только с использованием
их упрощённых моделей, 
в которых некоторые 
компоненты состояний 
этих систем  игнорируются.
В частности, анализ поведения сложной
программной системы (например, операционной 
системы компьютера) в большинстве случаев возможен 
только с использованием
таких упрощённых математических моделей этих систем, 
в которых принимаются во внимание значения лишь
некоторых программных переменных, от которых 
существенно зависит поведение анализируемой 
программной системы. 
\en

Как правило, моделирование  систем 
при помощи ВА 
производится: \bi\i либо с целью анализа свойств
этих систем (к числу которых относятся, например, корректность,
безопасность, надёжность, устойчивость функционирования 
в непредусмотренных ситуациях и т. д.), 
\i либо с целью вычисления различных количественных 
характеристик анализируемых систем, среди
которых могут быть, например, следующие:
\bi
\i частота выполнения тех или иных действий или переходов в анализируемых системах,
\i вероятность отказа компонентов анализируемых систем,
\i вероятность вторжения злоумышленника в компьютерную сеть, \i математическое ожидание времени отклика веб-сервиса.
\ei
\ei


Первоначальное понятие ВА, введённое
в работе М. Рабина \cite{rabin}, было предназначено главным образом для изучения вопросов представимости регулярных языков вероятностными автоматами. Затем оно было обобщено до такого понятия, которое позволило моделировать вероятностные преобразователи информации. Определение ВА в общей форме 
было введено независимо в работах 
Дж. Карлайла \cite{245}, Р. Г. Бухараева \cite{21} и П. Штарке \cite{415}. 

С  начала возникновения  понятия  ВА
исследовательская деятельность в этой области 
отличалась высокой активностью.
Результаты первых лет исследований в области ВА 
были систематизированы в книге \cite{paz}.
Подробный список  (около 500)
ссылок на работы 
с наиболее существенными теоретическими
и практическими результатами по ВА, 
полученными до 1985 г., 
можно найти в фундаментальной монографии Р. Г. Бухараева \cite{buharaevk}, которую можно рассматривать как 
итог первого периода развития теории ВА, 
продолжавшегося более двух десятилетий.

В последующие годы произошло некоторое снижение  
активности исследований в этой области, 
но в настоящее время теория ВА
вновь находится в состоянии подъёма. 
Возрождение исследовательской 
активности в области ВА
в значительной степени 
связано с тем, что в связи с бурным развитием современных информационных технологий возник широкий круг новых задач, в решении которых ВА могут служить
  эффективным инструментом. К числу таких задач 
  относятся задачи в следующих областях:
\bi
\i верификация программ и протоколов передачи данных в компьютерных сетях, 
\i информационный поиск в Интернете, 
\i финансово-экономический анализ,
\i  обработка и извлечение знаний из больших массивов данных  
(data mining и process mining),
 в частности, в задачах анализа биз\-нес-\-про\-цес\-сов, 
 биоинженерии и биоинформатики, 
\i  извлечение смысла из текстов на естественных языках,
\i  машинное зрение и обработка 
изображений
  и др.
\ei

Началом современного этапа развития 
теории ВА можно считать работу 
\cite{segala},
в которой рассмотрены ВА,
возникающие при 
моделировании 
параллельных вычислительных систем с асинхронным
взаимодействием.
В качестве вводных текстов в современную теорию 
ВА можно назвать работы
\cite{Stoelinga} и \cite{SokolovaVink}.

Главное отличие нового понятия ВА 
от того, которое изучалось в 
предшествующий период, заключается в
том, что в новом понимании ВА определяется  
как {\bf система переходов} ({\bf transition system}),
с которой связано некоторое множество переменных.
ВА функционирует путём 
выполнения переходов, после каждого из которых
происходит обновление значений переменных этого ВА.
Можно доказать, что если множество переменных
ВА конечно и множества
значений этих переменных
тоже конечны, то новое и старое понятия
ВА будут эквивалентны.

Наряду с упомянутыми выше понятиями ВА 
существуют и другие модели
динамических систем со случайным поведением, например
скрытые марковские модели (hidden Markov models) \cite{hmm},
байесовские сети (Bayesian networks) \cite{bayes}, 
вероятностные графические модели \cite{gms},
марковские решающие процессы (Mar\-kov 
decision processes) \cite{mdp},
вероятностные I/O автоматы (pro\-ba\-bi\-li\-s\-tic
   I/O automata) \cite{ioauto}.
Все эти модели являются частными случаями
исходного понятия ВА общего вида \cite{buharaevk}.

Наряду с перечисленными выше моделями 
в последние годы изучаются модели
динамических систем со случайным поведением, 
переходы в которых могут быть ассоциированы
не только с вероятностями их выполнения, 
но и с модальностями must и may, которые позволяют 
существенно усилить выразительные возможности 
этих моделей 
по сравнению с другими упомянутыми выше моделями.
Основные концепции и методы, относящиеся
к таким моделям,
содержатся в статье \cite{apapaper}.

Также изучаются и другие обобщения понятия ВА, в частности вероятностные сети Петри
\cite{petri1}, \cite{petri2}, ВА  с непрерывным временем 
\cite{conttime}, вероятностные процессные алгебры 
\cite{probprocalg}.

\chapter{Вспомогательные понятия}

\section{Случайные функции}

\subsection{Понятие случайной функции}\label{defsdffg}

Пусть задана пара множеств $X,Y$.

{\bf Случайной функцией
(СФ)} 
из $X$ в $Y$
называется
произвольная функция
$f$ вида 
\be{dfgfdsgfds}f: X\times Y \to [0,1],\ee
удовлетворяющая 
условиям: 
\bi
\i $\forall\,x\in X$ множество $\{y\in Y\mid
f(x,y)>0\}$  конечно или счётно,
\i $\forall\,x\in X\quad \sum\limits_{y\in Y}f(x,y)=1$.
\ei

Для любых $x\in X$ и $y\in Y$
значение $f(x,y)$ можно интерпретировать
как вероятность того, что СФ $f$
отображает $x$ в $y$.

Если $f$ -- СФ из $X$ в $Y$,
то мы будем обозначать этот факт 
записью $f: X\vo Y$.
Мы будем называть $X$
{\bf областью определения} СФ $f$, 
а $Y$ -- {\bf областью значений}
СФ $f$.

Если $f$ и $g$ -- СФ
вида $f:X\vo Y,\quad
g: Y\vo Z$ то 
их {\bf композицией} называется 
СФ $f g: X\vo Z$, определяемая следующим
образом:
\be{dsfgdsfgdsf}\forall\,x\in X,\;
\forall\,z\in Z\quad
(f g)(x,z)\eam \sum\limits_{y\in Y}f(x,y)
g(y,z)\ee

%

СФ \re{dfgfdsgfds}
называется {\bf детерминированной},
если для каждого $x\in X$ существует
единственный $y\in Y$, такой, что 
$f(x,y)=1$. Если $f$ -- детерминированная СФ
вида \re{dfgfdsgfds}, и $x,y$ -- такие элементы
$X$ и $Y$ соответственно, что $f(x,y)=1$, 
то мы будем говорить,
что {\bf $f$ отображает $x$ в $y$}.


\subsection{Матрицы, соответствующие 
конечным случайным функциям}

СФ 
называется {\bf конечной (КСФ)}, 
если её область определения и область значений являются конечными множествами.

Пусть задана КСФ 
$f: X\vo Y$,  и на $X$ и $Y$
заданы  упорядочения их элементов,
которые имеют вид 
$(x_1,\ldots. x_m)$ и
$(y_1,\ldots, y_n)$
соответственно.
Тогда $f$ можно представить в виде матрицы
(обозначаемой тем же символом $f$)
\be{dfgdsghdsfgdsg}
f=\;\left(
\matrix{
f(x_1,y_1)&\ldots&f(x_1,y_n)\cr 
\ldots&\ldots&\ldots\cr
f(x_m,y_1)&\ldots&f(x_m,y_n)\cr 
}
\right).\ee

Ниже мы будем отождествлять каждую КСФ
$f$ 
с соответствующей ей матрицей  \re{dfgdsghdsfgdsg}.

Мы будем предполагать, что для каждого
множества $X$, 
 являющегося 
областью определения или областью значений
какой-либо из  
рассматриваемых КСФ, на $X$ 
задано фиксированное упорядочение
его элементов. Таким образом, 
для каждой рассматриваемой КСФ
соответствующая  ей матрица
определена однозначно.

Согласно определению произведения матриц, 
из \re{dsfgdsfgdsf} следует, что 
матрица $f g$
является произведением матриц $f$ и $g$.


\subsection{Вероятностные распределения}

{\bf Вероятностным распределением} 
(или просто {\bf распределением})
на множестве $X$
называется произвольная СФ $\xi$ вида
$$\xi:\;{\bf 1} \vo X$$ 
где {\bf 1} -- множество, 
состоящее из одного элемента, который
мы будем обозначать символом $e$.
Совокупность всех распределений на 
$X$ мы будем 
обозначать записью $X^\bullettri$.
Для каждого $x\in X$ и каждого $\xi\in X^\bullettri$
значение $\xi(e,x)$  мы будем обозначать более коротко
записью 
$x^\xi$.
Для каждого $x\in X$ мы будем обозначать записью
$\xi_x$ распределение из $X^\bullettri$, 
определяемое следующим образом:
$\forall\, y\in X\;\;y^{\xi_x}\eam 1$, если $y=x$, и 
$y^{\xi_x}\eam 0$, если 
$y\not =x$.

\section{Строки и функции на строках}

\subsection{Строки и связанные с ними понятия}

Для каждого множества $X$
мы будем обозначать записью $X^*$
совокупность всех конечных
строк, компонентами 
которых являются элементы $X$. Множество
$X^*$ содержит {\bf пустую строку}, она обозначается
символом $\varepsilon$.

Для каждого $x\in X$ строка, состоящая из одного
этого элемента, обозначается той же записью $x$.

Для каждой строки $u\in X^*$
её {\bf длиной} называется количество компонентов
этой строки. Длина пустой строки равна нулю.
Длина строки $u$ обозначается
записью $|u|$. 

Для каждого целого числа $k\geq 0$ записи $X^k$,
$X^{\leq k}$, 
$X^{< k}$, 
$X^{\geq k}$, 
$X^{> k}$, 
обозначают 
совокупности всех строк из $X^*$, длина которых 
равна $k$, меньше или равна $k$, и т.д., соответственно.

Для каждой пары строк $u,v\in X^*$ 
их {\bf конкатенацией}
называется строка, обозначаемая записью $uv$,
и определяемая следующим образом:
\bi\i $u \varepsilon\eam \varepsilon u\eam u$,
и \i если $u=x_1\ldots x_n$ и $v=x'_1\ldots x'_m$, 
то $u v\eam x_1\ldots x_nx'_1\ldots x'_m$.\ei

Для каждой строки $u\in X^*$ запись $\tilde u$
обозначает строку $u$, записанную в обратном порядке,
т.е. $\tilde \varepsilon=\varepsilon$, и если $u=x_1\ldots x_n$, 
то $\tilde u=x_n\ldots x_1$.

\subsection{Функции 
на  строках}\label{sdvsdfdsfvzdxvfrr}

Пусть задано конечное множество $X$.

{\bf Функцией на строках из $X^*$}
мы будем называть
произвольную
функцию вида
$f:X^*\to{\bf R}$
(где символ {\bf R} обозначает множество
действительных чисел). Совокупность всех функций на строках
из $X^*$ мы будем обозначать записью 
${\bf R}^{X^*}$.

На множестве ${\bf R}^{X^*}$ 
определены следующие операции.
\bi
\i Для функций
   $f_1,f_2\in {\bf R}^{X^*}$ их {\bf сумма}
   $f_1+f_2$ и {\bf разность}
   $f_1-f_2$ 
 определяются следующим образом:
   \be{sdfdsafdsaf55566}
   \forall\,u\in X^*
   \left\{
   \by 
   (f_1+f_2)(u)\eam
   f_1(u)+f_2(u),\\
   (f_1-f_2)(u)\eam
   f_1(u)-f_2(u).
 \ey
   \right.\ee
\i Для каждого $a\in {\bf R}$
   и каждой функции
     $f\in {\bf R}^{X^*}$
     {\bf произведение $af$}
определяется следующим образом:
   \be{sdfdsafdsaf55566111}
   \forall\,u\in X^*\quad
   (af)(u)\eam
   af(u).
   \ee

\ei

Множество ${\bf R}^{X^*}$ можно рассматривать
как  векторное пространство над {\bf R} относительно
определённых выше операций сложения и умножения на числа из {\bf R}.


\section{Автоматы Мура}

\subsection{Понятие автомата Мура}

{\bf Автомат Мура} --
это совокупность объектов 
\be{asdfasdfasdf}M=(X, Y, S, \delta, \lambda, s^0)\ee
(называемая в этом 
параграфе просто {\bf автоматом}), 
компоненты
которой имеют следующий смысл:
\bi
\i $X$, $Y$, $S$ -- множества, элементы которых
называются соответственно {\bf входными
сигналами}, {\bf выходными 
сигналами}, и
{\bf состояниями} автомата $M$,
\i $\delta:S\times X \to S$ 
и $\lambda: S \to Y$ -- 
  отображения, 
называемые соответственно
{\bf отображением перехода} и
{\bf отображением
выхода} автомата $M$,
\i $s^0$ -- элемент  $S$, 
называемый {\bf начальным состоянием} автомата $M$.
\ei

Автомат является моделью динамической системы, работа
которой происходит в дискретном времени и заключается в
\bi
\i изменении состояний под воздействием входных сигналов,
поступающих на её вход, и
\i выдаче в каждый момент времени $t=0,1,\ldots$
некоторого выходного
сигнала.\ei

Функционирование автомата  $M$ вида \re{asdfasdfasdf}
происходит следующим образом. В каждый момент времени
$t=0,1,\ldots$ автомат
$M$ 
находится в некотором состоянии $s(t)$, 
причем $s(0) \eam
s^0$. 
В каждый момент времени $t$ автомат $M$
\bi\i получает  входной сигнал $x(t)\in X$,
\i переходит в состояние $s(t + 1) \eam \delta(s(t), x(t))$, и
\i выдаёт выходной сигнал $y(t) \eam\lambda (s(t))$.\ei

\subsection{Достижимые состояния и реакция автомата}

Пусть $M$ -- автомат вида \re{asdfasdfasdf}.
Для каждого $s \in S$ 
и каждой строки  $u\in X^*$
запись $s  u$ обозначает состояние, 
определяемое индуктивно следующим образом: 
$s\varepsilon\eam s$,
и если $u = v  x$, где $v\in X^*$ и $x\in X$, то $su\eam
\delta(sv,x)$.
Нетрудно видеть, что если строка
$u$ имеет вид $x_0 \ldots x_{n}$ то 
$s  u$ -- это состояние, 
в которое перейдёт
 $M$ через $n+1$ тактов времени, при условии, что
\bi\i в текущий момент времени $t$ он находился в состоянии $s$, и
\i в моменты $t, t+1,\ldots, t+n$ 
на вход $M$ подавались сигналы
 $x_0, x_1, \ldots, x_{n}$ соответственно.\ei


Состояние $s \in S$ 
называется {\bf достижимым}, если оно имеет
вид $s^0 u$ 
для некоторого $u \in X^*$. 

{\bf Реакция} автомата $M$ -- 
 это отображение $f_M:X^*\to Y,$
определяемое следующим образом:
$$\forall\,u\in X^*\quad f_M(u)\eam \lambda(s^0 u).$$

Нетрудно видеть, что если строка $u\in X^*$ имеет вид
$x_0 \ldots x_{n}$, то 
$f_M(u)$ --  это выходной сигнал, который выдает $M$ 
в момент $n+1$, если в моменты $0,1,\ldots,n$
на его вход подавались сигналы $x_0, x_1, \ldots, x_{n}$
соответственно.

Автоматы называются {\bf эквивалентными}, 
если их реакции совпадают.

\subsection{Достижимая часть автомата}
\label{Mooredd}

Пусть $M$ -- автомат вида \re{asdfasdfasdf}.
Обозначим
\bi
\i символом $S'$ множество всех достижимых состояний $M$, и 
\i символом $M'$ автомат, получаемый из $M$ заменой
$S$ на $S'$, и
отображений $\delta$ и $\lambda$
 на ограничения этих отображений
на подмножества $S'\times  X$ и $S'$ соответственно.\\
(нетрудно видеть, что $\forall\,s\in S',\;\forall\,x\in X\;\;\delta(s,x)\in S'$)\ei

Автомат $M'$ называется {\bf достижимой частью}
автомата $M$.
Очевидно, что  $M$ и $M'$ эквивалентны. 

Если  
 $X$ и $S$ конечны, то 
$S'$ 
может
быть найдено следующим образом:
определим последовательность
$S_0\subseteq  S_1\subseteq   S_2\subseteq  \ldots$, где
\bi
\i $S_0\eam\{s^0\}$,
\i $\forall\,i\geq 0\quad S_{i+1}\eam S_i\cup\{sx\mid s\in S_i,\;x\in X\}$.
\ei

Т.к. все члены последовательности $S_0,S_1,\ldots$ --
подмножества конечного множества $S$, то
$\exists\,k<|S|: S_k=S_{k+1}$.
Нетрудно видеть, что $S_k=S'$. 

\section{Линейные автоматы
}
\label{linearautomaton}

Пусть заданы конечное множество $X$ и натуральное число $n$.

{\bf Линейным автоматом (ЛА)} размерности $n$ над $X$
мы будем называть тройку $L$ вида
\be{vdsvczvzxcvzx}L=(\xi^0, \{L^x\mid x\in  X\}, \lambda),\ee
где 
\bi
\i $\xi^0$ -- вектор-строка размерности $n$ над  ${\bf R}$,
\i $\forall\,x\in X\;\;L^x$ -- квадратная
матрица размерности $n$ над ${\bf R}$, и 
\i $\lambda$ -- вектор-столбец размерности $n$ над  ${\bf R}$.
\ei

Для каждого ЛА $L$
мы будем обозначать записью $\dim L$
размерность этого ЛА.

ЛА \re{vdsvczvzxcvzx} определяет
 автомат Мура, обозначаемый
тем же символом $L$, 
\bi
\i множествами входных и выходных сигналов которого
являются $X$ и {\bf R} соответственно, 
\i множеством состояний 
которого является совокупность ${\bf R}^n$
всех вектор--строк размерности $n$ над ${\bf R}$, 
\i начальным состоянием --
вектор-строка  $\xi^0$, 
\i отображение перехода сопоставляет  паре 
$(\xi, x)\in {\bf R}^n\times X$ вектор-строку
$\xi L^x$, и
\i отображение выхода сопоставляет 
состоянию $\xi\in {\bf R}^n$ число $\xi \lambda\in {\bf R}$.
\ei

Нетрудно видеть, что   реакция
$f_L$ данного автомата
сопоставляет каждой строке $u\in X^*$ число 
$\xi^0 L^u \lambda$,
где 
$L^\varepsilon\eam E$ (единичная матрица размерности $n$),
и если строка $u$ имеет вид $x_1\ldots x_k$, то $L^u\eam L^{x_1}\ldots L^{x_k}$.



Пусть $f$ -- функция из ${\bf R}^{X^*}$.
Мы будем 
называть её 
{\bf ли\-ней\-но-\-ав\-то\-ма\-т\-ной функцией (ЛАФ)}, если для некоторого ЛА
$L$ над $X$ верно равенство $f=f_L$.

\chapter{Вероятностные автоматы и вероятностные 
реакции}\label{vaobshchegovida}

\section{Вероятностные автоматы}

\subsection{Понятие вероятностного автомата}

{\bf Вероятностный автомат
(ВА)} -- это пятерка $A$ 
вида 
\be{dfsgdsfgdsffd44555}A=(X,Y,S,P, 
\xi^0)\ee
компоненты которой имеют следующий смысл.
\bn
\i $X$, $Y$ и $S$ -- конечные множества, элементы
которых называются соответственно 
{\bf входными сигналами},
{\bf выходными сигналами} и
{\bf состояниями} ВА $A$.
\i $P$ -- СФ вида $P:S\times X\vo S\times Y$, называемая {\bf поведением} ВА $A$.
$\forall\,(s,x,s',y) \in S\times X\times S\times Y$
значение $P(s,x,s',y)$ понимается как 
   вероятность того, что \bi\i если  в текущий 
   момент времени $(t)$ 
    $A$ находится в состоянии
   $s$, и 
   в этот момент времени на его вход
   поступил сигнал $x$, \i то
   в следующий момент времени $(t+1)$
      $A$ будет находиться в состоянии 
     $s'$, и 
  в момент времени $t$
   выходной сигнал $A$ 
  равен $y$.\ei
\i $\xi^0$  -- распределение на $S$,
называемое
 {\bf начальным распределением} ВА $A$.
$\forall\,s\in S$  
значение $s^{\xi^0}$ понимается как
вероятность того, что в начальный момент времени $(t=0)$
ВА $A$ находится в состоянии 
$s$.
\en

ВА \re{dfsgdsfgdsffd44555} называется 
{\bf детерминированным}, если $\xi^0=\xi_s$
для некоторого $s\in S$, 
и СФ $P$ является детерминированной.

\subsection{Матрицы, связанные с 
вероятностными автоматами}
\label{fdgjhglrereerrer}

Пусть $A$ -- ВА вида \re{dfsgdsfgdsffd44555}, 
и упорядочение множества $S$ его состояний 
имеет вид 
$(s_1,\ldots, s_n)$. Для любых
$x\in X$ и $y\in Y$
мы будем обозначать
записью $A^{xy}$ матрицу порядка $n$
\be{dsfdsafdsf5566}
\left(
\by
P(s_1,x,s_1,y)&\ldots&P(s_1,x,s_n,y)\\
\ldots&\ldots&\ldots\\
P(s_n,x,s_1,y)&\ldots&P(s_n,x,s_n,y)
\ey
\right)\ee
и для любой пары строк
$u\in X^*,v\in Y^*$
мы будем обозначать
записью $A^{u,v}$ 
(запятая в этой записи может опускаться)
матрицу
порядка $n$, определяемую
 следующим образом:
\bi
\i $A^{\varepsilon,\varepsilon} = E$ (единичная матрица),
 \i если $|u|\neq |v|$, 
 то $A^{u,v} = 0$ (нулевая матрица), и 
\i если $u=x_1\ldots x_k$ и $v=y_1\ldots y_k$,
то $A^{u,v} = A^{x_1y_1}\ldots A^{x_ky_k}$.
\ei

Пусть $s$ -- произвольное состояние из $S$,
и в упорядочении элементов $S$
данное состояние имеет
номер $i$ (т.е. $s=s_i$).
Мы будем называть
\bi
\i строку номер $i$ матрицы $A^{u,v}$ --
{\bf строкой $s$}, и обозначать её записью
$\vec A^{u,v}_{s}$
\i столбец номер $i$ матрицы $A^{u,v}$ --
{\bf столбцом $s$}, и обозначать его записью
${A^{u,v}_{s}}^\downarrow$
\ei

Для любых $s,s'\in S$ мы будем обозначать
записью 
$A^{u,v}_{s,s'}$
элемент  
матрицы $A^{u,v}$, находящийся в строке
$s$ столбце $s'$.

Если строки $u\in X^*$ и $v\in Y^*$ 
имеют вид 
$x_0\ldots x_k$ и  $y_0\ldots y_k$ соответственно, 
то  $A^{u,v}_{s,s'}$
можно понимать как
   вероятность того, что \bi\i если   в текущий 
   момент  ($t$) 
    $A$ находится в состоянии
   $s$, и, 
начиная с этого момента,
      на вход $A$ последовательно
   поступали элементы строки $u$
   (т.е. 
      в момент $t$ поступил сигнал $x_0$,
   в момент $t+1$ поступил сигнал $x_1$,
   и т.д.)
\i то
в моменты $t,t+1,\ldots, t+k$
   выходные сигналы  $A$ 
   равны $y_0,\ldots, y_k$ соответственно, и
     в момент $t+k+1$
      $A$ будет находиться в состоянии 
     $s'$.
\ei
\subsection{Реакция вероятностного автомата}
\label{asdfasdfsafa}

Пусть заданы ВА $A$ вида \re{dfsgdsfgdsffd44555}
и распределение $\xi\in S^\bullettri$.

Мы будем говорить, что  {\bf ВА $A$  в момент времени  $t$
 имеет распределение $\xi$}, 
 если  для каждого состояния $s\in S$ вероятность того, что 
$A$ в момент времени  $t$ находится в состоянии $s$, 
 равна $s^\xi$.

{\bf Реакцией} ВА $A$ в распределении $\xi$
называется функция
$$A^\xi: X^*\times Y^*\to{\bf R}$$
определяемая следующим образом:
$$\forall\,u\in X^*, \,\forall\,v \in Y^*\quad
A^\xi(u,v)\eam \xi A^{u,v} I$$
где запись $I$ обозначает 
вектор-столбец порядка $|S|$, все компоненты
которого равны 1.

{\bf Реакцией ВА $A$} мы будем называть реакцию этого
ВА в его начальном распределении. Мы будем обозначать
реакцию ВА $A$ записью $f_A$.
Нетрудно доказать, что если ВА $A$ детерминированный, 
то СФ $f_A$ -- детерминированная.

Если строки $u\in X^*$ и $v\in Y^*$ 
имеют вид 
$x_0\ldots x_k$ и  $y_0\ldots y_k$
соответственно, то  $f_A(u,v)$
можно понимать как
   вероятность того, что если, начиная с момента 0,
      на вход $A$ последовательно
   поступали элементы строки $u$
   (т.е. 
      в момент $0$ поступил сигнал $x_0$,
   в момент $1$ поступил сигнал $x_1$,
   и т.д.), то
в моменты $0,1,\ldots, k$
   выходные сигналы  $A$ 
   равны $y_0,\ldots, y_k$ соответственно.\\

\refstepcounter{theorem}
{\bf Теорема \arabic{theorem}\label{th13332}}.
Если $A$ --  ВА вида \re{dfsgdsfgdsffd44555} и 
 $\xi\in S^\bullettri$, 
то $A^\xi$  -- СФ.\\

{\bf Доказательство}. 

Поскольку 
$\forall\,u\in X^*, \forall\,v\in X^*$ значение
$A^\xi(u,v)$ неотрицательно, 
то для доказательства теоремы
достаточно доказать, что 
$\forall\,u\in X^*\;\;\sum\limits_{v\in Y^*}
A^\xi(u,v)=1$, т.е.
\be{dsfdsdsafs33}
\forall\,u\in X^*\quad\sum\limits_{v\in Y^*}\xi A^{u,v} I=1.\ee

Поскольку $A^{u,v}=0$ при $|u|\neq|v|$, то
\re{dsfdsdsafs33} эквивалентно условию: $\forall\,k\geq 0$
\be{dsfdsdsafs331}
\forall\,u\in X^k\quad
\sum\limits_{v\in Y^k}\xi A^{u,v} I =1.\ee

Докажем \re{dsfdsdsafs331} индукцией по $k$.
Если $k=0$, то \re{dsfdsdsafs331}  следует из того, что
$A^{\varepsilon,\varepsilon}=E$
и $\xi E I=\xi I=1$ (т.к. $\xi\in S^\bigtriangleup$).

Пусть \re{dsfdsdsafs331} верно для некоторого $k$.
Докажем, что 
\be{dsfdsdsafs333331}
\forall\,u\in X^{k+1}\quad
\sum\limits_{v\in Y^{k+1}}\xi A^{u,v} I =1.\ee
\re{dsfdsdsafs333331} эквивалентно соотношению
\be{dsfdsdsafs333332}
\forall\,u\in X^{k}, \;\forall\,x\in X\quad
\sum\limits_{v\in Y^{k},\;y\in Y}\xi A^{ux,vy} I =1.\ee
Т.к. $A^{ux,vy} = A^{u,v}  A^{xy}$, то 
\re{dsfdsdsafs333332} можно переписать в виде
\be{dsfdsdsafs333333}
\forall\,u\in X^{k}, \;\forall\,x\in X\quad
\sum\limits_{v\in Y^{k}}\xi A^{u,v}  
(\sum\limits_{y\in Y}A^{xy}
 I) =1.\ee
\re{dsfdsdsafs333333} следует из \re{dsfdsdsafs331}
и равенства
\be{dsfdsdsafs333334}
\sum\limits_{y\in Y}A^{xy} I=I\ee 
которое 
верно потому, что если $A^{xy}$ имеет вид \re{dsfdsafdsf5566},
то $\forall\,i\in\{1,\ldots,n\}$
элемент с индексом 
$i$ столбца $\sum\limits_{y\in Y}A^{xy} I$
равен сумме $$\sum\limits_{y\in Y,\;j=1,\ldots,n}P(s_i,x,s_j,y)$$
которая равна 1, т.к. 
$P$ -- СФ вида $P:S\times X\vo S\times Y$.
$\blackbox$\\


Распределения $\xi_1,\xi_2 \in S^\bullettri$
называются {\bf эквивалентными относительно $A$}, если 
реакции $A^{\xi_1}$ и $A^{\xi_2}$ совпадают, т.е.
$$\forall\,u\in X^*, \,\forall\,v \in Y^*\qquad
\xi_1 A^{u,v} I= \xi_2 A^{u,v} I.$$
Если распределения 
$\xi_1$ и $\xi_2$ эквивалентны относительно $A$, 
то мы будем обозначать этот факт записью 
$\xi_1\dequiv{A}\xi_2$.

\subsection{Базисные матрицы вероятностных автоматов}
\label{sadfasdfasdgfdgaadas}

Ниже мы будем использовать следующее обозначение: 
для каждого множества $W$ элементов какого-либо 
линейного пространства мы будем обозначать записью
$\langle W\rangle$ подпространство этого линейного пространства,
порожденное векторами из  $W$.

Пусть $A$ -- ВА вида \re{dfsgdsfgdsffd44555}.
Обозначим  записью $AI$ совокупность
всех  вектор-столбцов  вида $A^{u,v} I$, где 
   $u\in X^*, v\in Y^*$.

{\bf Базисной матрицей} ВА  $A$
называется  матрица,  обозначаемая
записью $[A]$, и удовлетворяющая
условиям:
\bi
\i каждый столбец матрицы $[A]$ является элементом
$AI$,
\i столбцы матрицы $[A]$
   образуют базис пространства $\langle AI\rangle$.  
\ei
 
 Нетрудно видеть, что для любых
$\xi_1,\xi_2\in
S^\bullettri$
$$\xi_1\dequiv{A}\xi_2
\quad
\Leftrightarrow
\quad
\xi_1 [A] = \xi_2 [A].$$

Для каждого $s\in S$ 
 мы будем называть {\bf строкой $s$} матрицы  $[A]$
 ту её строку, которая содержит значения вида $\vec A^{u,v}_{s}
  I$.  Мы будем обозначать эту строку записью $[A]_s$.

Матрица $[A]$  м.б. построена при помощи 
излагаемого ниже алгоритма.

Пусть $k\geq 0$.
Обозначим 
записью $AI_k$ 
совокупность вектор-столбцов вида 
$A^{u,v} I$, где $u\in X^*$, 
$v\in Y^*$,
$|u|=|v|\leq k$.
Нетрудно видеть, что 
\be{dsfsdafdsafsadfas}
\langle AI_0\rangle \subseteq 
\langle AI_1\rangle\subseteq
\langle AI_2\rangle\subseteq \ldots\quad\mbox{и} \quad
\bigcup\limits_{k\geq 0}\langle AI_k\rangle =
\langle AI\rangle.\ee
Поскольку все пространства $\langle AI_k\rangle$
являются подпространствами конечномерного
линейного пространства (размерности $|S|$), то, следовательно, 
последовательность включений в \re{dsfsdafdsafsadfas}
не может неограниченно возрастать, т.е. 
для некоторого $k$ верны равенства 
\be{dsafsadfasdferrrr}
\langle AI_k\rangle = 
\langle AI_{k+1}\rangle=\langle AI_{k+2}\rangle=
\ldots = \langle AI\rangle.\ee

Алгоритм построения матрицы $[A]$
 основан на следующей теореме.\\

\refstepcounter{theorem}
{\bf Теорема \arabic{theorem}\label{th1}}.
Если для некоторого $k$ верно равенство
\be{dsafdsafsadfsdfsfs}\langle AI_{k+1}\rangle = 
\langle AI_{k}\rangle\ee то  
$k$ обладает
свойством
\re{dsafsadfasdferrrr}. \\

{\bf Доказательство}.

Достаточно доказать равенство 
\be{dsfdsfsdfseeeeee}\langle AI_{k+2}\rangle = 
\langle AI_{k}\rangle.\ee

Пусть $V\in AI_{k+2} \setminus AI_{k+1}$, 
тогда $V$ имеет вид 
$A^{xy} A^{u,v} I$, где $x\in X, y\in Y$ и 
$|u|=|v|=k+1$.
Поскольку $A^{u,v} I \in AI_{k+1} \subseteq
\langle AI_{k}\rangle$, то, следовательно, $A^{u,v} I$
является линейной комбинацией вида
$$A^{u,v} I 
=\sum\limits_{i=1}^m\lambda_i V_i\quad
(\forall\,i=1,\ldots, m\quad
\lambda_i\in {\bf R},\;
V_i \in AI_{k}).$$
Следовательно, 
\be{dsafdsafsadfsdfsf33s}V=
A^{xy} \Big(\,\sum\limits_{i=1}^m\lambda_i V_i\,\Big)=
\sum\limits_{i=1}^m\lambda_i (A^{xy} V_i)=
\sum\limits_{i=1}^m\lambda_i W_i\ee
где $W_i\in  AI_{k+1}\subseteq  
\langle AI_{k}\rangle$. 
 откуда на основании 
\re{dsafdsafsadfsdfsf33s} заключаем, что
$V$ является линейной комбинацией элементов
$\langle AI_{k}\rangle$, поэтому $V\in \langle AI_{k}\rangle$.
Таким образом, $AI_{k+2} \subseteq \langle AI_{k}\rangle$, 
откуда следует \re{dsfdsfsdfseeeeee}. $\blackbox$\\

Из теоремы \ref{th1}
непосредственно следует, что
если $k$  --  наименьший номер, 
для которого верно \re{dsafdsafsadfsdfsfs}, 
то $k\leq |S|-1$.

Используя теорему \ref{th1}, можно определить
следующий алгоритм построения матрицы $[A]$.
Мы будем обозначать записью $CA$ переменную,
значениями которой являются множества вектор-столбцов
порядка $|S|$. 
Алгоритм состоит из перечисленных ниже трёх шагов.
Шаг 2 может выполняться несколько раз.
\bn
\i Значение $CA$ полагается равным $\{I\}\;(=AI_0)$.
\i Пусть $V_1,\ldots, V_m$ -- список всех столбцов вида
$A^{xy} V$, где $x\in X, y\in Y$ и $V\in CA$.
Выполняется цикл: 
\bc
\verb'for i=1 to m do {'\\
\verb'  if'  $V_i\not \in \langle CA\rangle$
\verb'then'  $V_i$ добавляется к $CA$\\
\verb'}'\ec
\i Если во время выполнения шага 2 множество
$CA$ изменилось, то шаг 2 выполняется ещё раз, 
иначе алгоритм заканчивает работу.
\en

Обоснуем корректность данного алгоритма.
Нетрудно видеть, что 
если перед 
выполнением шага 2 было верно равенство
$\langle CA\rangle=\langle AI_k\rangle$ для некоторого $k\geq 0$,
то после выполнения этого шага 
будет верно равенство
$\langle CA\rangle=\langle AI_{k+1}\rangle$.
Следовательно, через не более чем $|S|-1$ 
выполнений  шага 2 будет верно равенство 
$\langle CA\rangle=\langle AI\rangle$,
и шаг 2 выполнится не более $|S|$ раз.
Поскольку каждый добавляемый к $CA$ вектор $V_i$ 
не принадлежит пространству $\langle CA\rangle$,
то, следовательно, в каждый момент времени 
$CA$ состоит из линейно независимых векторов, т.е.
после завершения работы алгоритма $CA$
является базисом пространства $\langle AI\rangle$.
$\blackbox$

\subsection{Матричные обозначения}
\label{fgdsfgdsfgsdgsfdghsfd}

Мы будем использовать следующие обозначения, 
связанные с матрицами.

\bi
\i Если $\xi_1$ и $\xi_2$ -- вектор-строки размерностей
   $n_1$ и $n_2$ соответственно, 
   то запись $(\xi_1, \xi_2)$ обозначает
   вектор-строку размерности $n_1+n_2$, первые
   $n_1$ компонентов которой совпадают с соответствющими
   компонентами $\xi_1$, а остальные компоненты -- 
   с соответствующими компонентами $\xi_2$.
\i Если $\lambda_1$ и $\lambda_2$ -- вектор-столбцы 
    размерностей
   $n_1$ и $n_2$ соответственно, 
   то запись $\left(\by \lambda_1\\ \lambda_2\ey\right)$ обозначает
   вектор-столбец размерности $n_1+n_2$, первые
   $n_1$ компонентов которого совпадают с соответствующими
   компонентами $\lambda_1$, а остальные компоненты -- 
   с соответствующими компонентами $\lambda_2$.
\i Если $A$ и $B$ -- матрицы
размерностей $(m,n)$ и $(k,l)$ соответственно, 
   то запись
$\left(\begin{array}{ccccc} A&{\bf 0}\cr
   {\bf 0}&B\cr\ey\right)$
   обозначает матрицу 
размерности $(m+k,n+l)$, определяемую естественным образом.
\i Для каждой матрицы $A$ запись $\tilde A$ (или $A^\sim$)
обозначает матрицу, транспонированную к матрице $A$.
\ei

\subsection{Эквивалентность
 вероятностных автоматов}

Пусть задана пара ВА $A_1, A_2$, 
у которых одинаковы
множества входных сигналов и 
множества выходных сигналов, т.е. 
$A_1$ и $A_2$ имеют вид
$$A_i=(X,Y,S_i, P_i, \xi^0_i)\quad(i=1,2).$$

$A_1$ и $A_2$ называются {\bf эквивалентными},
если их реакции совпадают, т.е. 
верно равенство
\be{sdfsadfsadfgsfdgfsdgf}
f_{A_1} = 
f_{A_2}.\ee

Нетрудно доказать, что равенство
\re{sdfsadfsadfgsfdgfsdgf} равносильно соотношению
$\xi_1\dequiv{A}\xi_2$, где $A$ имеет вид
$(X,Y,S_1\sqcup S_2, P,\xi^0)$, и
\be{sdvdsvfdvdfvdsfvd}
\by 
\forall\,x\in X,\;y\in Y\quad
A^{xy}=\left(\begin{array}{ccccc} A_1^{xy}&{\bf 0}\\{\bf 0}&A_2^{xy}
\ey\right),\\
\xi_1=(\xi^0_1,{\bf 0}),\;\;
\xi_2=({\bf 0},\xi^0_2)\ey\ee
(символы {\bf 0} в \re{sdvdsvfdvdfvdsfvd} изображают 
нулевые матрицы
или вектор-строки соответствующих размеров).


Если ВА $A_1$ и $A_2$ эквивалентны, то мы будем обозначать
этот факт записью $A_1\sim A_2$.

\section{Редукция вероятностных автоматов}
\label{reddsdsafsafa}

{\bf Редукция} ВА заключается в построении 
по заданному ВА $A$  такого ВА, который был бы 
эквивалентен $A$, и содержал меньше состояний, чем $A$
(если это возможно).
Мы будем рассматривать два метода редукции:
выделение достижимой части и удаление выпуклых 
комбинаций.

\subsection{Выделение достижимой части}
\label{dsfasdfdsaf}

Пусть $A$ -- ВА вида \re{dfsgdsfgdsffd44555}.
Понятие {\bf достижимого состояния} ВА $A$ определяется
рекурсивно: состояние $s\in S$ достижимо, если 
\bi
\i либо $s^{\xi^0}\neq 0$, 
\i либо существует достижимое состояние $s'\in S$, такое, что 
\be{sdfsdfsdfsdgfsdgsd}\exists\, x\in X, y\in Y:\;\;
P(s',x,s,y)>0.\ee
\ei

Нетрудно доказать, что 
$A\sim A_r\eam (X,Y,S_r, P_r, \xi^0_r)$, где 
\bi
\i $S_r$ состоит из 
всех достижимых состояний ВА $A$, и 
\i    $P_r$
и  
 $\xi^0_r$ являются 
соответствующими ограничениями
   $P$
и   $\xi^0$.
\ei

ВА $A_r$ называется
{\bf достижимой частью} ВА $A$. 
Алгоритм построения по заданному ВА 
его достижимой части аналогичен соответствующему
алгоритму для детерминированных автоматов
(см. конец пункта \ref{Mooredd}).

\subsection{Удаление выпуклых комбинаций}
\label{gfdsfdgsfdgsgee}

Пусть $A$ -- ВА вида $(X,Y,S,P,\xi^0)$.
Мы будем говорить, что 
состояние $s\in S$ является {\bf выпуклой комбинацией}
других состояний ВА $A$, если 
строка $s$ матрицы $[A]$
является выпуклой комбинацией других строк этой матрицы, 
т.е. существует распределение $\xi\in (S\setminus \{s\})^\bullettri$,
удовлетворяющее условию
\be{fdsadfsdfdsgfsdg}
[A]_s=\sum\limits_{s'\in S\setminus \{s\}}(s')^\xi [A]_{s'}.\ee

Если в множестве $S$ состояний ВА $A$ есть состояние $s$, являющееся
выпуклой комбинацией других состояний этого ВА, то можно
определить ВА $B$, который эквивалентен  $A$,
и множество состояний которого имеет вид
$S\setminus \{s\}$.  Мы будем говорить, что $B$ получается
из $A$  путем удаления выпуклой комбинации $s$.

Автомат $B$ определяется следующим 
образом. Пусть упорядочение множества $S$ имеет вид
$(s_1, \ldots, s_n)$, и вышепомянутое состояние $s$
является последним в этом упорядочении (т.е. $s=s_n$).
Обозначим символом $M$ матрицу
$$\left(\begin{array}{ccccccc}
1&0&\ldots&0&0\\
&\ldots\\
0&0&\ldots&1&0\\
s_1^\xi&s_2^\xi&\ldots&s^\xi_{n-1}&0
\ey\right)$$
и обозначим символом $C$ ВА  $(X,Y,S,Q,\xi^0 M)$,
где 
\be{asdfsadfsadfee22}
\forall\,x\in X,\;y\in Y\quad C^{xy}=A^{xy} M.\ee

Докажем, что $\forall\,u\in X^*, v\in Y^*$
верно равенство 
\be{dfgdfgf}C^{u,v} I=
A^{u,v} I.\ee
 \re{dfgdfgf} верно, 
когда $u$ и $v$ имеют разную длину.
Для $u$ и $v$ одинаковой длины 
будем доказывать \re{dfgdfgf} индукцией по длине $u$.
\bn\i
 \re{dfgdfgf} верно, когда
$u=v=\varepsilon$.
\i Пусть \re{dfgdfgf} верно для некоторых 
$u,v$. Тогда $\forall\,x\in X, y\in Y$
\be{asdfsadfsadfee}C^{xu,yv} I=C^{xy} C^{u,v} I=
A^{xy} M A^{u,v} I\ee
(второе равенство в \re{asdfsadfsadfee} следует из 
\re{asdfsadfsadfee22} и \re{dfgdfgf}).

Докажем, что верно равенство 
\be{sdfsadfasdgfsdgfsdgsd}M A^{u,v} I = A^{u,v} I.\ee
Из \re{fdsadfsdfdsgfsdg}  следует. что
$$(s_1^\xi\;\ldots\;s_{n-1}^\xi\;0) [A] =[A]_{s_n}=
(0\;\ldots\;0\;1) [A]
$$
откуда следует
\be{sdfsadf}
M [A]=\left(\begin{array}{ccccccc}
1&0&\ldots&0&0\\
&\ldots\\
0&0&\ldots&1&0\\
0&0&\ldots&0&1
\ey\right) [A]=[A]
\ee
Из \re{sdfsadf} следует, что для каждого столбца  $V$,
матрицы   $[A]$ верно равенство 
\be{sdfsadfdsfdsfdsfds}M  V=V.\ee
Поскольку столбцы    $[A]$ образуют 
базис     $\langle AI \rangle$, 
то, следовательно, \re{sdfsadfdsfdsfdsfds}
 верно
в том случае, когда $V$ 
является  произвольным
элементом $\langle AI \rangle$.
В частности, \re{sdfsadfdsfdsfdsfds} верно для всех векторов
из $AI$.
 Таким образом, равенство 
\re{sdfsadfasdgfsdgfsdgsd} доказано.

Из \re{sdfsadfasdgfsdgfsdgsd} и из \re{asdfsadfsadfee}
следует, что 
\be{a33sdfsadfsadfee}C^{xu,yv} I=
A^{xy} M A^{u,v} I=
A^{xy} A^{u,v} I=
A^{xu,yv} I.
\ee
  Таким образом, если  \re{dfgdfgf} верно,
  то будет верно равенство, получаемое из  \re{dfgdfgf}
  заменой $u$ на  $xu$, а $v$ -- на   $yv$.
\en  Следовательно, 
 \re{dfgdfgf} верно для всех $u\in X^*$, $v\in Y^*$.
 
 Докажем, что ВА $A$  и $C$ эквивалентны, т.е. 
$f_A=f_C$.
Данное равенство равносильно утверждению
 \be{dsfsdfsdfgagfsdgsd}\forall\,u\in X^*, \,\forall\,v \in Y^*\qquad
\xi^0 A^{u,v} I= \xi^0 M  C^{u,v} I.\ee
\re{dsfsdfsdfgagfsdgsd}
 следует из  \re{dfgdfgf} и из \re{sdfsadfasdgfsdgfsdgsd}.
 
 Заметим, что состояние $s_n$ ВА $C$ не является 
 достижимым. Действительно,  т.к.
 последний столбец матрицы  $M$
   является нулевым, то 
\bi\i значение 
  $s_n^{\xi^0 M}$, которое является 
  последним элементом век\-тор-\-стро\-ки  $\xi^0 M$, равно 0, и 
\i для каждого $x\in X$ и каждого $y\in Y$
последний столбец матрицы $C^{xy}= A^{xy} M$
является нулевым, поэтому 
неравенство \re{sdfsdfsdfsdgfsdgsd}, в котором
$P$ заменено на  $Q$, и $s$ -- на $s_n$, неверно
для каждого  $s'\in S$.
\ei 
 
Искомый ВА $B$ определяется как ВА, 
 получаемый  из ВА $C$ удалением
 недостижимого состояния
 $s_n$ и соответствующим ограничением поведения
 и начального распределения ВА  $C$.
 Из утверждения в пункте \ref{dsfasdfdsaf} следует, 
 что $C\sim B$.
 Поскольку свойство эквивалентности ВА является транзитивным,
 то из $A\sim C$ и $C\sim B$ следует, что 
 $A\sim B$. $\blackbox$

\subsection{Метод распознавания выпуклых комбинаций
состояний}
\label{sifgawerfasdfa}

Для реализации изложенного в предыдущем пункте
метода редукции ВА путем удаления выпуклых 
комбинаций состояний
 необходимо иметь алгоритм решения следующей задачи:
пусть задан ВА $A$, 
и $s$ -- одно из состояний этого ВА, требуется 
\bi\i определить, 
является ли  состояние $s$ выпуклой комбинацией
других состояний ВА $A$, т.е. является ли строка $[A]_s$
выпуклой комбинацией других строк  матрицы $[A]$, и
\i если ответ на этот вопрос положителен, то найти  коэффициенты
этой выпуклой комбинации.\ei

Данную задачу можно свести к задаче линейного программирования (ЗЛП), на основе нижеследующей теоремы.\\

\refstepcounter{theorem}
{\bf Теорема \arabic{theorem}\label{th012323331}}.

Пусть задан ВА $A$, и
$s$ -- одно из состояний этого ВА.
Обозначим записью $\{W_1,\ldots, W_m\}$ 
совокупность строк матрицы $[A]$, за исключением строки $[A]_s$.

Тогда следующие утверждения эквивалентны.
\bn
\i  $[A]_s$ является выпуклой комбинацией строк $\{W_1,\ldots, W_m\}$, т.е. 
\be{dsafasdfasdfdsaf}
\exists\,(\xi_1,\ldots, \xi_m)\in \{1,\ldots, m\}^\bigtriangleup:
[A]_s=\sum\limits_{i=1}^m \xi_i W_i.\ee
\i Существует решение ЗЛП,
в которой
\bi
\i множество переменных имеет вид
  $$\{x_1,\ldots, x_m, y_1,\ldots, y_n\}$$
  где $n$ -- число состояний ВА $A$,
  \i ограничения в форме неравенств имеют вид
$x_{i}\geq 0$ и $y_{j}\geq 0$, где 
$i\in \{1,\ldots, m\}$,
$j\in \{1,\ldots, n\}$,
\i ограничения в форме равенств выражаются
в виде матричного равенства
$(X,\;Y)\left(\by 
W\\E_n
\ey
\right)=[A]_s$, где \bi
\i $X$ и $Y$ -- 
вектор-строки переменных:
$$X = (x_{1},\ldots, x_{m}),\quad
    Y = (y_{1},\ldots, y_{n}).$$
\i $W$ -- матрица, получаемая из $[A]$
путем удаления строки $[A]_s$,
\i $E_n$ -- единичная матрица порядка $n$,
\ei
а также
равенства $\sum\limits_{i=1}^m x_i+
\sum\limits_{i=1}^n y_i=1$,
\i целевая функция имеет вид $\sum\limits_{i=1}^n y_{i}\to\min$,
\ei
и значение целевой функции на этом решении равно 0.
\en

{\bf Доказательство}.

Пусть верно утверждение 1.
Тогда решение ЗЛП имеет вид
$x_1=\xi_1,\ldots, x_m=\xi_m$, 
$y_1=0,\ldots, y_n=0$.

Обратно, пусть верно утверждение 2, т.е.
существует решение ЗЛП, 
значение целевой функции на котором равно 0.
Тогда из ограничения  $y_j\geq 0\;(j=1,\ldots, n)$
следует, что 
значения переменных $y_1,\ldots, y_n$ на этом 
решении равны 0. Нетрудно видеть, что 
совокупность $(\xi_1,\ldots, \xi_m)$
значений переменных $x_1,\ldots, x_m$ 
на этом решении удовлетворяет условиям в 
соотношении \re{dsafasdfasdfdsaf}. $\blackbox$\\

Отметим, что одно из опорных решений 
ЗЛП, сформулированной в теореме \ref{th012323331},
имеет вид $X={\bf 0}$, $Y=[A]_s$.

\section{Вероятностные реакции}

\subsection{Понятие вероятностной реакции}

Пусть $X$ и $Y$ -- конечные множества.

{\bf Вероятностной реакцией (ВР)} из $X$ в $Y$ называется СФ
$f: X^*\vo Y^*$, удовлетворяющая
условию: $\forall\,u\in X^*,\forall\,v\in Y^*$
\be{safdsdfdsafsafarrr}
\by
\mbox{если }|u|\neq |v|,
\mbox{то }f(u,v)=0\\
\forall\,x\in X\quad
f(u,v)=\sum\limits_{y\in Y}f(u x,v y).
\ey\ee
Запись $R(X,Y)$ обозначает совокупность
всех ВР из $X$ в $Y$.\\

%
%
%

\refstepcounter{theorem}
{\bf Теорема \arabic{theorem}\label{th0121}}.

Для каждого ВА 
$A=(X,Y,S,P, \xi^0)$ 
и каждого $\xi\in S^\bigtriangleup$
$$A^{\xi} \in R(X,Y).$$

{\bf Доказательство}.

Докажем, что  $\forall\,u\in X^*,\forall\,v\in Y^*$
СФ $f\eam A^{\xi}$
удовлетворяет  условию
\re{safdsdfdsafsafarrr}.
\bi
\i Если $|u|\neq |v|$, то $A^{u,v}=0$,
поэтому $A^{\xi}(u,v) = \xi A^{u,v} I=0$,
\i $\forall\,x\in X$
\be{dsvsdfsdr444}\by\sum\limits_{y\in Y}A^{\xi}(u x,v y)=
\sum\limits_{y\in Y}\xi A^{ux,vy} I=\\=
\sum\limits_{y\in Y}\xi A^{u,v} A^{xy} I=
\xi A^{u,v} (\sum\limits_{y\in Y}A^{xy} I)=\\=
\xi A^{u,v} I = A^\xi(u,v)\ey\ee
(в \re{dsvsdfsdr444} используется равенство
\re{dsfdsdsafs333334}). $\blackbox$
\ei

\subsection{Остаточные вероятностные реакции}

Пусть заданы 
\bi\i конечные множества $X$ и $Y$, 
\i ВР $f\in R(X,Y)$,
и 
\i строки $u\in X^*, v\in Y^*$, такие, что $f(u,v)\neq 0$.
\ei

Обозначим записью $f_{u,v}$
функцию вида $f_{u,v}:  X^* \times Y^*\to {\bf R}$,
определяемую следующим образом:
\be{dsfsadfsadfs5556666}
\forall\,u'\in X^*,\forall\,v'\in Y^*\quad
f_{u,v}(u',v')\eam \frac{f(uu',vv')}{f(u,v)}.\ee

\refstepcounter{theorem}
{\bf Теорема \arabic{theorem}\label{th0122}}.

Если $f\in R(X,Y)$ и $f(u,v)\neq 0$, то функция
$f_{u,v}$, определяемая соотношением
\re{dsfsadfsadfs5556666}, является СФ.\\

{\bf Доказательство}.

Поскольку все значения функции $f_{u,v}$
неотрицательны, то 
достаточно доказать, что 
\be{sdfsdfsadffds444}
\forall\,u'\in X^*\quad
\sum\limits_{v'\in Y^*}
f_{u,v}(u',v')=
 1.\ee
\re{sdfsdfsadffds444} эквивалентно соотношению
\be{sdfsdfsadffds4441}
\forall\,u'\in X^*\quad
\sum\limits_{v'\in Y^*}f(uu',vv')=
 f(u,v).\ee

Из предположения $f(u,v)\neq 0$ следует, что $|u|=|v|$.
Поэтому $f(uu',vv') =0$ при $|u'|\neq |v'|$, и, следовательно,
\re{sdfsdfsadffds4441} эквивалентно условию: $\forall\,k\geq 0$
\be{dsfdsdsafs33123}
\forall\,u'\in X^k\quad
\sum\limits_{v'\in Y^k}f(uu',vv')=
 f(u,v).\ee

Докажем \re{dsfdsdsafs33123} индукцией по $k$.
Если $k=0$, то \re{dsfdsdsafs33123}, очевидно, верно.
 Пусть \re{dsfdsdsafs33123} верно для некоторого $k$.
Докажем, что 
\be{dsfdsdsafs33333333221}
\forall\,u'\in X^{k+1}\quad
\sum\limits_{v'\in Y^{k+1}}f(uu',vv') =f(u,v).\ee
\re{dsfdsdsafs33333333221} эквивалентно утверждению:
$\forall\,u'\in X^{k}, \;\forall\,x\in X$
\be{ds332fdsdsafs333332}
\sum\limits_{v'\in Y^{k},\;y\in Y}f(uu'x,vv'y) =f(u,v).\ee

Поскольку $\forall\,x\in X,\;\forall\,u'\in X^{k}, \;
\forall\,v'\in Y^{k}$
$$f(uu',vv')=\sum\limits_{y\in Y}f(uu' x,vv' y)$$
то \re{ds332fdsdsafs333332} можно переписать 
следующим образом:
\be{ds332fdsdsafs33333233}
\by \sum\limits_{v'\in Y^{k},\;y\in Y}f(uu'x,vv'y) =
\sum\limits_{v'\in Y^{k}}
(\sum\limits_{y\in Y}
f(uu'x,vv'y)) =\\=
\sum\limits_{v'\in Y^{k}}
f(uu',vv') =
f(u,v).\ey\ee
Последнее равенство в \re{ds332fdsdsafs33333233}
совпадает с равенством в \re{dsfdsdsafs33123},
и оно верно по индуктивному предположению.
$\blackbox$\\

\refstepcounter{theorem}
{\bf Теорема \arabic{theorem}\label{th013322}}.

Если $f\in R(X,Y)$ и $f(u,v)\neq 0$, то 
$f_{u,v}\in R(X,Y)$.\\

{\bf Доказательство}.

Требуется доказать, что 
$\forall\,u'\in X^*,\forall\,v'\in Y^*$
\be{safdsdfdsafsafarrr33}
\by
\mbox{если }|u'|\neq |v'|,
\mbox{то }f_{u,v}(u',v')=0\\
\forall\,x\in X\quad
f_{u,v}(u',v')=\sum\limits_{y\in Y}f_{u,v}(u' x,v' y).
\ey\ee
Из условия $f(u,v)\neq 0$ следует, что $|u|=|v|$, поэтому,
согласно определению функции $f_{uv}$, 
можно переписать \re{safdsdfdsafsafarrr33} 
следующим образом:
\be{safdsdfdsafsafarrr33333}
\by
\mbox{если }|uu'|\neq |vv'|,
\mbox{то }f(uu',vv')=0\\
\forall\,x\in X\quad
f(uu',vv')=\sum\limits_{y\in Y}f(uu' x,vv' y).
\ey\ee

Первое утверждение  в 
\re{safdsdfdsafsafarrr33333} верно потому, что $f$ -- ВР,
а второе утверждение следует из
доказанного выше соотношения \re{dsfdsdsafs33123}
(в  данном случае $k=1$). $\blackbox$\\

\refstepcounter{theorem}
{\bf Теорема \arabic{theorem}\label{th0133224}}.

Если $f\in R(X,Y)$, $f(u,v)\neq 0$ и $f(uu',vv')\neq 0$, то 
$$f_{uu',vv'}=(f_{u,v})_{u',v'}.$$

{\bf Доказательство}.

$\forall\,u''\in X^*,\forall\,v''\in Y^*$
\bi\i[(\rm{a})] $f_{uu',vv'}(u'',v'')=\frac{f(uu'u'',vv'v'')}{f(uu',vv')}$
\i[(\rm{b})]
$(f_{u,v})_{u',v'}(u'',v'')=\frac{f_{u,v}(u'u'',v'v'')}{f_{u,v}(u',v')}=
\frac{f(uu'u'',vv'v'')/f(u,v)}{f(uu',vv')/f(u,v)}$
\ei
Нетрудно видеть, что правые части в (\rm{a}) и (\rm{b}) совпадают.
 $\blackbox$\\

Если $f\in R(X,Y)$ и  $u,v$ -- строки из $X^*$ и $Y^*$
соответственно, такие, что
$f(u,v)\neq 0$, то 
 $f_{u,v}$ называется {\bf остаточной ВР} для $f$.
 Мы будем обозначать записью 
$S_f$ совокупность всех 
остаточных ВР для $f$.
Отметим, что $f\in S_f$, т.к. $f(\varepsilon, \varepsilon)=1$,
поэтому $f_{\varepsilon, \varepsilon}=f$. 

Обозначим записью $A_f$ пятерку 
$(X,Y,S_f,P_f, \xi_f)$,
где
$P_f$ -- СФ вида 
$$P_f:S_f\times X\vo S_f\times Y,$$ определяемая
следующим образом: 
\be{safsfer5556}
\by
\forall\,g,g'\in S_f, \,\forall\,x\in X,\,
\forall\,y\in Y\\
P_f(g,x,g',y)\eam\left\{\by
g(x,y),&\mbox{если } 
g'=g_{x,y},\\
0,&\mbox{иначе.}
\ey\right.
\ey\ee

\refstepcounter{theorem}
{\bf Теорема \arabic{theorem}\label{th015}}.

Если $f\in R(X,Y)$ и  $|S_f|<\infty$, то 
$A_f$ -- ВА, и 
$f_{A_f}=f$.\\

{\bf Доказательство}.

Из определения СФ $P_f$ следует, что 
для любых
$g\in S_f$,
$x\in X$, 
$y \in Y$, таких, что 
$g(x,y)\neq 0$, верно равенство
\be{asdfsdfdsgfsdgfd}
\xi_g  A_f^{xy}=g(x,y) \xi_{g_{x,y}}.\ee
(напомним, что
  $\xi_g$
 --
 распределение, такое, что
$\forall\,h\in S_f\;$
$h^{\xi_g}= 1$, если $h=g$,
и $h^{\xi_g}=0$, если $h\neq g$).

Докажем, что 
$\forall\,u\in X^*,\;\forall\,v \in Y^*:
f(u,v)\neq 0$
\be{sdfsadfsdfsaf}
\xi_f  A_f^{u,v}=f(u,v) \xi_{f_{u,v}}.\ee
Доказательство будем вести индукцией по длине $u$.

Если $|u|=0$, т.е. $u=v=\varepsilon$,
то обе части \re{sdfsadfsdfsaf} равны $\xi_f$.

Иначе $u$ и $v$ имеют вид 
$u'x$ и $v'y$ соответственно, причём $f(u',v')\neq 0$,
(т.к. если $f(u',v')= 0=\sum\limits_{y'\in Y}f(u'x,v'y')$, 
то $f(u,v)= f(u'x,v'y)= 0$), 
и,
по индуктивному предположению, верно равенство
\be{sd4fsadfsdfsaf}
\xi_f  A_f^{u',v'}=f(u',v') \xi_{f_{u',v'}}.\ee
Используя \re{asdfsdfdsgfsdgfd},
\re{sd4fsadfsdfsaf} и теорему \ref{th0133224},
получаем цепочку равенств
\be{sd444fsadfsdfsaf}
\by\xi_f  A_f^{u,v} =\\= \xi_f  A_f^{u'x,v'y}=
\xi_f  A_f^{u',v'} A_f^{x,y}=\\=
f(u',v') \xi_{f_{u',v'}} A_f^{x,y}=\\=
f(u',v') f_{u',v'}(x,y) \xi_{(f_{u',v'})_{x,y}}=\\=
f(u',v') \frac{f(u'x,v'y)}{f(u',v')}
  \xi_{f_{u'x,v'y}}=\\=
f(u'x,v'y)
  \xi_{f_{u'x,v'y}}=
f(u,v)
  \xi_{f_{u,v}}.  \ey\ee
Таким образом, для любых
$u\in X^*$, 
$v \in Y^*$, таких, что 
$f(u,v)\neq 0$,
верно равенство \re{sdfsadfsdfsaf}, из которого
следует цепочка равенств
$$\by 
A_f^{\xi_f}(u,v)=\xi_f  A_f^{u,v}  I=\\=
f(u,v)
  \xi_{f_{u,v}} I=f(u,v)\cdot 1 = f(u,v).\ey$$

Следовательно, в случае $f(u,v) \neq 0$
верно равенство 
\be{asdfsdfsdfdsf}f_{A_f}(u,v)\eam
A_f^{\xi_f}(u,v) = f(u,v).\ee
Докажем, что \re{asdfsdfsdfdsf}  верно и в случае 
$f(u,v) = 0$. 
\bi\i Если $|u|\neq |v|$, то левая часть \re{asdfsdfsdfdsf}
 равна 0 по определению матриц вида $A^{u,v}$. 
 \i Пусть $|u|=|v|>0$, и 
$u$ и $v$ имеют вид
$x_1\ldots x_n$ и $y_1\ldots y_n$ соответственно.
Существуют  
номер $k\in \{1,\ldots, n\}$ и строки $u',v'$, такие, что
\be{sfsdfdsfgsdgf}\by
u=u'x_k\ldots x_n, \;v=v'y_k\ldots y_n,\\
f(u', v')\neq 0,\;
f(u'x_{k}, v' y_{k})=0.\ey\ee
Имеем цепочку равенств
\be{asdfsdfasd667775}\by
A_f^{\xi_f}(u,v) =\xi_f  A_f^{u',v'} 
 A_f^{x_{k},y_{k}}\ldots
 A_f^{x_n,y_n}
 I
=\\=f(u',v')\xi_{f_{u',v'}}
 A_f^{x_{k},y_{k}}\ldots
 A_f^{x_n,y_n}
 I\ey\ee
Строка $\xi_{f_{u',v'}} A_f^{x_{k},y_{k}}$
является нулевой, поскольку её
элементы 
имеют вид
\be{sadfsadfsdaf}P_f(g,x_k,g',y_k)\ee
где $g=f_{u',v'}$, и,
согласно определению \re{safsfer5556}
СФ $P_f$, 
элемент \re{sadfsadfsdaf} отличен от 0 
если и только если $f_{u',v'}(x_k,y_k)\neq 0$, т.е.
$f(u'x_k,v'y_k)\neq 0$. Учитывая \re{sfsdfdsfgsdgf}, получаем, что 
все 
элементы \re{sadfsadfsdaf} равны 0, т.е. 
$\xi_{f_{u',v'}} A_f^{x_{k},y_{k}}$
является нулевой строкой. 
Таким образом, правая часть в 
\re{asdfsdfasd667775} равна 0, откуда следует, что 
равенство \re{asdfsdfsdfdsf}
в рассматриваемом случае также верно.
$\blackbox$
\ei

\subsection{Реализуемость вероятностных реакций}

ВР $f$ называется {\bf реализуемой}, если 
$\exists$ ВА $A: \;f_A=f$.

Согласно теореме \ref{th015},
если $|S_f|<\infty$, то $f$ реализуема.
Обращение этого утверждения неверно:
согласно нижеследующей теореме,
существует реализуемая ВР $f$, такая, что 
$|S_f|=\infty$. \\

\refstepcounter{theorem}
{\bf Теорема \arabic{theorem}\label{th0115}}.

Пусть $f=f_A$, 
где $A$ -- ВА вида $(X,Y,S,P, \xi^0)$, компоненты которого удовлетворяют условиям:
$$\by |S|=2,\;
\xi^0=(1,0),\; \\ \exists\,x\in X, \exists\,y\in Y:
A^{x,y}=\alpha\left(\begin{array}{ccccc} 1&\beta\cr
0&1\cr\ey\right)\quad(\alpha, \beta\in (0,1)).\ey$$
Тогда $|S_{f}|=\infty$.\\

{\bf Доказательство}.

Если $u_k=\underbrace{x\ldots x}_k$,
$v_k=\underbrace{y\ldots y}_k$, то 
$A^{u_k,v_k}=\alpha^k\left(\begin{array}{ccccc} 1&k\beta\cr
0&1\cr\ey\right)$, поэтому $f(u_k,v_k)=\alpha^k(1+k\beta)$.
$\forall\,k\geq 1$ определена остаточная ВР $f_{u_k,v_k}$,
и нетрудно видеть, что $\forall\,s\geq 1$
$$\by
f_{u_k,v_k}(u_s,v_s)=
\frac{f(u_ku_s,v_kv_s)}{f(u_k,v_k)}=
\frac{\alpha^{k+s}(1+(k+s)\beta)}{\alpha^k(1+k\beta)}=
\frac{\alpha^{s}(1+(k+s)\beta)}{1+k\beta}.
\ey$$
Если для некоторых 
$k_1, k_2\geq 1$ функции 
$f_{u_{k_1},v_{k_1}}$ и
$f_{u_{k_2},v_{k_2}}$ совпадают, то 
$$\by\forall\,s\geq 1\quad 
\frac{\alpha^{s}(1+(k_1+s)\beta)}{1+k_1\beta}=
\frac{\alpha^{s}(1+(k_2+s)\beta)}{1+k_2\beta},\ey$$
откуда следует, что  $k_1=k_2$.
Таким образом, при различных $k$ функции 
$f_{u_k,v_k}$ различны, т.е. $|S_f|=\infty$.
$\blackbox$\\

Пусть $X$ и $Y$ -- конечные множества.
Мы будем использовать следующие определения
и обозначения.

\bi
\i Запись $[0,1]^{X^*\times Y^*}$ обозначает множество
 функций вида $$f:X^*\times Y^*\to [0,1].$$
\i Для каждого $\Gamma\subseteq [0,1]^{X^*\times Y^*}$
{\bf конусом} над 
$\Gamma$ называется подмножество
$C_0(\Gamma)\subseteq [0,1]^{X^*\times Y^*}$, 
состоящее из функций вида $\sum\limits_{i=1}^n
a_if_i$, где \bi\i $\forall\,i=1,\ldots, n\quad a_i\in [0,1],\;
f_i\in\Gamma,\quad
\sum\limits_{i=1}^na_i\leq 1$, и \i $\forall\,(u,v)\in 
X^*\times Y^*\quad\Big(\sum\limits_{i=1}^n
a_if_i\Big)(u,v)\eam \sum\limits_{i=1}^n
a_if_i(u,v)$.\ei
\i $\forall\,x\in X, \forall\,y \in Y$ запись $D^{xy}$
обозначает отображение вида 
$$D^{xy}: [0,1]^{X^*\times Y^*}\to [0,1]^{X^*\times Y^*},$$
называемое {\bf  сдвигом},
сопоставляющее каждой функции $f$ из 
$[0,1]^{X^*\times Y^*}$
функцию, обозначаемую записью $fD^{xy}$, где
\be{sadfsadfdsfrrrr5556}
\forall\,u\in X^*, \forall\,v \in Y^*
\quad
(fD^{xy})(u,v)\eam f(xu,yv).\ee

\i Подмножество 
$\Gamma\subseteq [0,1]^{X^*\times Y^*}$
называется {\bf устойчивым относительно сдвигов}, 
если 
$$\forall\,f\in \Gamma,\;
\forall\,x\in X, \forall\,y\in Y\quad fD^{xy}\in C_0(\Gamma).
$$

\ei

\refstepcounter{theorem}
{\bf Теорема \arabic{theorem}\label{th01315}}.

Пусть $X$ и $Y$ -- конечные множества, и 
$f\in R(X,Y)$. Следущие условия эквивалентны:
\bi
\i $f$ реализуема,
\i $\exists$ конечное  $\Gamma\subseteq 
R(X,Y)$,
устойчивое относительно сдвигов, и
такое, что $f\in C_0(\Gamma)$.
\ei

{\bf Доказательство}.

Пусть $f$ реализуема, т.е. $\exists$ ВА $A=(X,Y,S,P, \xi^0)$:
$$\forall\,u\in X^*, \forall\,v\in Y^*\quad
f(u,v)=\xi^0A^{u,v}I.$$

$\forall\,s\in S$ обозначим записью $A_s$
ВА $(X,Y,S,P, \xi_s)$.
В качестве искомого $\Gamma$ можно взять множество
$\{f_{A_s}\mid s\in S\}$. 

$f\in C_0(\Gamma)$, т.к. $f=\sum\limits_{s\in S}s^{\xi^0}
f_{A_{s}}$, и $\Gamma\subseteq R(X,Y)$ (по теореме \ref{th0121}).

Докажем, что 
$\Gamma$ устойчиво относительно сдвигов, т.е. 
$\forall\,s\in S$, $\forall\,x\in X, \;\forall\,y\in Y$
$f_{A_s}D^{xy}\in C_0(\Gamma)$.
Согласно
\re{sadfsadfdsfrrrr5556},
\be{fdsafrefrefwerfewrf}\by
\forall\,u\in X^*, \forall\,v \in Y^*\\
(f_{A_s}D^{xy})(u,v)= f_{A_s}(xu,yv)=
\xi_sA^{xu,yv}I=
\xi_sA^{xy}A^{u,v}I.\ey\ee
Нетрудно видеть, что 
$$\xi_sA^{xy}A^{u,v}I= \sum\limits_{s'\in S}a_{s'}
f_{A_{s'}}(u,v)$$ где $\forall\,s'\in S\;\; a_{s'}$ -- 
компонента вектор-строки $\xi_sA^{xy}$,
соответствующая состоянию $s'$
(т.е. элемент матрицы $A^{xy}$, находящийся 
в строке $s$ столбце $s'$).
Свойства 
$\forall\,s'\in S\;\; a_{s'}\in [0,1]$ и 
$\sum\limits_{s'\in S}a_{s'}\leq 1$ являются
следствием соответствующих свойств матрицы $A^{xy}$.

Обратно, пусть $f\in C_0(\Gamma)$, где $\Gamma
=\{f_1,\ldots, f_n\}\subseteq R(X,Y)$, 
и $\Gamma$ устойчиво относительно сдвигов.
Определим $A$ как ВА 
\be{sfddsfsafsadgs}A\eam (X,Y,S,P, \xi^0),\ee
компоненты которого имеют следующий вид.
\bi
\i $S\eam \{1,\ldots, n\}$.
\i $\xi^0=(a_1,\ldots, a_n)$,
где $a_1,\ldots, a_n$ -- коэффициенты
представления $f$ в виде суммы
\be{sadfsadfasdfdsa34344367}
f = \sum\limits_{i=1}^na_if_i,\quad \mbox{где }
\forall\,i=1,\ldots, n\;\;a_i\geq 0,\; 
\sum\limits_{i=1}^na_i
\leq 1.\ee
По предположению, $f\in R(X,Y)$, в частности,
 $f(\varepsilon, \varepsilon)=1$,
откуда следует равенство $\sum\limits_{i=1}^na_i=1$,
поэтому $\xi^0\in S^\bigtriangleup$.
\i Поведение $P:S\times X\times S\times Y\to[0,1]$
ВА \re{sfddsfsafsadgs}
определяется матрицами $A^{xy}$ порядка $n$ ($x\in X, y\in Y$):
$$P(i,x,j,y)\eam A^{xy}_{ij},$$ 
где $\forall\,x\in X,\,\forall\,y\in Y,\,\forall\,i=1,\ldots, n$
строка $i$ матрицы $A^{xy}$  состоит из элементов 
$a_{i1},\ldots, a_{in}$ 
представления функции $f_iD^{xy}$ в виде суммы
$\sum\limits_{j=1}^na_{ij}f_j$
$(\forall\,i,j\;\;a_{ij}\geq 0,\;
\sum\limits_{j=1}^na_{ij}
\leq 1$).

Докажем, что $P$ является СФ вида $S\times X\vo S\times Y$.
Данное утверждение эквивалентно соотношению
$\Big(\sum\limits_{y\in Y}A^{xy}\Big)I=I.$

$\forall\,i=1,\ldots, n$ из
\be{sdfsdgsfdg55566}
f_iD^{xy}=\sum\limits_{j=1}^nA^{xy}_{ij}f_j\ee
следует, что 
\be{sdfsdfsdfdsaf}(f_iD^{xy})(\varepsilon, \varepsilon)=
\sum\limits_{j=1}^nA^{xy}_{ij}f_j(\varepsilon, \varepsilon).\ee
Т.к. $\forall\,i=1,\ldots, n\;\;f_j\in R(X,Y)$, то 
$f_j(\varepsilon, \varepsilon)=1$.
Кроме того, левая часть \re{sdfsdfsdfdsaf} равна $f_i(x,y)$.
Поэтому \re{sdfsdfsdfdsaf}
можно переписать 
 в виде
$f_i(x,y)=
\sum\limits_{j=1}^nA^{xy}_{ij}$, откуда следует соотношение
\be{fdgfsdgfsdgsd5544}
\sum\limits_{y\in Y}f_i(x,y)=
\sum\limits_{y\in Y}\sum\limits_{j=1}^nA^{xy}_{ij}.\ee
Т.к. $f_i\in R(X,Y)$, то, согласно 
второму соотношению в \re{safdsdfdsafsafarrr},
левая часть \re{fdgfsdgfsdgsd5544} равна $f_i(\varepsilon,
\varepsilon)$, т.е. равна 1. Учитывая это, и поменяв порядок
суммирования в правой части \re{fdgfsdgfsdgsd5544}, получаем
соотношение
\be{fdgf4sdgfsdgsd5544}
\sum\limits_{j=1}^n\sum\limits_{y\in Y}A^{xy}_{ij}=1.\ee
Нетрудно видеть, что истинность 
\re{fdgf4sdgfsdgsd5544} $\forall\,i=1,\ldots, n$
эквивалентна доказываемому равенству 
$\Big(\sum\limits_{y\in Y}A^{xy}\Big)I=I.$
\ei

Докажем, что реакция ВА \re{sfddsfsafsadgs}
совпадает с $f$, т.е. 
\be{sfdgfsdfdsf555}
\forall\,u\in X^*, \forall\,v\in Y^*\quad
\xi^0A^{u,v}I = f(u,v).\ee
Если $|u|\neq |v|$, то 
левая часть равенства в 
\re{sfdgfsdfdsf555} равна 0 по 
определению матриц вида $A^{u,v}$, и 
правая часть равенства в 
\re{sfdgfsdfdsf555} равна 0 согласно 
предположению $f\in R(X,Y)$ и 
первому соотношению в \re{safdsdfdsafsafarrr}.

Пусть $|u|=|v|$. Докажем (индукцией по $|u|$), что
\be{sadfdsafsadgfsadfsa}A^{u,v}I=\left(\by f_1(u,v)\\\ldots\\f_n(u,v)
\ey\right).\ee

Если $u=v=\varepsilon$, то обе части \re{sadfdsafsadgfsadfsa}
равны $I$.

Если $u=xu'$ и $v=yv'$, то, предполагая верным равенство
\re{sadfdsafsadgfsadfsa}, в котором $u$ и $v$ заменены на $u'$ и $v'$,
имеем:
\be{sdafsadfasd55566}
\by
A^{u,v}I = A^{xu',yv'}I = A^{xy}A^{u',v'}I =A^{xy}
\left(\by f_1(u',v')\\\ldots\\f_n(u',v')
\ey\right)=\\=
\left(\by \sum\limits_{i=1}^n A^{xy}_{1j}f_j(u',v')\\\ldots\\
\sum\limits_{i=1}^n A^{xy}_{nj}f_j(u',v')
\ey\right).
\ey\ee
Из 
\re{sdfsdgsfdg55566} следует, что правую часть 
в \re{sdafsadfasd55566} можно переписать в виде
\be{sdfsadfg2255566}\left(\by (f_1D^{xy})(u',v')\\\ldots\\
(f_nD^{xy})(u',v')
\ey\right).
\ee
Согласно определению \re{sadfsadfdsfrrrr5556}
функций вида $fD^{xy}$, столбец \re{sdfsadfg2255566}
совпадает с правой частью
доказываемого равенства \re{sadfdsafsadgfsadfsa}.

Таким образом, равенство
\re{sadfdsafsadgfsadfsa} доказано. Согласно этому равенству,
левая часть доказываемого равенства
\re{sfdgfsdfdsf555} равна
\be{sdfsdfdsfs336667}\xi^0 \left(\by f_1(u,v)\\\ldots\\f_n(u,v)\ey\right)
=
\sum\limits_{i=1}^n\xi^0_if_i(u,v).\ee

По определению $\xi^0$ (см. \re{sadfsadfasdfdsa34344367}), 
правая часть \re{sdfsdfdsfs336667} равна $f(u,v)$,
т.е.  правой части доказываемого равенства 
\re{sfdgfsdfdsf555}.
$\blackbox$

\section{Случайные последовательности}

\subsection{Понятие случайной последовательности}

Пусть $X$ -- конечное множество.

{\bf Случайной последовательностью
(СП)} над $X$ 
называется функция $\zeta\in [0,1]^{X^*}$, удовлетворяющая
условиям
\be{vdsfdgvfgsdgsdgsderr}
\by\zeta(\varepsilon)=1,\\
\forall\,u\in X^*\quad
\zeta(u)=\sum\limits_{x\in X}\zeta(ux).\ey\ee

Множество всех СП над $X$
будет обозначаться записью $R({\bf 1}, X)$.

Если $\zeta\in R({\bf 1}, X)$, то запись $D_\zeta$
обозначает множество $$\{u\in X^*\mid \zeta(u)\neq 0\}.$$

\subsection{Остаточные случайные последовательности}

Если $\zeta\in R({\bf 1}, X)$, 
и $u$ -- строка из $X^*$, такая, что $\zeta(u)\neq 0$,
то запись $\zeta_u$ обозначает СП из $R({\bf 1}, X)$,
называемую {\bf остаточной СП} для $\zeta$, и
  определяемую 
следующим образом:
$$\by\forall \,u'\in X^*\quad
\zeta_u(u')\;\eam \;\frac{\zeta(uu')}{\zeta(u)}.\ey$$

$\forall\,\zeta\in R({\bf 1}, X)$ запись 
$S_\zeta$ обозначает множество всех остаточных СП для $\zeta$.
Отметим, что $\zeta\in S_\zeta$, т.к. $\zeta=\zeta_\varepsilon$.

Обозначим записью $A_\zeta$ пятерку 
$({\bf 1},X,S_\zeta,P_\zeta, \xi_\zeta)$,
где  {\bf 1} -- одноэлементное множество,
единственный элемент которого будет 
обозначаться символом $e$, и $P_\zeta$ -- СФ вида 
$$P_\zeta:S_\zeta\times {\bf 1}\vo S_\zeta\times X,$$ 
определяемая
следующим образом: 
$\forall\,\zeta',\zeta''\in S_\zeta$,
$\forall\,x\in X$
\be{safsfe3r3325556}P_\zeta(\zeta',e,\zeta'',x)\eam\left\{\by
\zeta'(x),&\mbox{если } 
\zeta''=\zeta'_{x},\\
0,&\mbox{иначе.}
\ey\right.\ee

\refstepcounter{theorem}
{\bf Теорема \arabic{theorem}\label{th30ew15}}.

Если $\zeta\in R({\bf 1},X)$ и  $|S_\zeta|<\infty$, то 
$A_\zeta$ -- ВА, и 
$$\forall\,u\in X^*\;\;f_{A_\zeta}(e^{|u|},u)=\zeta(u)$$
(запись $e^{|u|}$ обозначает строку из $|u|$ символов $e$,
где $\{e\}={\bf 1}$).

{\bf Доказательство}.

Утверждение теоремы следует из того, что 
если строка $u$  имеет вид $x_1\ldots x_k$, где
$x_1,\ldots, x_k\in X$, то 
    $$f_{A_\zeta}(u)=
  \zeta(x_1) \zeta_{x_1}(x_2)
  \ldots \zeta_{x_1\ldots x_{k-1}}(x_k).\quad
\blackbox$$


\refstepcounter{theorem}
{\bf Теорема \arabic{theorem}\label{th30ew13335}}.

Если $\zeta\in R({\bf 1},X)$ и  $|S_\zeta|<\infty$, то 
$\zeta$ полностью определяется своими значениями 
на строках из $X^{\leq 2(|S_\zeta|-|X|+1)}.\;\;\blackbox$

\subsection{Парные случайные последовательности}

Пусть $X,Y$ -- пара конечных множеств.

{\bf Парной случайной последовательностью
(ПСП)} над парой $(X,Y)$ 
называется функция $\eta\in [0,1]^{X^*\times Y^*}$, 
удовлетворяющая условиям:
\be{asdfdsafasdg3545}\by\eta(\varepsilon,\varepsilon)=1,\\
\forall\,u\in X^*,\,\forall\,v\in Y^*\quad\left\{\by
\mbox{если } |u|\neq |v|, \mbox{ то }\eta(u,v)=0,\\
\eta(u,v)=\sum\limits_{x\in X,y\in Y}\eta(ux,vy).\ey\right.
\ey\ee
Запись $R({\bf 1}, X\times Y)$ обозначает
множество всех ПСП над $(X,Y)$.

Если $\eta\in R({\bf 1}, X\times Y)$,
то записи $\eta^X$ и $\eta^Y$ обозначают СП
из $R({\bf 1}, X)$ и $R({\bf 1}, Y)$ соответственно, 
определяемые следующим образом:
\be{fvfvxcbcxvb}
\by \forall\,u\in X^*\quad
\eta^X(u)=\sum\limits_{v\in Y^*}\eta(u,v),\\
\forall\,v\in Y^*\quad
\eta^Y(v)=\sum\limits_{u\in X^*}\eta(u,v).\ey\ee

Если $\eta\in  R({\bf 1}, X\times Y)$, 
и $u,v$ -- строки из $X^*$ и $Y^*$ соответственно,
 такие, что $\eta(u,v)\neq 0$,
то запись $\eta_{u,v}$ обозначает ПСП из 
$R({\bf 1}, X\times Y)$, 
называемую {\bf остаточной ПСП} для $\eta$ и 
 определяемую 
следующим образом:
$$\by\forall \,u'\in X^*,\;\forall\,v'\in Y^*\quad
\eta_{u,v}(u',v')\;\eam \;\frac{\eta(uu',vv')}{\eta(u,v)}.\ey$$

$\forall\,\eta\in R({\bf 1}, X\times Y)$ запись 
$S_\eta$ обозначает множество 
всех остаточных ПСП для $\eta$. Отметим, что 
$\eta\in S_{\eta}$, т.к. $\eta=\eta_{\varepsilon,\varepsilon}$.\\

\refstepcounter{theorem}
{\bf Теорема \arabic{theorem}\label{2t33h4443s}}.

Пусть $X,Y$ -- пара конечных множеств, и 
$\eta\in R({\bf 1}, X\times Y)$. 

Тогда  следующие условия эквивалентны:
\bn
\i $|S_\eta|<\infty$,
\i $|S_{\eta^X}|<\infty$, и существуют 
\bi
\i ВА $A$ вида 
   $(X,Y,S,P, \xi_{s^0})$, 
\i функция $\delta: S\times (X\times Y)\to S$, и
\i совокупность $\{f_s\mid s\in S\}$
функций вида $X\times Y\to [0,1]$,
\ei
удовлетворяющие условиям:
\be{sdfdsdfgfsdgsd}
\by \forall\,s,s'\in S, \,\forall\,x\in X,\,
\forall\,y\in Y\\
P(s,x,s',y)\eam\left\{\by
f_s(x,y),&\mbox{если } 
s'=s(x,y),\\
0,&\mbox{иначе,}
\ey\right.\ey\ee
и
\be{fsdfdsgfsdgfsdg}
\by \forall\,x\in X,\,
\forall\,y\in Y,\,\forall\,u\in X^*, \,\forall\,v\in Y^*\\
\eta(ux,vy)=\eta(u,v)\cdot f_{s^0(u,v)}(x,y),\ey\ee

где $\forall\,s\in S, \,\forall\,x\in X,\,
\forall\,y\in Y,\,\forall\,u\in X^*, \,\forall\,v\in Y^*$
$$\by  s(\varepsilon,\varepsilon)\eam s,\;\;
s(ux,vy)\eam \delta(s(u,v),(x,y)).\quad\blackbox
\ey$$
\en
(отметим, что из \re{sdfdsdfgfsdgsd} и
 \re{fsdfdsgfsdgfsdg} следует равенство  
$f_A=\eta$).

%
%

\subsection{Автоматные преобразования
 случайных последовательностей}

Для любой   СП $\zeta\in R({\bf 1},X)$
 и 
любого ВА $A$ вида $(X,Y,\ldots)$
запись $\eta_{\zeta,A}$
обозначает функцию из $[0,1]^{X^*\times Y^*}$,
определяемую следующим образом:
$\forall\,u\in X^*,\,\forall\,v\in Y^*$
\be{dfasfasdfw4}
\eta_{\zeta,A}(u,v)\eam \zeta(u)f_A(u,v).\ee

\refstepcounter{theorem}
{\bf Теорема \arabic{theorem}\label{th4443303315ee3s}}.

$\forall\,\zeta\in R({\bf 1},X),\;\forall\,$
 ВА $A$ вида $(X,Y,\ldots)$
$$\eta_{\zeta,A}\in R({\bf 1},X\times Y).$$

{\bf Доказательство}.

Докажем, что  
выполнены условия \re{asdfdsafasdg3545} для $\eta\eam
\eta_{\zeta,A}$.

\bi
\i $\eta_{\zeta,A}(\varepsilon, \varepsilon) =  \zeta(\varepsilon)
f_A(\varepsilon,\varepsilon)  = 1\cdot 1=1$.
\i Если $|u|\neq |v|$, то $\eta_{\zeta,A}(u,v)= \zeta(u) f_A(u,v)  = 
\zeta(u)\cdot 0 = 0$.
\i Равенство
$\eta_{\zeta,A}(u,v)=\sum\limits_{x\in X,y\in Y}\eta_{\zeta,A}(ux,vy)$, т.е.
\be{445vdsfdgvfgsdgsdgsderr}
 \zeta(u)f_A(u,v) = \sum\limits_{x\in X,y\in Y}\zeta(ux)f_A(ux,vy) \ee
верно потому, что, согласно 
\re{vdsfdgvfgsdgsdgsderr}, левая часть 
\re{445vdsfdgvfgsdgsdgsderr} равна
\be{dfsdgsfdgsdfgsgfsd45t454}
\!\!\!\! 
\zeta(u)f_A(u,v) = \Big(\sum\limits_{x\in X}\zeta(ux)\Big)f_A(u,v)  = 
\sum\limits_{x\in X}\Big(\zeta(ux) f_A(u,v)  \Big)\ee
и, поскольку $f_A\in R(X,Y)$, то, согласно 
\re{safdsdfdsafsafarrr},
$$
\forall\,x\in X\quad
f_A(u,v)=\sum\limits_{y\in Y}f_A(u x,v y),
$$ поэтому правая часть \re{dfsdgsfdgsdfgsgfsd45t454}
равна
\be{d34fsdgsfdgsdfgsgfsd45t45444}
\sum\limits_{x\in X}\Big( \zeta(ux)f_A(u,v) \Big)=
\sum\limits_{x\in X}\Big(\big(\sum\limits_{y\in Y}\zeta(ux)
f_A(u x,v y)\big) \Big).
\ee
Нетрудно видеть, что правая часть 
\re{d34fsdgsfdgsdfgsgfsd45t45444}
совпадает с правой частью  \re{445vdsfdgvfgsdgsdgsderr}.
$\blackbox$

\ei

%
%
%

С $\eta_{\zeta,A}$ 
связаны две СП, определяемые 
соотношениями 
\re{fvfvxcbcxvb}:
\bi
\i  $\eta_{\zeta,A}^X$, нетрудно видеть, что эта СП 
совпадает с $\zeta$, и 
\i  $\eta_{\zeta,A}^Y$, мы будем 
обозначать эту СП 
 записью $\zeta_A$, и
называть её 
{\bf результатом
преобразования СП $\zeta$ вероятностным автоматом $A$}.\ei

Будем говорить, что СП $\zeta$ и $\zeta'$ {\bf эквивалентны}
(и обозначать это записью $\zeta\sim \zeta'$),
если $\exists$ ВА $A$ и $B$, такие, что $\zeta'=\zeta_A$
и $\zeta=\zeta'_B$. \\


\refstepcounter{theorem}
{\bf Теорема \arabic{theorem}\label{22th4434s}}.

\bn
\i Если СП $\zeta\in  R({\bf 1},X)$ такова, что 
множество $S_\zeta$ конечно,  то $\forall$ ВА $A=(X,\ldots)$
множество $S_{\zeta_A}$ конечно.
\i  Если СП $\zeta$ и $\zeta'$
таковы, что множества
 $S_\zeta$ и $S_{\zeta'}$ конечны, то 
 $\exists$  ВА $A: \zeta'=\zeta_A$.



\i  \label{4th4443s} 
Для любых СП 
$\zeta\in  R({\bf 1},X)$ и $\zeta'\in  R({\bf 1},Y)$
следующие условия эквивалентны:
\bn\i \label{dsfafsaa} $\zeta\sim \zeta'$ 
  \i \label{dsfafsab} 
  $\exists$ детерминированный ВА $A$: $\zeta'=\zeta_A$ и
  его реакция $f_A$ биективно отображает
    $D_{\zeta}$ на $D_{\zeta'}$. 
  \en
\en
  {\bf Доказательство}.
      
Обоснуем лишь импликацию  \ref{dsfafsab} $\Rightarrow$
\ref{dsfafsaa}. Из \ref{dsfafsab} следует 
существование детерминированного ВА $B$, 
такого, что ограничения $f_A$ и $f_B$ на 
    $D_{\zeta}$ и $D_{\zeta'}$ соответственно
    являются взаимно обратными отображениями,
    откуда нетрудно вывести равенство $\zeta'_B=\zeta$:
  $$\by\forall\,u\in X^*\;\;
  \zeta'_{B}(u)=
  \sum\limits_{v\in Y^*}\zeta'(v)f_{B}(v,u)=
  \sum\limits_{v\in D_{\zeta'}}\zeta'(v)f_{B}(v,u)=\\
  \sum\limits_{v\in f_B^{-1}(u)}\zeta'(v)=
  \sum\limits_{v\in f_B^{-1}(u)}\zeta_A(v)=
  \sum\limits_{v\in f_B^{-1}(u)}\sum\limits_{r\in X^*}\eta_{\zeta,A}(r,v)=\\=
  \sum\limits_{v\in f_B^{-1}(u)}\sum\limits_{r\in X^*}\zeta(r)\cdot f_A(r,v)=
  \sum\limits_{v\in f_B^{-1}(u)}\sum\limits_{r\in D_\zeta}\zeta(r)\cdot f_A(r,v)=\\=
   \sum\limits_{v\in f_B^{-1}(u)}\sum\limits_{r\in f_A^{-1}(v)}\zeta(r)=
  \sum\limits_{r\in f_A^{-1}f_B^{-1}(u)
  }\zeta(r)=\zeta(u). \quad\blackbox\ey$$



\subsection{Цепи Маркова}

Один из способов задания СП связан с понятием 
цепи Маркова.


{\bf Цепь Маркова (ЦМ)} -- это пятёрка $M$ вида 
$$M=(X,S,P,\lambda,\xi^0),$$
компоненты которой имеют следующий смысл.
\bn
\i $X$ и $S$ -- конечные множества, элементы
которых называются соответственно 
{\bf  сигналами} и
{\bf состояниями} ЦМ $M$.
\i $P$ -- СФ вида $S\vo S$, 
называемая {\bf функцией перехода}.
$\forall\,(s,s') \in S\times S$
значение $P(s,s')$ понимается как 
   вероятность того, что если  в текущий 
   момент времени $(t)$ 
    $M$ находится в состоянии
   $s$,  то
   в следующий момент времени $(t+1)$
      $M$ будет находиться в состоянии 
     $s'$.
\i $\lambda$ -- функция вида $S\to X$,
$\forall\,s \in S\;
\lambda(s)$ понимается как 
сигнал, который ЦМ $M$ выдаёт
в текущий 
   момент времени, если в этот момент времени
    $M$ находится в состоянии
   $s$.
   \i $\xi^0$  -- распределение на $S$,
называемое
 {\bf начальным распределением}.
$\forall\,s\in S\;  
s^{\xi^0}$ понимается как
вероятность того, что в начальный момент времени $(t=0)$
ЦМ $M$ находится в состоянии $s$.
\en



{\bf Функция ЦМ}  $M=(X,\ldots)$ -- это 
функция $f_M\in [0,1]^{X^*}$, значение которой 
\bi\i на пустой строке равно 1, и 
\i на непустой строке $u=x_0\ldots x_k\in X^*$ 
равно вероятности того, что 
в моменты $0,\ldots, k$
    сигналы, выдаваемые  $M$, 
   совпадают с $x_0,\ldots, x_k$, соответственно.\ei

Пусть  $M=(X,S,P,\lambda,\xi^0)$ -- ЦМ, и
упорядочение множества $S$ её состояний 
имеет вид 
$(s_1,\ldots, s_n)$. 
Мы будем использовать следующие обозначения.
\bi
\i Будем обозначать
тем же символом $M$ матрицу порядка $n$
\be{dsfdsafdsf5566123}
\left(
\begin{array}{ccccc}
P(s_1,s_1)&\ldots&P(s_1,s_n)\\
\ldots&\ldots&\ldots\\
P(s_n,s_1)&\ldots&P(s_n,s_n)
\ey
\right).\ee
\i Для любого $x\in X$
будем обозначать записью $E^x$
квадратную матрицу порядка $n$, каждый элемент которой
равен 0 или 1, и элемент $E^x$ в строке $i$
 и столбце $j$ равен 1 тогда и только тогда, когда
 $i=j$ и $\lambda(s_i)=x$.\ei

\refstepcounter{theorem}
{\bf Теорема \arabic{theorem}\label{22th443344435s}}.
  
Пусть $M$ -- ЦМ вида $(X,S,P,\lambda,\xi^0)$. 
Если строка $u\in X^*$
непуста и имеет вид $x_0\ldots x_k$, то 
\be{adfsdwewewed}
f_M(u)=\xi^0 E^{x_0}M E^{x_1}\ldots M E^{x_k}I,\ee
где $I$ -- 
столбец порядка $n$, все элементы
которого равны 1.\\

{\bf Доказательство}. 

Равенство \re{adfsdwewewed}
следует из правил вычисления вероятностей
несовместных и независимых событий.
 $\blackbox$\\

\refstepcounter{theorem}
{\bf Теорема \arabic{theorem}\label{22th443344433422235s}}.
  
Если $M$ -- ЦМ, то $f_M\in R({\bf 1}, X)$.\\

{\bf Доказательство}. 

Первое соотношение в \re{vdsfdgvfgsdgsdgsderr}
выполняется по определению, а второе
соотношение в \re{vdsfdgvfgsdgsdgsderr}
для $\zeta\eam f_M$ следует из \re{adfsdwewewed},
равенства
$MI=I$ и того, что $\sum\limits_{x\in X}E^x$ является
единичной матрицей. 
$\blackbox$\\

\refstepcounter{theorem}
{\bf Теорема \arabic{theorem}\label{22th443343134s}}.
  
 Пусть $X$ -- конечное множество, и $f\in {[0,1]}^{X^*}$. 
 Тогда следующие утверждения эквивалентны:
 \bn
 \i $\exists$ ЦМ $M: f=f_M$,
\i $\exists$ ВА $A$ вида $({\bf 1}, X,\ldots): f=f_A\circ in$, \\
где $in: X^*\to {\bf 1}^*\times X^*, \forall\,u\in X^*\;\;
in(u)=(e^{|u|}, u)$,
\i $\exists\,\zeta\in R({\bf 1}, Y):\;\forall\,y_1,\ldots,y_n\in Y\quad
   \zeta(y_1\ldots y_n)=\zeta(y_1)\ldots \zeta(y_n)$, и
   $\exists$ детерминированный ВА $A$: $f=\zeta_A$.
   $\blackbox$
\en


\chapter{Вероятностные автоматы Мура
с числовым выходом}

\section{Вероятностные автоматы Мили и Мура}

С каждым ВА $A=(X,Y,S,P,\xi^0)$
связаны СФ 
 $$\delta: S\times X\vo S,\;\;
   \lambda: S\times X\vo Y,$$
 называемые соответственно {\bf функцией перехода}
 и {\bf функцией выхода},
 и определяемые 
следующим образом:
\bi\i
$\forall\,s,s'\in S, 
x\in X\quad
\delta(s,x,s')\eam \sum\limits_{y\in Y}P(s,x,s',y)$,
\i $\forall\,s\in S, 
x\in X, y\in Y\quad
\lambda(s,x,y)\eam \sum\limits_{s'\in S}P(s,x,s',y).$
\ei

ВА $A=(X,Y,S,P,\xi^0)$  называется {\bf ВА Мили}, если
$$\forall\,s,s'\in S, 
x\in X, y\in Y
\quad
P(s,x,s',y) = \delta(s,x,s') \cdot
\lambda(s,x,y).
$$

ВА $A=(X,Y,S,P,\xi^0)$ называется 
{\bf ВА Мура}, если он является ВА Мили, 
и его 
функция выхода $\lambda$ не зависит от $x$
(т.е. имеет вид $\lambda:S\vo Y$).

Пусть $A=(X,Y,S,P,\xi^0)$ -- ВА Мура, 
и упорядочение множества $S$ его состояний 
имеет вид 
$(s_1,\ldots, s_n)$. 
$\forall\,x\in X$ 
мы будем обозначать
записью $A^{x}$ матрицу порядка $n$,
называемую {\bf матрицей перехода}, соответствующей
входному сигналу $x$ и имеющую вид
\be{dsfdsafdsf55661}
\left(
\by
\delta(s_1,x,s_1)&\ldots&\delta(s_1,x,s_n)\\
\ldots&\ldots&\ldots\\
\delta(s_n,x,s_1)&\ldots&\delta(s_n,x,s_n)
\ey
\right),\ee
где $\delta$ -- функция перехода ВА $A$.

$\forall\,x\in X$,
$\forall\,s,s'\in S$ мы будем обозначать
записью 
$A^{x}_{s,s'}$
элемент  
матрицы $A^{x}$, находящийся в строке
$s$ столбце $s'$ (т.е. $A^{x}_{s,s'}=\delta(s,x,s')$).
Из того, что $\delta$ -- СФ, следует, что элементы 
матрицы $A^x$ обладают свойствами
\be{fvfdvdfbvfdxbxgbxfd}\by
\forall\,s,s'\in S\;\;A^{x}_{s,s'}\geq 0,\\
\forall\,s\in S\quad \sum\limits_{s'\in S}A^{x}_{s,s'}=1\;\;
\mbox{(т.е. $A^{x}I=I$)}.\ey\ee
(Матрица, обладающая такими свойствами, называется
{\bf стохастической}.)

$\forall\,u\in X^*$
мы будем обозначать
записью $A^{u}$ 
матрицу
порядка $n$, определяемую
 следующим образом:
$A^{\varepsilon} \eam E$, и
если $u=x_1\ldots x_k$,
то $A^{u} \eam A^{x_1}\ldots A^{x_k}$.

Нетрудно доказать, что 
матрица $A^{u}$ --  стохастическая.

Пусть $s$ -- произвольное состояние из $S$,
и в упорядочении элементов $S$ 
данное состояние имеет
номер $i$ (т.е. $s=s_i$). Будем называть
\bi
\i строку номер $i$ матрицы $A^{u}$ --
{\bf строкой $s$}, и обозначать её записью $\vec A^{u}_{s}$
\i столбец номер $i$ матрицы $A^{u}$ --
{\bf столбцом $s$}, и обозначать его записью ${A^{u}_{s}}^\downarrow$.
\ei

$\forall\,u\in X^*$,
$\forall\,s,s'\in S$ мы будем обозначать
записью 
$A^{u}_{s,s'}$
элемент  
матрицы $A^{u}$, находящийся в строке
$s$ столбце $s'$.

Если строка $u\in X^*$
имеет вид 
$x_0\ldots x_k$, 
то  $A^{u}_{s,s'}$
можно понимать как
   вероятность того, что \bi\i если   в текущий 
   момент  ($t$) 
  ВА  $A$ находится в состоянии
   $s$, и, 
начиная с этого момента,
      на вход $A$ последовательно
   поступают элементы строки $u$
   (т.е. 
      в момент $t$ поступил сигнал $x_0$,
   в момент $t+1$ поступил сигнал $x_1$,
   и т.д.)
\i то
в момент $t+k+1$
      $A$ будет находиться в состоянии 
     $s'$.
\ei


{\bf ВА Мура с детерминированным выходом} -- 
это ВА Мура $(X,Y,S,P,\xi^0)$, функция выходов $\lambda$
которого является детерминированной
(т.е. можно считать, что $\lambda$ имеет вид $S\to Y$).

Если ВА $A=(X,Y,S,P,\xi^0)$ является
ВА Мура с детерминированным выходом, то 
для обозначения такого ВА
мы будем использовать запись
\be{asdfsafgfsgsdfgsdf}
(\xi^0, \{A^x\mid x\in  X\}, \lambda),\ee
компоненты которой определяются следующим образом.
\bi
\i Компонента $\xi^0$ в \re{asdfsafgfsgsdfgsdf}
является вектор-строкой, 
соответствующей начальному распределению $\xi^0$ ВА $A$.
\i Компонента $\{A^x\mid x\in  X\}$ в \re{asdfsafgfsgsdfgsdf}
является совокупностью матриц перехода  ВА $A$.
\i Компонента $\lambda$  в \re{asdfsafgfsgsdfgsdf} является
вектор-столбцом $\left(\by \lambda(s_1)\\\ldots\\\lambda(s_n)
\ey\right)$
значений функции выхода $\lambda:S\to Y$
 ВА $A$
(где $(s_1,\ldots, s_n)$ -- фиксированное 
упорядочение множества $S$.)
\ei

Если $A$ -- ВА Мура с детерминированным выходом,
то запись $S_A$ обозначает множество состояний этого ВА.

\section{Вероятностные автоматы 
Му\-ра с числовым выходом}

\subsection{Понятие вероятностного автомата 
Му\-ра с числовым выходом}

{\bf ВА Мура с числовым выходом} -- это
ВА Мура с детерминированным выходом,
множество выходных сигналов которого
является подмножеством 
множества ${\bf R}$ действительных чисел.

Для ВА Мура с числовым выходом можно определить
понятие реакции, отличное от того понятия реакции
ВА,
которое было определено в пункте \ref{asdfasdfsafa}.
Мы будем называть это понятие {\bf усреднённой реакцией}.

\subsection{Усреднённые реакции}

Пусть $A$ -- ВА Мура с числовым выходом вида
 \re{asdfsafgfsgsdfgsdf},
и $\xi\in S_A^\bigtriangleup$.

{\bf Усреднённая реакция} ВА $A$ в распределении $\xi$ --
это функция
$A^\xi: X^*\to{\bf R},$
определяемая следующим образом:
$$\forall\,u\in X^*\;\;
 A^\xi(u)\eam \xi A^{u}\lambda.$$
Усреднённую реакцию  ВА $A$ в его начальном распределении
мы будем называть просто {\bf  усреднённой реакцией}  ВА $A$,
и  будем обозначать её
записью $ f_{A}$.

Если строка $u\in X^*$ 
имеет вид 
$x_0\ldots x_k$, то  
значение $ A^\xi(u)$
можно понимать следующим образом:
\bi\i если $A$
в некоторый момент времени $t$
имеет распределение $\xi$,
и, начиная с этого момента,
на его вход последовательно
   поступают элементы строки $u$
   (т.е. 
      в момент $t$ поступает сигнал $x_0$,
   в момент $t+1$ поступает сигнал $x_1$,
   и т.д.),
\i то $ A^\xi(u)$ -- это
среднее значение (т.е. математическое ожидание)
выходного сигнала $A$ 
в момент времени $t+k+1$.
\ei

Мы будем говорить, что
распределения $\xi_1,\xi_2 \in S_A^\bullettri$
{\bf эквивалентны по усреднению
относительно $A$} (и обозначать это записью
$\xi_1\dapprox{ A}\xi_2$),
если усреднённые реакции 
$A^{\xi_1}$ и $A^{\xi_2}$ совпадают, т.е.
$$\forall\,u\in X^*\qquad
\xi_1 A^{u} \lambda= \xi_2 A^{u} \lambda.$$

Пусть задана пара $A_1,A_2$
ВА Мура с числовым выходом,
у которых одинаковы
множества входных сигналов, т.е. 
$A_1$ и $A_2$ имеют вид
$$A_i=(\xi^0_i,\{A_i^x\mid x\in X\},\lambda_i)\quad
(i=1,2).$$
Мы будем говорить, что 
$A_1$ и $A_2$ 
{\bf эквивалентны по усреднению}
(и обозначать это записью $A_1\approx A_2$),
если их усреднённые реакции совпадают,
т.е.
$$\forall\,u\in X^*\qquad
\xi^0_1 A_1^{u} \lambda_1= \xi^0_2 A_2^{u} \lambda_2.$$

\subsection{Усреднённые базисные матрицы}

Для ВА Мура с числовым выходом можно ввести 
понятие усреднённой базисной матрицы, 
аналогично тому, как было введено понятие 
базисной матрицы в пункте
\ref{sadfasdfasdgfdgaadas}.

Пусть $A=(\xi^0,\{A^x\mid x\in X\},\lambda)$ -- 
 ВА Мура с числовым выходом.
  Обозначим записью $\hat A$ совокупность
всех  вектор-столбцов  вида $A^{u} \lambda$, где 
   $u\in X^*$.
 
 {\bf Усреднённой базисной матрицей} ВА  $A$
называется  матрица,  обозначаемая
записью $[\![A]\!]$, и удовлетворяющая
условиям:
\bi
\i каждый столбец матрицы $[\![A]\!]$ является элементом
$\hat A$,
\i столбцы матрицы $[\![A]\!]$
   образуют базис линейного пространства $\langle \hat A\rangle$.  
\ei

  Нетрудно видеть, что 
$$
\forall\,\xi_1,\xi_2\in
S_A^\bullettri
\quad
\xi_1\dapprox{A}\xi_2
\quad
\Leftrightarrow
\quad
\xi_1 [\![A]\!] = \xi_2 [\![A]\!].$$

  Матрицу $[\![A]\!]$ можно построить при помощи
 алгоритма, аналогичного соответствующему алгоритму 
из пункта \ref{sadfasdfasdgfdgaadas}.

Для каждого $s\in S_A$ 
 мы будем называть {\bf строкой $s$} матрицы  $[\![A]\!]$
 ту её строку, которая содержит значения вида $\vec A^{u}_{s}
  \lambda$.  Мы будем обозначать эту строку записью $[\![A]\!]_s$.

Мы будем говорить, что 
состояние $s\in S_A$ является {\bf выпуклой комбинацией}
других состояний ВА $A$, если 
строка $s$ матрицы $[\![A]\!]$
является выпуклой комбинацией других строк этой 
матрицы, 
т.е. существует распределение $\xi\in 
(S_A\setminus \{s\})^
\bullettri$,
удовлетворяющее условию
\be{fdsadfsdfdsgfsdg1}
[\![A]\!]_s=\sum\limits_{s'\in S_A
\setminus \{s\}}(s')^\xi [\![A]\!]_{s'}.\ee

\subsection{Редукция  вероятностных автоматов Мура 
с числовым выходом}

Пусть $A$ -- ВА Мура с числовым выходом.

{\bf Редукция} ВА $A$
заключается в построении 
такого ВА Мура с числовым выходом $B$, который 
\bi\i был бы 
эквивалентен  $A$ по усреднению, 
и \i содержал бы меньше состояний, чем $A$
(если это возможно).\ei

К вероятностным автоматам Мура с числовым выходом можно применять те же методы редукции, которые
были изложены в пункте \ref{reddsdsafsafa}.
Мы рассмотрим лишь метод редукции 
путем удаления выпуклых комбинаций. 
Данный метод основан на нижеследующей теореме.\\

\refstepcounter{theorem}
{\bf Теорема \arabic{theorem}\label{th0442211331576}}.

Пусть $A=(\xi^0_A,\{A^x\mid x\in X\},\lambda_A)$ -- ВА  Мура с числовым выходом, и состояние $s\in S_A$
является выпуклой комбинацией других состояний,
т.е. 
$\exists \,\xi\in (S_A\setminus \{s\})^\bigtriangleup$:
верно \re{fdsadfsdfdsgfsdg1}.
Будем считать, что 
упорядочение   $S_A$ имеет вид
$(s_1, \ldots, s_n)$ и $s=s_n$. 

Обозначим символом $B$ ВА  Мура с числовым выходом,
который имеет вид
$(\xi^0_B,\{B^x\mid x\in X\},\lambda_B),$
   где $S_B=S_A\setminus \{s_n\}$, и
\be{fsddsagfasdfgsdgsfdgfsdgsdgfsd}
\by
\xi^0_B= \xi^0_A \left(\begin{array}{ccccccc}E_{n-1}\cr
\xi\ey\right),\\
B^x= (E_{n-1}\;\;{\bf 0}) A^x 
\left(\begin{array}{ccccccc}E_{n-1}\cr
\xi\ey\right),\\
\lambda_B = (E_{n-1}\;\;{\bf 0})\lambda_A,\ey\ee
где $E_{n-1}$ -- единичная матрица
порядка $n-1$,
$\xi=(s_1^\xi,\ldots,s^\xi_{n-1})$,
{\bf 0} -- столбец порядка $n-1$ с нулевыми
компонентами.

Тогда
$A\approx B$.\\

{\bf Доказательство}.

С учетом предположений, изложенных 
в формулировке теоремы, 
можно переписать \re{fdsadfsdfdsgfsdg1}  в виде
\be{gfsdgfsdgsfdgsdg3}
[\![A]\!]_{s_n}=\sum\limits_{i=1}^{n-1}
s_i^\xi [\![A]\!]_{s_i}.\ee

Согласно определению матрицы $[\![A]\!]$,
равенство \re{gfsdgfsdgsfdgsdg3}
равносильно соотношению
\be{sadfsadsdvfsd}\forall\,u\in X^*\quad
A_n^u\lambda_A=
\sum\limits_{i=1}^{n-1}
s_i^\xi A_i^u\lambda_A,
\ee
где $\forall\,i=1,\ldots, n\;\;A_i^u$ обозначает
$i$--ю строку матрицы $A^u$.

Можно переписать \re{sadfsadsdvfsd} в матричном виде:
\be{sadf3sadsdvfsd}\forall\,u\in X^*\quad
\vec e_n A^u\lambda_A=
\xi (E_{n-1}\;\;{\bf 0})A^u
\lambda_A
\ee
(где $\vec e_n$ -- вектор-строка 
длины $n$, 
у которой
$n$--я компонента равна 1, а остальные компоненты равны 0).

В частности, \re{sadf3sadsdvfsd} верно при $u=
 \varepsilon$, т.е.
\be{sadf33sadsdvfsd}
\vec e_n \lambda_A=
\xi (E_{n-1}\;\;{\bf 0})
\lambda_A.
\ee
 
Докажем, что $A\approx B$, т.е. $\forall\,u\in X^*\;
\xi^0_A A^{u} \lambda_A= \xi^0_B B^{u} \lambda_B$.
Согласно \re{fsddsagfasdfgsdgsfdgfsdgsdgfsd}, 
для этого достаточно доказать, что 
$\forall\,u\in X^*$
 \be{dsfsdfsdfgagfsdgsd1}
A^{u} \lambda_A= \left(\begin{array}{ccccccc}E_{n-1}\cr
\xi\ey\right) B^{u} (E_{n-1}\;\;{\bf 0})\lambda_A.\ee

Докажем \re{dsfsdfsdfgagfsdgsd1} индукцией по $|u|$.

Если $u=\varepsilon$, то \re{dsfsdfsdfgagfsdgsd1}
имеет вид
\be{sadfdsgfdgsfdgsd}
\lambda_A=
\left(\begin{array}{ccccccc}E_{n-1}\cr
\xi\ey\right) (E_{n-1}\;\;{\bf 0})\lambda_A.
\ee
Согласно правилам матричного умножения, 
и учитывая \re{sadf33sadsdvfsd},
можно переписать правую часть  
\re{sadfdsgfdgsfdgsd}  следующим образом:
$$\by\left(\begin{array}{ccccccc}E_{n-1}(E_{n-1}\;\;{\bf 0})\lambda_A\cr
\xi(E_{n-1}\;\;{\bf 0})\lambda_A\ey\right)=
\left(\begin{array}{ccccccc}(E_{n-1}\;\;{\bf 0})\lambda_A\cr
\vec e_n \lambda_A\ey\right)=
\lambda_A.\ey$$
Таким образом, в случае $u=\varepsilon$ равенство
\re{dsfsdfsdfgagfsdgsd1} верно.

Пусть  \re{dsfsdfsdfgagfsdgsd1} верно для некоторого $u$.
Докажем, что $\forall\,x\in X$
 \be{dsfsdfsdfgagfsdgsd21}
A^{xu} \lambda_A= \left(\begin{array}{ccccccc}E_{n-1}\cr
\xi\ey\right) B^{xu} (E_{n-1}\;\;{\bf 0})\lambda_A.\ee
Используя определение  $B^x$ из 
\re{fsddsagfasdfgsdgsfdgfsdgsdgfsd},  перепишем
 \re{dsfsdfsdfgagfsdgsd21} в виде
 \be{dsfsdfsdfgagfsdgsd231}
A^{xu} \lambda_A= \left(\begin{array}{ccccccc}E_{n-1}\cr
\xi\ey\right)  (E_{n-1}\;\;{\bf 0}) A^x 
\left(\begin{array}{ccccccc}E_{n-1}\cr
\xi\ey\right)
B^{u} (E_{n-1}\;\;{\bf 0})\lambda_A.\ee

Учитывая индуктивное предположение
\re{dsfsdfsdfgagfsdgsd1}, и используя правила
матричного умножения  перепишем
 \re{dsfsdfsdfgagfsdgsd231} в виде
 \be{3dsfsdfsdfgagfsdgsd231}
\by
A^{xu} \lambda_A= \left(\begin{array}{ccccccc}
E_{n-1}(E_{n-1}\;\;{\bf 0})\cr
\xi(E_{n-1}\;\;{\bf 0})\ey\right)   A^x 
A^{u} \lambda_A=\\=
\left(\begin{array}{ccccccc}
(E_{n-1}\;\;{\bf 0})\cr
\xi(E_{n-1}\;\;{\bf 0})\ey\right)  
A^{xu} \lambda_A=
\left(\begin{array}{ccccccc}
(E_{n-1}\;\;{\bf 0})A^{xu} \lambda_A\cr
\xi(E_{n-1}\;\;{\bf 0})A^{xu} \lambda_A\ey\right).\ey\ee

Используя 
\re{sadf3sadsdvfsd}, перепишем правую часть 
 \re{3dsfsdfsdfgagfsdgsd231} в виде
 \be{23dsfsdfsdfgagfsdgsd231}
\left(\begin{array}{ccccccc}
(E_{n-1}\;\;{\bf 0})A^{xu} \lambda_A\cr
\vec e_n A^{xu}\lambda_A\ey\right)=
\left(\begin{array}{ccccccc}
(E_{n-1}\;\;{\bf 0})\cr
\vec e_n \ey\right)A^{xu} \lambda_A=
A^{xu} \lambda_A.
\ee

Таким образом,  \re{dsfsdfsdfgagfsdgsd21} верно.

Следовательно, \re{dsfsdfsdfgagfsdgsd1} верно 
для любого $u\in X^*$.
$\blackbox$\\

Отметим, что задачу распознавания того, является
 ли какое-либо из состояний ВА Мура с числовым выходом
 выпуклой комбинацией других состояний этого ВА, можно
 решать методом, аналогичным методу,
 изложенному в пункте 
 \ref{sifgawerfasdfa}.
 
 \subsection{Соглашение}
 
 Начиная со следующего пункта и до конца книги,  
все рассматриваемые ВА по умолчанию 
(т.е. если их вид не указан особо) 
предполагаются ВА Мура с числовым выходом.
Мы будем обозначать эти ВА записями вида
\re{asdfsafgfsgsdfgsdf},
и для каждого такого ВА $A$ запись $f_A$
обозначает его усреднённую реакцию
(которую мы будем называть просто {\bf реакцией}).
Те ВА, которые были введены в главе \ref{vaobshchegovida},
мы будем называть {\bf ВА общего вида}.
Для каждого рассматриваемого ВА $A$ запись $S_A$
обозначает множество состояний этого ВА.

\section{Вероятностная реализуемость функций на строках}

Пусть $X$  -- конечное множество.

Функция на строках $f\in {\bf R}^{X^*}$ называется {\bf 
вероятностно реализуемой}, если $\exists$ ВА, 
реакция которого совпадает с $f$.

Будем использовать следующие определения
и обозначения.
\bi
\i Для каждого $\Gamma\subseteq {\bf R}^{X^*}$
{\bf выпуклой оболочкой} множества
$\Gamma$ называется подмножество
$C(\Gamma)\subseteq {\bf R}^{X^*}$, 
состоящее из фун\-кций вида $\sum\limits_{i=1}^n
a_if_i$ (называемых {\bf выпуклыми комбинациями}
фун\-к\-ций $f_1,\ldots, f_n$), где 
\bi\i $\forall\,i=1,\ldots, n\quad a_i\in [0,1],\;
f_i\in\Gamma,\quad
\sum\limits_{i=1}^na_i= 1$, и \i $\forall\,u\in 
X^*\quad\Big(\sum\limits_{i=1}^n
a_if_i\Big)u\eam \sum\limits_{i=1}^n
a_if_i(u)$.\ei
\i $\forall\,x\in X$ запись $D^{x}$
обозначает отображение вида 
$$D^{x}: {\bf R}^{X^*}\to {\bf R}^{X^*},$$
называемое {\bf  сдвигом},
сопоставляющее каждой функции $f$ из 
${\bf R}^{X^*}$
функцию, обозначаемую записью $fD^{x}$, где
\be{1sadfsadfdsfrrrr5556}
\forall\,u\in X^*
\quad
(fD^{x})(u)\eam f(xu).\ee

\i Подмножество 
$\Gamma\subseteq {\bf R}^{X^*}$
называется {\bf устойчивым относительно сдвигов}, 
если 
$$\forall\,f\in \Gamma,\;
\forall\,x\in X\quad fD^{x}\in C(\Gamma).
$$

\ei

\refstepcounter{theorem}
{\bf Теорема \arabic{theorem}\label{th02211315}}.

Пусть $X$  -- конечное множество, и 
$f\in {\bf R}^{X^*}$. Следущие условия эквивалентны:
\bi
\i $f$ вероятностно реализуема,
\i $\exists$ конечное  $\Gamma\subseteq 
{\bf R}^{X^*}$,
устойчивое относительно сдвигов, и
такое, что $f\in C(\Gamma)$.
\ei

{\bf Доказательство}.

Пусть $f$ вероятностно реализуема, т.е. $\exists$ ВА 
$A=(\xi^0,\{A^x\mid x\in X\},\lambda)$:
$$\forall\,u\in X^*\quad
f(u)=\xi^0A^{u}\lambda.$$
Будем считать, что множество состояний $S_A$ этого ВА имеет вид 
$\{1,\ldots, n\}$, и $\forall\, i\in S_A$ запись
$\xi_i$ обозначает распределение из $S_A^\bigtriangleup$, представляемое
вектор-строкой порядка $n$, 
$i$--я компонента которой равна 1, 
а остальные компоненты равны 0.

$\forall\,i=1,\ldots, n$ определим $A_i\eam
(\xi_i,\{A^x\mid x\in X\},\lambda)$.

Полагаем $\Gamma\eam \{ f_{A_i}\mid i=1,\ldots, n\}$. 
$f\in C(\Gamma)$, т.к. $f=\sum\limits_{i=1}^n i^{\xi^0}
 f_{A_{i}}$.

Докажем, что 
$\forall\,i=1,\ldots, n$, $\forall\,x\in X\quad
 f_{A_i}D^{x}\in C(\Gamma)$.

Согласно
\re{1sadfsadfdsfrrrr5556},
\be{1fdsafrefrefwerfewrf}\by
\forall\,u\in X^*\quad
( f_{A_i}D^{x})(u)=  f_{A_i}(xu)=
\xi_iA^{xu}\lambda=
\xi_iA^{x}A^{u}\lambda.\ey\ee
Нетрудно видеть, что 
$$\xi_iA^{x}A^{u}\lambda= \sum\limits_{j=1}^nA^x_{ij}
 f_{A_{j}}(u).$$ 
Таким образом, $f_{A_i}D^{x} = \sum\limits_{j=1}^nA^x_{ij}
 f_{A_{j}}\in C(\Gamma)$.

Обратно, пусть $f\in C(\Gamma)$, где $\Gamma
=\{f_1,\ldots, f_n\}$, 
и $\Gamma$ устойчиво относительно сдвигов.
Определим ВА 
\be{1sfddsfsafsadgs}A\eam (\xi^0,\{A^x\mid x\in X\},\lambda),\ee
где 
\bi\i $\xi^0$ -- вектор-строка 
коэффициентов представления 
$f$ в виде выпуклой комбинации функций из $\Gamma$, т.е.
\be{1sadfsadfasdfdsa34344367}
f = \sum\limits_{i=1}^ni^{\xi^0}f_i,\ee

\i $\forall\,x\in X,\,\forall\,i=1,\ldots, n$
строка $i$ матрицы $A^{x}$  состоит из коэффициентов
представления функции $f_iD^{x}$ в виде выпуклой комбинации
функций из $\Gamma$, т.е. 
\be{sdfsdgfsdfgegerrttr}
f_iD^{x}=\sum\limits_{j=1}^nA^x_{ij}f_j,
\ee
\i $\lambda\eam \left(\by f_1(\varepsilon)\\\ldots\\f_n(\varepsilon)
\ey\right)$.
\ei
Докажем, что реакция ВА \re{1sfddsfsafsadgs}
совпадает с $f$, т.е. 
\be{1sfdgfsdfdsf555}
\forall\,u\in X^*\quad
\xi^0A^{u}\lambda = f(u).\ee

Для этого сначала докажем (индукцией по $|u|$), что
\be{1sadfdsafsadgfsadfsa}A^{u}\lambda=
\left(\by f_1(u)\\\ldots\\f_n(u)
\ey\right).\ee

Если $u=\varepsilon$, то обе части \re{1sadfdsafsadgfsadfsa}
совпадают по определению $\lambda$.

Если $u=xu'$, то, предполагая верным равенство
\re{1sadfdsafsadgfsadfsa}, в котором $u$ заменено на $u'$,
имеем:
\be{1sdafsadfasd55566}
\by
A^{u}\lambda = A^{xu'}\lambda = A^{x}A^{u'}\lambda 
=A^{x}
\left(\by f_1(u')\\\ldots\\f_n(u')
\ey\right)=\\=
\left(\by \sum\limits_{i=1}^n A^{x}_{1j}f_j(u')\\\ldots\\
\sum\limits_{i=1}^n A^{x}_{nj}f_j(u')
\ey\right).
\ey\ee
Из \re{sdfsdgfsdfgegerrttr}
следует, что правую часть 
в \re{1sdafsadfasd55566} можно переписать в виде
\be{1sdfsadfg2255566}\left(\by (f_1D^{x})(u')\\\ldots\\
(f_nD^{x})(u')
\ey\right).
\ee
Согласно определению \re{1sadfsadfdsfrrrr5556}
функций вида $fD^{x}$, столбец \re{1sdfsadfg2255566}
совпадает с правой частью
доказываемого равенства \re{1sadfdsafsadgfsadfsa}.

Таким образом, равенство
\re{1sadfdsafsadgfsadfsa} доказано. Согласно этому равенству,
левая часть доказываемого равенства
\re{1sfdgfsdfdsf555} равна
\be{1sdfsdfdsfs336667}\xi^0 \left(\by f_1(u)\\\ldots\\f_n(u)
\ey\right)
=
\sum\limits_{i=1}^ni^{\xi^0}f_i(u).\ee

По определению $\xi^0$ (см. \re{1sadfsadfasdfdsa34344367}), 
правая часть \re{1sdfsdfdsfs336667} равна $f(u)$,
т.е.  правой части доказываемого равенства 
\re{1sfdgfsdfdsf555}.
$\blackbox$

\section{Связь между линейно-автоматными функциями
и реакциями вероятностных автоматов}

\refstepcounter{theorem}
{\bf Теорема \arabic{theorem}\label{th0131335}}.

Пусть $L=(\xi_L^0, \{L^x\mid x\in  X\}, \lambda_L)$ 
-- ЛА размерности $n$.

Тогда существуют ВА 
$A$ и
 число 
$a>0$, 
такие, что $\forall\,u\in X^*$
\be{dcsdcvzdvzxc}\by
 f_A(u)=a^{|u|+1} f_L(u)+\frac{1}{n+2}.\ey\ee

{\bf Доказательство.}

Если все компоненты $\lambda_L$ равны нулю, то
все значения функции $f_L$ равны нулю, в этом случае
искомый ВА может иметь вид 
$$\by(\vec e_1,\{E\mid x\in X\}, \frac{1}{n+2}e_1^\downarrow),\ey$$
где $\vec e_1$ и
$e_1^\downarrow$ -- 
вектор-строка и вектор-столбец соответственно,
у которых
первая компонента равна 1, а остальные равны 0.

Пусть не все компоненты $\lambda_L$ равны нулю.
Можно считать, что $\lambda_L=e_1^\downarrow$
(а если  $\lambda_L\not=e_1^\downarrow$, то заменим $L$ на 
эквивалентный ему ЛА
$$
\Big(\xi^0_LP, \{P^{-1}L^xP\mid x\in X\},
e_1^\downarrow \Big),$$
где $P$ -- обратимая матрица, первый столбец которой
равен $\lambda_L$).

$\forall\,x\in X$ 
определим $A^x_1$  как 
матрицу порядка $n+2$ вида
$$\left(\by 
&&&a_1&0\cr
&L^x&&\ldots&\ldots\cr
&&&a_n&0\cr
0&\ldots&0&0&0\cr
b_1&\ldots&b_n&b_0&0\cr
\ey\right)$$
в которой \bi\i
левая верхняя  подматрица порядка $n$
совпадает с $L^x$, и \i  компоненты $a_1,\ldots, a_n$
и $b_1,\ldots, b_n, b_0$ выбраны так, что
сумма компонентов в каждой строке и в каждом
столбце матрицы $A^x_1$
равна нулю.\ei

Второе из этих свойств можно выразить в виде
равенств 
\be{dfasdfeferrrrrr}
A^x_1I=0^\downarrow, \quad \tilde I A^x_1=\vec
0,\ee где
$I$ и $\tilde I$ --  
вектор-столбец и вектор-строка порядка $n+2$,
каждый элемент которых равен 1, и 
$0^\downarrow$ и $\vec 0$ -- 
вектор-столбец и вектор-строка порядка $n+2$,
каждый элемент которых равен  0.

Нетрудно видеть, что $\forall\,u\neq\varepsilon$
левая верхняя  подматрица порядка $n$ матрицы $A^u_1$
совпадает с $L^u$, и все компоненты строки $n-1$
и столбца $n$ матрицы $A^u_1$ равны нулю.
Кроме того, будут верны равенства 
\re{dfasdfeferrrrrr}, в которых $x$ заменено на $u$.

Обозначим записью $\xi^0_1$ вектор-строку 
$(\xi^0_L,c,0)$
порядка $n+2$,
в которой
левая подстрока порядка $n$
совпадает с $\xi^0_L$, и число $c$
выбрано так, что
сумма компонентов $\xi^0_1$ равна нулю
(т.е. $\xi^0_1I=0$).

Выберем число $a>0$ так, чтобы модуль каждого элемента
строки $a\xi^0_1$ и матрицы $aA^x_1$ был 
меньше $\frac{1}{n+2}$.

Определим матрицу $B$ 
порядка $n+2$ и вектор-строку $\xi$ порядка $n+2$
следующим образом:
$$\by B\eam
\frac{1}{n+2}\left(\by 1&\ldots&1\cr\ldots&\ldots&\ldots\cr
1&\ldots&1\cr
\ey\right),\quad \xi\eam
\frac{1}{n+2}(1\ldots 1).\ey$$

Нетрудно видеть, что 
\be{asdfsfsadf345345342}
\by
B^2=B,\; BI=I,\; \xi I=1,\; 
\xi B=\xi, \\
A^x_1B=0,\; BA^x_1=0,\;\xi^0_1B=0,\;\xi A^x_1=0,\ey\ee 
где символ $0$ в \re{asdfsfsadf345345342}
обозначает нулевую матрицу или вектор-строку
порядка $n+2$.

Искомый ВА имеет вид
$A=(\xi_A^0, \{A^x\mid x\in  X\}, \lambda_A),$ где
$$\xi_A^0:=a\xi_1^0+\xi,\;\;
\forall\,x\in X\;\;A^x:=aA^x_1+B,\;\;
\lambda_A:= \left(\by \lambda_L\\
0\\
0
\ey\right).$$

Нетрудно видеть, что все элементы вектор-строки 
$\xi_A^0$ и матриц $A^x$ положительны, $\xi_AI=1$,
и $\forall\,x\in  X\;\;A^xI=I$ (т.е. $\xi_A$ -- распределение, и 
$\forall\,x\in  X$ матрица $A^x$
определяет СФ).

Докажем, что верно равенство \re{dcsdcvzdvzxc}.

Если $u=\varepsilon$, то левая часть \re{dcsdcvzdvzxc}
равна $$\by
a\xi^0_L\lambda_L+\xi\lambda_A=
a\xi^0_L\lambda_L+\frac{1}{n+2},\ey$$ что равно правой части \re{dcsdcvzdvzxc}.

Если $u\not=\varepsilon$, то 
из \re{asdfsfsadf345345342} следует, что 
$A^u=a^{|u|}A_1^u+B$, и, используя \re{asdfsfsadf345345342},
получаем:
$$\by
 f_{A}(u)=\xi_A^0A^u \lambda_A=
(a\xi_1^0+\xi)(a^{|u|}A_1^u+B)\lambda_A=\\=
(a^{|u|+1}\xi^0_1 A^u_1+\xi) \lambda_A=
a^{|u|+1} \xi^0_1A^u_1 \lambda_A+\frac{1}{n+2}=\\=
a^{|u|+1} f_L(u)+\frac{1}{n+2}.\;\blackbox\ey$$

\section{Эргодичные  автоматы}


\subsection{Вспомогательные понятия и результаты}

Мы будем использовать следующие обозначения:
\bi
\i если $v=(v_1,\ldots, v_n)\in {\bf R}^n$,
то записи  $|v|$ 
и $|\!|v|\!|$ обозначают
соответственно числа 
$$\by
\max\limits_{i=1\ldots n}|v_i|\quad\mbox{и}\quad
\max\limits_{i=1\ldots n} v_i - \min\limits_{i=1\ldots n} v_i,\ey$$
\i если $A$ -- квадратная матрица порядка $n$, то записи $|A|$ и $|\!|A|\!|$  обозначают соответственно
числа $$\max\limits_{i,j=1\ldots n}|a_{ij}|\quad\mbox{и}\quad
 \max\limits_{i=1\ldots n} |\!|A^\downarrow_i|\!|,$$ 
 где $A^\downarrow_i$ -- $i$--й столбец
 матрицы $A$,

\i если $A$ -- матрица (в частности, вектор-строка 
или вектор-столбец), то запись $A>0$ 
означает, что каждый элемент этой матрицы
положителен,

 \i если $A$ -- квадратная матрица, 
 то запись $Q(A)$ обозначает 
 {\bf булев шаблон} матрицы $A$, т.е.
   матрицу, получаемую из $A$ заменой каждого ненулевого
   элемента на 1, 
 
 \i если $\{A^x\mid x\in X\}$ совокупность матриц
порядка $n$, индексированных элементами множества $X$, то 
\bi\i $\forall\,u\in X^*$
запись $A^{u}$ обозначает матрицу, определяемую
 следующим образом:
$A^{\varepsilon}$ -- единичная матрица порядка $n$,
и если $u=x_1\ldots x_k$,
то $A^{u} \eam A^{x_1}\ldots A^{x_k}$, и 
\i $\forall\,u\in X^*,\forall\,i,j\in \{1,\ldots, n\}$ 
запись
$A^{u}_{ij}$ обозначает 
элемент 
матрицы $A^{u}$ в строке
$i$ и столбце $j$.
\ei

\ei 

Нетрудно доказать, что если $\{A^x\mid x\in X\}$  --
совокупность стохастических матриц одинакового порядка,
то 
$$\forall\,u,v\in X^*\;\;
Q(A^{uv})=Q(A^{u})Q(A^{v}),$$
где произведение булевых
   шаблонов определяется 
   аналогично обычному произведению
  матриц, 
   с единственным отличием: сумма  
   $1+1$ считается равной $1$ (мы будем называть
   такое произведение {\bf булевым произведением}).\\

\refstepcounter{theorem}
{\bf Теорема \arabic{theorem}.\label{da2asdfifraf1}}

Пусть заданы  конечное множество $X$, 
натуральное число $n$, и 
совокупность $\{A^x\mid x\in X\}$
стохастических матриц порядка $n$. 

Следующие условия эквивалентны:
\bi
\i[(\rm{a})] $\lim\limits_{|u|\to\infty}|\!|A^u|\!|=0$,
\i[(\rm{b})] $\exists\,k>0:\forall\,u\in X^k\;
\exists\,i\in\{1,\ldots, n\}:$
${A^{u}_{i}}^\downarrow>0$, 
\i[(\rm{c})] $\exists\,k>0:\forall\,u\in X^k$
\be{adsfsafasdfsadfs}
\forall\,i,i'\in \{1,\ldots, n\}\;\exists \,j:A^u_{ij}>0,A^u_{i'j}>0.\ee
\ei
{\bf Доказательство}.

Схема доказательства: $(\rm{a})\Rightarrow(\rm{b})
\Rightarrow
(\rm{c})\Rightarrow(\rm{a})$.

\bi
\i $(\rm{a})\Rightarrow (\rm{b})$:
выберем $k$ так, что $\forall\,u\in X^k\;|\!|A^u|\!|<\frac{1}{n}$.

Поскольку $\forall\,u\in X^k$ матрица $A^u$ стохастическая,
то в любой её строке существует элемент $\geq \frac{1}{n}$.
Столбец ${A^{u}_{i}}^\downarrow$, в котором содержится этот элемент,
обладает свойством $|\!|{A^{u}_{i}}^\downarrow|\!|<\frac{1}{n}$, поэтому ${A^{u}_{i}}^\downarrow>0$.

\i $(\rm{b})\Rightarrow (\rm{c})$:
очевидно.

%
%
%
   
\i   $(\rm{c})\Rightarrow (\rm{a})$:
пусть верно (\rm{c}), т.е. 
$\exists\,k>0:\forall\,u\in X^k$ верно
\re{adsfsafasdfsadfs}.
Обозначим символом $c$
минимальный положительный элемент матриц вида $A^u$,
где $u\in X^k$. 

\begin{description}
\i[Лемма.]$\;$\\
Для любой матрицы $B$
порядка $n$ верно неравенство
\be{dsfdsagafsdg33}
|\!|A^uB|\!|\leq (1-c)\cdot |\!|B|\!|.\ee

\i[Доказательство.]$\;$\\
Обозначим матрицу $A^uB$
символом $D$, и элементы матриц
$B$ и $D$ --  записями
$b_{ij}$, $d_{ij}\;\;(i,j=1,\ldots, n)$ соответственно.

Для доказательства
неравенства \re{dsfdsagafsdg33}
достаточно доказать, что 
$\forall\,i,i',j\in\{1,\ldots, n\}$ 
верно неравенство
\be{dfdsafsdafa44}
|d_{ij}-d_{i'j}|\leq (1-c)(M_j-m_j),\ee
где $M_j:=\max\limits_tb_{tj}$,
$m_j=\min\limits_tb_{tj}$.

Пусть $i,i'\in\{1,\ldots, n\}$.
Обозначим записью $j_0$ индекс, удовлетворяющий
условию $A^u_{ij_0}\geq c,A^u_{i'j_0}\geq c$.

Верны равенства
$$\by 
d_{ij}=\sum\limits_{t\neq j_0}A^u_{it}b_{tj}+
\Big(A^u_{ij_0}M_j-A^u_{ij_0}(M_j-b_{j_0j})
\Big)\\
d_{i'j}=\sum\limits_{t\neq j_0}A^u_{i't}b_{tj}+
\Big(A^u_{i'j_0}m_j+A^u_{i'j_0}(b_{j_0j}-m_j)
\Big)\ey$$
из которых следуют соотношения
$$\by 
d_{ij}\leq \Big(\sum\limits_{t}A^u_{it}\Big)M_j
-A^u_{ij_0}(M_j-b_{j_0j})=
M_j
-A^u_{ij_0}(M_j-b_{j_0j})\\
d_{i'j}\geq \sum\limits_{t}A^u_{i't}m_j+
A^u_{i'j_0}(b_{j_0j}-m_j)=m_j+
A^u_{i'j_0}(b_{j_0j}-m_j)
\ey$$
из которых следует неравенство
\be{11224fsdfa}d_{ij}- d_{i'j}\leq M_j
-A^u_{ij_0}(M_j-b_{j_0j}) - 
m_j-
A^u_{i'j_0}(b_{j_0j}-m_j).
\ee
Поскольку $M_j-b_{j_0j}\geq 0$
и $b_{j_0j}-m_j\geq 0$, то, используя определение $c$,
можно оценить сверху 
правую часть \re{11224fsdfa}
значением
$$M_j
-c(M_j-b_{j_0j}) - 
m_j-
c(b_{j_0j}-m_j)=(1-c)(M_j-m_j).$$
Таким образом, доказано неравенство
\be{fasdfas3dfsaf}d_{ij}- d_{i'j} \leq (1-c)(M_j-m_j).\ee
В силу произвольности выбора индексов $i,i'$
верно неравенство  
\be{fasdfasd44fsaf}d_{i'j}- d_{ij} \leq (1-c)(M_j-m_j).\ee
Из \re{fasdfas3dfsaf} и \re{fasdfasd44fsaf}
следует  \re{dfdsafsdafa44}.
$\blackbox$
\end{description}

Из леммы следует, что 
\be{adsfsadfsadfsadfsa}
\forall\,u\in X^*\;|\!|A^u|\!|\leq (1-c)^{[|u|/k]}\cdot |\!|A^{l}|\!|\quad
(|l|<k).\ee
Правая часть \re{adsfsadfsadfsadfsa}
стремится к нулю при $|u|\to \infty$, поэтому 
условие (a) верно. $\blackbox$
\ei

\subsection{Понятие эргодичного вероятностного автомата и 
критерий эргодичности}
\label{ergodich}

ВА 
 $A=(\xi^0,\{A^x\mid x\in X\},\lambda)$ 
  называется
{\bf эргодичным}, если 
\be{dfasdfsadfdsagfsd}\forall\,s,s'\in S_A
\quad |\vec A_{s}^u-
\vec A_{s'}^u|\to 0\quad(|u|\to\infty).\ee

\refstepcounter{theorem}
{\bf Теорема \arabic{theorem}\label{th04432331c431}}.

ВА $A=(\xi^0,\{A^x\mid x\in X\},\lambda)$ 
эргодичен тогда и только тогда, когда 
$\forall\,u\in X^*\setminus \{\varepsilon\}$
матрица $A^u$ {\bf регулярна} 
(т.е. $\exists\,n:\;(A^u)^n>0$).\\

{\bf Доказательство.}

Если $A$ эргодичен, но для некоторого 
$u\in X^*\setminus \{\varepsilon\}$ матрица  
$A^u$ нерегулярна, то 
$\forall\,k\geq 1$  
$(A^u)^k$  нерегулярна.
Из  эргодичности $A$ следует, что 
выполнено условие  (a)
в теореме \ref{da2asdfifraf1}
($\lim\limits_{|u|\to\infty}|\!|A^u|\!|=0$).
Поэтому выполнено условие (b)
в этой теореме, т.е.
$$\exists\,l>0:\forall\,u\in X^l\;
\exists\,i:\;
{A^{u}_{i}}^\downarrow>0.$$ 
Можно доказать, что 
это противоречит нерегулярности матриц 
вида $(A^u)^k$, где 
$k$ -- произвольное натуральное число.

Обратно, пусть $\forall\,u\in X^*\setminus \{\varepsilon\}$
 $\exists\,n:\; (A^u)^n>0$.

Мы будем использовать следующие обозначения.
\bi
\i Запись ${\cal Q}_A$ обозначает множество 
  $\{Q(A^u)\mid u\in X^*\}$ булевых шаблонов матриц вида
   $A^u$. 
   
   Нетрудно видеть, что множество ${\cal Q}_A$
   конечно.

\i Символ ${\cal Q}_A^I$ обозначает
  множество всех матриц из ${\cal Q}_A$, 
  содержащие столбец  $I$ (все его элементы равны 1).
\i Символ $k$ обозначает число различных 
  матриц из ${\cal Q}_A\setminus {\cal Q}_A^I$.
\ei

Отметим, что 
\be{vfvdsfvfsdsdvfd}
P\in  {\cal Q}_A^I\;\Rightarrow\;
\forall\,Q\in {\cal Q}_A\;\;QP\in {\cal Q}_A^I.\ee

Докажем, что 
\be{dfdsfasdfdsafda333}
\forall\,u\in X^{k+1}\;\;Q(A^{u})\in {\cal Q}_A^I.\ee

Пусть $u=(x_1\ldots x_{k+1})$ 
и $Q(A^{u})\not\in {\cal Q}_A^I$.  
Из \re{vfvdsfvfsdsdvfd} следует, что 
\be{zfdsdgvfdgvfd}\forall\,i=1,\ldots, k+1\quad
Q(A^{x_{1}})\cdot\ldots \cdot Q(A^{x_{k+1}})\in {\cal Q}_A\setminus {\cal Q}_A^I.
\ee
Поскольку $|{\cal Q}_A\setminus {\cal Q}_A^I|=k$,
то из \re{zfdsdgvfdgvfd} следует, что
\be{dsfsadf4444222333}\by
\exists\,i,j\in \{1,\ldots, k+1\}:i< j,\\
Q(A^{x_{i}})\cdot\ldots \cdot Q(A^{x_{k+1}})=
Q(A^{x_{j}})\cdot\ldots \cdot Q(A^{x_{k+1}})\not\in {\cal Q}_A^I.\ey\ee

Определим $v\eam (x_i\ldots x_{j-1}),\;w\eam (x_j\ldots x_{k+1})$.
Из \re{dsfsadf4444222333} следует, что
\be{afsdfasdsew}\forall\,t\geq 1\quad
Q((A^v)^t A^w)=Q(A^w)\not\in {\cal Q}_A^I.\ee
Поскольку 
\bi\i из регулярности $A^v$ следует, что $\exists\,t\geq 1:
(A^v)^t>0$, и 
\i из регулярности $A^w$ следует, что $\exists\,s:$
${A^w_s}^\downarrow\neq {\bf 0}^\downarrow$,
\ei
то $(A^v)^t{A^w_s}^\downarrow>0$,
т.е. $Q((A^v)^t A^w)\in {\cal Q}_A^I$,
что противоречит \re{afsdfasdsew}. 

Таким образом,  свойство \re{dfdsfasdfdsafda333} верно.
Это свойство совпадает с
условием (b) теоремы \ref{da2asdfifraf1}
для ВА $A$. Следовательно, верно условие 
(a) этой теоремы,  
откуда следует эргодичность $A$.
$\blackbox$

\section{Устойчивость вероятностных автоматов}

\subsection{Вспомогательные утверждения}


%
%

Мы будем использовать следующие
обозначения: если $A$ -- матрица, и $c\in {\bf R}$, то записи $A>c$ 
и $A\geq c$ означают, что каждый элемент этой матрицы
$>c$ или $\geq c$ соответственно. \\

\refstepcounter{theorem}
{\bf Теорема \arabic{theorem}\label{dasfsadfasdfasdfsaf}}.

Если $A$  -- стохастическая матрица порядка $n$, $A\geq c\geq 0$,
и $\lambda\in {\bf R}^n$ -- 
вектор-столбец,  
то $$|\!|A\lambda|\!|\leq (1-2c)|\!|\lambda|\!|.$$ 

{\bf Доказательство}.

Обозначим элементы
матрицы $A$
и вектор-столбца $\lambda$  
записями $a_{ij}$ и
$\lambda_i$
$(i,j=1,\ldots, n)$ соответственно.
Кроме того, 
обозначим записью $b=(b_1\ldots b_n)^\sim$
вектор-столбец
$A\lambda$. 
Будем считать, что
 $$b_1=\max b_i, \quad b_2=\min b_i,\quad 
\lambda_1=\max \lambda_i,\quad 
\lambda_2=\min b_i$$
(если это не так, то соответствующим образом
переставим в матрице $A$
 столбцы и строки).

Надо доказать, что $b_1-b_2\leq
(1-2c)(\lambda_1-\lambda_2)$.

Мы докажем более сильное неравенство:
$$b_1-b_2\leq
(1-a_{12}-a_{21})(\lambda_1-\lambda_2).$$

Нетрудно видеть, что верны следующие соотношения:
\bi\i
$b_1=\sum\limits_j a_{1j}\lambda_j=(1-\sum\limits_{j\geq 2} a_{1j})\lambda_1+\sum\limits_{j\geq 2} a_{1j}
\lambda_j =\\=\lambda_1-\sum\limits_{j\geq 2}a_{1j}(\lambda_1-\lambda_j)\leq \lambda_1-a_{12}(\lambda_1-\lambda_2)$,
\i
$b_2=\sum\limits_j a_{2j}\lambda_j=a_{21}\lambda_1+
(1-a_{21}-\sum\limits_{j\geq 3} a_{2j})\lambda_2+
\sum\limits_{j\geq 3} a_{2j}\lambda_j=\\=
\lambda_2+a_{21}(\lambda_1-\lambda_2)+
\sum\limits_{j\geq 3}a_{2j}(\lambda_j-\lambda_2)\geq
\lambda_2+a_{21}(\lambda_1-\lambda_2)$.
\ei

Таким образом,
$$\by
b_1-b_2\leq \lambda_1-a_{12}(\lambda_1-\lambda_2)
-\lambda_2-a_{21}(\lambda_1-\lambda_2)=\\=
(\lambda_1-\lambda_2)-(a_{12}+a_{21})(\lambda_1-\lambda_2)=
(\lambda_1-\lambda_2)(1-a_{12}-a_{21}).\;\;
\blackbox\ey$$

\refstepcounter{theorem}
{\bf Теорема \arabic{theorem}\label{dasfsadfasdfasd3334fsaf}}.

Если $\exists\,c>0$:
$\forall\,x\in X$ матрица $A^{x}$ удовлетворяет неравенству
$A^{x}\geq c$ и является
стохастической, то $\forall\,u\in X^*$
\be{adfsadfdsafsafs}
|\!|A^u|\!|\leq(1-2 c)^{|u|-1}.\ee

{\bf Доказательство}.

Обозначим символом $n$  порядок матриц $A^x$.

Сначала докажем, что $\forall\,u\in X^*$ 
верно соотношение
\be{sdfdsafsafads44}\forall\,\lambda\in {\bf R}^n\quad
|\!|A^u\lambda|\!|\leq(1-2 c)^{|u|}|\!|\lambda|\!|.\ee

\re{sdfdsafsafads44} верно для $u=\varepsilon$.

Если \re{sdfdsafsafads44} верно для некоторого $u\in X^*$,
то $\forall\,x\in X$, используя предположение 
\re{sdfdsafsafads44} и 
теорему 
\ref{dasfsadfasdfasdfsaf}, получаем:
$$\by
|\!|A^{ux}\lambda|\!|=|\!|A^{u}A^{x}\lambda|\!|\leq
(1-2 c)^{|u|}|\!|A^x \lambda|\!|\leq\\\leq
(1-2 c)^{|u|}(1-2c) |\!|\lambda|\!|=
(1-2 c)^{|ux|}|\!|\lambda|\!|.\ey$$
Таким образом, \re{sdfdsafsafads44}  верно для всех 
$u\in X^*$.

Теперь
докажем \re{adfsadfdsafsafs} индукцией по $|u|$.
\bi
\i Если $|u|=0$ или 1, то \re{adfsadfdsafsafs} верно.
\i Пусть \re{adfsadfdsafsafs} верно для некоторой строки
$u\in X^*$,
и пусть 
$x\in X$. 

Обозначим совокупность столбцов 
$A^x$ записью $(\lambda_1\ldots \lambda_n)$. 

Используя \re{sdfdsafsafads44}
и свойство $\forall\,i=1,\ldots, n\;|\!|\lambda_i|\!|\leq 1$,
получаем:
$$\by
|\!|A^{ux}|\!|=|\!|A^{u}A^{x}|\!|=
|\!|A^{u}(\lambda_1\ldots \lambda_n)|\!|=
|\!|(A^{u}\lambda_1\ldots A^{u}\lambda_n)|\!|=\\=
\max\limits_i|\!|A^{u}\lambda_i|\!|\leq
(1-2c)^{|u|}\max\limits_i|\!|\lambda_i|\!|\leq 
(1-2c)^{|u|}=\\=
(1-2c)^{|ux|-1}.
\ey$$
\ei
Таким образом, \re{adfsadfdsafsafs} верно для любой строки
$u\in X^*.\quad \blackbox$\\

\refstepcounter{theorem}
{\bf Теорема \arabic{theorem}\label{da2asdfa554sd3334fsaf}}.

Если $A$  -- стохастическая матрица
 порядка $n$, то 
для любой матрицы $B$ порядка $n$ верно 
неравенство $$|AB-B|\leq |\!|B|\!|.$$

{\bf Доказательство}.

Пусть представление матрицы $B$  в виде 
последовательности столбцов 
имеет вид $(B_1^\downarrow\ldots
B_n^\downarrow)$, тогда 
$$\by
|AB-B|=|A(B_1^\downarrow\ldots
B_n^\downarrow)-(B_1^\downarrow\ldots
B_n^\downarrow)|=\\=
|(AB^\downarrow_1-B^\downarrow_1\ldots AB^\downarrow_n-B^\downarrow_n)|=\\=
\max\limits_i|AB^\downarrow_i-B^\downarrow_i|\leq
\max\limits_i|\!|B^\downarrow_i|\!|= |\!|B|\!|,
\ey$$
где последнее неравенство следует из нижеслежующей леммы.

\begin{description}
\i[Лемма.]$\;$\\ 
Если $A$  -- стохастическая матрица порядка $n$,
и $\lambda\in {\bf R}^n$  -- 
ве\-к\-тор-\-сто\-л\-бец, то 
$|A\lambda - \lambda|
\leq |\!|\lambda|\!|.$

\i[Доказательство.]$\;$\\
Пусть элементы  $A$
и $\lambda$  
имеют вид $a_{ij}$ и
$\lambda_i$
$(i,j=1,\ldots, n)$.
Поскольку $\forall\,i=1,\ldots, n\;\;\sum\limits_{j}a_{ij}=1$,
то 
$$\by
|A\lambda - \lambda|=
\max\limits_{i}|\sum\limits_{j}(a_{ij}\lambda_j)
-\lambda_i|=\\=
\max\limits_{i}|\sum\limits_{j}(a_{ij}\lambda_j)
-\sum\limits_{j}(a_{ij}\lambda_i)|=
\max\limits_{i}|\sum\limits_{j}a_{ij}(\lambda_j-
\lambda_i)|\leq\\\leq
\max\limits_{i}\sum\limits_{j}a_{ij}|\lambda_j-
\lambda_i|\leq  \max\limits_{i}\sum\limits_{j}a_{ij}|\!|\lambda|\!|=
|\!|\lambda|\!|.\;\blackbox\;
\ey$$
\end{description}

\refstepcounter{theorem}
{\bf Теорема \arabic{theorem}.\label{da2asdfifraf}}

Пусть $P,Q$ -- стохастические матрицы порядка $n$,
тогда для любой матрицы $B$ порядка $n$
верно неравенство
\be{dfsasdfsaaeee3}|PB-QB|\leq |\!|B|\!|.\ee

{\bf Доказательство}.

Обозначим матрицу $P-Q$ символом $A$,
и элементы матриц $A$, $P$, $Q$ -- записями
$a_{ij}$, $p_{ij}$, $q_{ij}\;\;(i,j=1,\ldots, n)$ соответственно.

Матрица $A$
удовлетворяет условию: 
\be{adsfsdfdsags4}\forall\,i=1,\ldots, n\quad
\sum\limits_{j=1}^n a_{ij}=0,\quad
\sum\limits_{j=1}^n a^{(+)}_{ij}\leq 1,
\ee
где $a^{(+)}_{ij}\eam \left\{\by
a_{ij},&\mbox{если }a_{ij}\geq 0\\
0,&\mbox{иначе}
\ey\right.$,
т.к. 
$\forall\,i=1,\ldots, n$
\bi
\i $\sum\limits_j a_{ij}=
\sum\limits_j (p_{ij}-q_{ij})=\sum\limits_j p_{ij} -
\sum\limits_j q_{ij} = 1-1=0$,
\i если $a_{ij_1},\ldots,a_{ij_k}$ -- список всех неотрицательных
элементов $i$--й строки матрицы $A$, то
$$\by
\sum\limits_{s} a_{ij_s}=
\sum\limits_{s} (p_{ij_s}-q_{ij_s})=
\sum\limits_{s} p_{ij_s} - \sum\limits_{s} q_{ij_s}\leq
1-\sum\limits_{s} q_{ij_s}\leq 1.\ey$$
\ei

Перепишем доказываемое неравенство \re{dfsasdfsaaeee3} в виде
\be{adsfasdfasdfs}|AB|\leq |\!|B|\!|.\ee

Обозначим матрицу $AB$ символом $C$,
и элементы матриц $A$, $B$, $C$ -- записями
$a_{ij}$, $b_{ij}$, $c_{ij}\;\;(i,j=1,\ldots, n)$ соответственно.

Пусть элемент $c_{kr}$ матрицы $C$ таков, что 
\be{safdgsdgfsd555}
|c_{kr}| = \max\limits_{i,j} |c_{ij}|\quad(=|AB|\,).\ee

Определим $M\eam \max\limits_j b_{jr}$, $m
\eam \min\limits_j b_{jr}$, 
 и $B^\downarrow_r\eam\;r$--й столбец
матрицы $B$. Из этого определения следует, что 
\be{sdfdsfdsgsdf}M-m=|\!|B_r^\downarrow|\!|\leq |\!|B|\!|.\ee

Из \re{safdgsdgfsd555}
и \re{sdfdsfdsgsdf} следует, что 
для доказательства неравенства
\re{adsfasdfasdfs} достаточно доказать неравенство
\be{dfasdgfsdgsdgfsd55}
|c_{kr}|\leq M-m.
\ee

   Обозначим записью $a^{(+)}$ ($a^{(-)}$) сумму 
   всех положительных 
   (отрицательных) чисел вида 
   $a_{kj}\;(j=1,\ldots, n)$. Заметим, что  
   \bi
   \i из 
   $\sum\limits_{j=1}^n a_{kj}=0$ следует 
   $a^{(+)}+a^{(-)}=0$,
   или $a^{(-)} = -a^{(+)}$,
   \i из  $\sum\limits_{j=1}^n a^{(+)}_{kj}\leq 1$ следует 
   $a^{(+)}\leq 1$.
   \ei
   
Для доказательства неравенства \re{dfasdgfsdgsdgfsd55}
рассмотрим отдельно случаи $c_{kr}\geq 0$
и $c_{kr}< 0$.
\bi\i
Если $c_{kr}\geq 0$,
то 
   $$\by |c_{kr}| = 
   c_{kr}=\sum\limits_j a_{kj}b_{jr}
   \leq a^{(+)}M+a^{(-)}m=\\=a^{(+)}(M-m)\leq M-m,\ey$$
   т.е. в данном случае неравенство
   \re{dfasdgfsdgsdgfsd55} верно.
\i Если $c_{kr}< 0$, то 
$$c_{kr}=\sum\limits_j a_{kj}b_{jr}\geq a^{(+)} m + 
a^{(-)}M =a^{(+)}(m-M),$$
поэтому $|c_{kr}| = - c_{kr} = a^{(+)}(M-m) \leq M-m$,
т.е. в данном случае неравенство
   \re{dfasdgfsdgsdgfsd55} также верно. $\blackbox$
\ei

\subsection{Понятие устойчивости вероятностных автоматов}

Пусть заданы ВА $A=(\xi^0,\{A^x\mid x\in X\},\lambda)$
и число $\varepsilon>0$.

Запись $O_\varepsilon(A)$ обозначает множество ВА, каждый из
которых получается из $A$ путем прибавления 
к элементам строки 
$\xi^0$ и матриц $A^x$  чисел из интервала 
$(-\varepsilon, \varepsilon)$.

ВА $A$ называется
{\bf устойчивым} относительно ИТС $a\in I(A)$,
если 
$\exists\,\varepsilon>0:
\forall\,B\in O_\varepsilon(A)\quad a\in I(B)$ и $A_a=B_a$.
\\

\refstepcounter{theorem}
{\bf Теорема \arabic{theorem}\label{th04433d2d733225t432}}.

Пусть задан ВА $A=(\xi_A^0,\{A^x\mid x\in X\},\lambda)$,
и 
$\forall\,x\in X\;\;A^x>0$.

Тогда $\forall\,a\in I(A)\;\;
A$ устойчив относительно $a$.\\

{\bf Доказательство}.

По предположению, $\exists\,\delta>0$: $\forall\,u\in X^*$
\be{dfdsafdsfdsfgfsde33}
|f_A(u)-a|>\delta.\ee 

Докажем, что $\exists\,\varepsilon>0$:
$\forall\,B
\in O_{\varepsilon}(A),\;\forall\,u\in X^*$ 
\be{dfasdfasdfsa}
\by 
|f_A(u)-f_B(u)|\leq \frac{\delta}{2}.
\ey\ee
Отметим, что из 
\re{dfasdfasdfsa} следуют  требуемые
 соотношения $a\in I(B)$ и $A_a=B_a$, т.к. 
\bi
\i $a\in I(B)$, верно потому, что 
$\forall\,u\in X^*$ из неравенства 
$$|f_A(u)-a|\leq |f_A(u)-f_B(u)|+|f_B(u)-a|$$
 согласно \re{dfasdfasdfsa} и \re{dfdsafdsfdsfgfsde33}
следуют соотношения
\be{dfdsafdsfdsfgfsde3333}\by
|f_B(u)-a|\geq |f_A(u)-a| - |f_A(u)-f_B(u)| > \\ >
\delta-\frac{\delta}{2}=\frac{\delta}{2},\ey\ee

\i $A_a=B_a$ верно потому, что 
если $\exists\,u\in X^*$:
 соотношение   $$f_A(u)>a\;\Leftrightarrow\;
f_B(u)>a$$ неверно, то, согласно 
\re{dfdsafdsfdsfgfsde33} и \re{dfdsafdsfdsfgfsde3333},
будет верно неравенство
$$|f_A(u)-f_B(u)|>\delta,$$ 
которое противоречит \re{dfasdfasdfsa}.
\ei

Пусть $B$ имеет вид
$(\xi_B^0,\{B^x\mid x\in X\},\lambda)$, тогда можно 
переписать 
\re{dfasdfasdfsa}  в виде
\be{dfe4asdfasdfs3345a}
\by 
|\xi_A^0 A^u\lambda-\xi_B^0 B^u\lambda|\leq \frac{\delta}{2}.
\ey\ee
Поскольку
$$\by 
|\xi_A^0 A^u\lambda-\xi_B^0 B^u\lambda|\leq \\\leq
|\xi_A^0 A^u\lambda-\xi_A^0 B^u\lambda|
+|\xi_A^0 B^u\lambda-
\xi_B^0 B^u\lambda|=\\=
|\xi_A^0 (A^u-B^u)\lambda|
+|(\xi_A^0-\xi_B^0) B^u\lambda|\leq\\\leq
|A^u-B^u|\cdot n \cdot |\lambda|+|\xi_A^0-\xi_B^0|
\cdot n\cdot  |\lambda|
\ey$$
(где $n=|S_A|$),
то, следовательно, неравенство \re{dfe4asdfasdfs3345a}
 будет верно, если будут верны неравенства
 \be{sfgsdfgfsd5555}\by
|A^u-B^u|\leq \delta_1,\quad
|\xi_A^0-\xi_B^0|\leq \delta_1,\quad
\mbox{где } \delta_1=\frac{\delta}{4n|\lambda|}.
\ey\ee

Таким образом, для  доказательства теоремы \re{th04433d2d733225t432} 
достаточно доказать, что $\exists\,\varepsilon\in(0,\delta_1)$:
$\forall\,B
\in O_{\varepsilon}(A),\;\forall\,u\in X^*$ 
\be{asdfasdf555334343}
\by|A^{u}-B^{u}|<\delta_1.\ey\ee

По предположению, $\exists\,c>0$:
$\forall\,x\in X\;\; A^x\geq c$.

Выберем 
$k$ так, чтобы было верно неравенство
\be{dafdsafsarr44eee4}\by
(1-c)^{k-1}<\frac{\delta_1}{3}.\ey\ee

\begin{description}
\i[Лемма.]$\;$\\
Пусть 
$\varepsilon\in (0,\frac{c}{2})$,
$B
\in O_{\varepsilon}(A)$,
и   
\be{asdfasdf55543}
\by\forall\,u\in X^{\leq k}\quad
|A^{u}-B^{u}|<\frac{\delta_1}{3}.\ey\ee

Тогда  $\forall\,u\in X^*$ верно неравенство
\re{asdfasdf555334343}.

\i[Доказательство.]$\;$\\
Если $|u|\leq k$, то \re{asdfasdf555334343}  следует из
\re{asdfasdf55543}.

Пусть $|u|>k$, тогда $u=u_1u_2$, где 
$|u_2|=k$, и
\be{dsafasdfdfsads44}
\by
|A^u-B^u|=|A^{u_1u_2}-B^{u_1u_2}|\leq \\\leq 
|A^{u_1}A^{u_2}-A^{u_2}|+
|B^{u_1}B^{u_2}-B^{u_2}|+
|A^{u_2}-B^{u_2}|.\ey\ee

Согласно теореме \re{da2asdfa554sd3334fsaf},
верны неравенства  
$$|A^{u_1}A^{u_2}-A^{u_2}|\leq |\!|A^{u_2}|\!|,\quad
|B^{u_1}B^{u_2}-B^{u_2}|\leq |\!|B^{u_2}|\!|,$$
поэтому из \re{dsafasdfdfsads44}  и 
из истинности  $\forall\,u\in X^k$ 
неравенства в \re{asdfasdf55543}
следует неравенство 
\be{asdfasdf3333355543}\by
|A^{u}-B^{u}|<|\!|A^{u_2}|\!|+|\!|B^{u_2}|\!|+
\frac{\delta_1}{3}.\ey\ee

Cогласно теореме \ref{dasfsadfasdfasd3334fsaf}
 и условию \re{dafdsafsarr44eee4}, 
\bi\i из условия $\forall\,x\in X\;\; A^{x}\geq c$
следует соотношение 
\be{asdfasdf33333555431}\by
\forall\,u\in X^k\\
|\!|A^u|\!|\leq(1-2 c)^{k-1}<(1-c)^{k-1}<\frac{\delta_1}{3},\ey\ee
\i из условий $B \in O_{\varepsilon}(A)$,
$\varepsilon\in (0,\frac{c}{2})$ и 
 $\forall\,x\in X\;\; A^{x}\geq c$ 
 следует условие 
 $\forall\,x\in X\;\; B^{x}\geq \frac{c}{2}$,
 из которого
следует соотношение 
\be{asdfasdf33333555432}\by
\forall\,u\in X^k\\
|\!|B^u|\!|\leq(1-2\cdot \frac{c}{2})^{k-1}=
(1-c)^{k-1}<\frac{\delta_1}{3}.\ey\ee
\ei 
 
Из \re{asdfasdf3333355543}, 
\re{asdfasdf33333555431},  и
\re{asdfasdf33333555432} следует, что
$$\by
|A^{u}-B^{u}|<\frac{\delta_1}{3}+\frac{\delta_1}{3}+
\frac{\delta_1}{3}=\delta_1.\;\;\blackbox\ey$$
\end{description}

Таким образом, для доказательства теоремы 
\ref{th04433d2d733225t432}
осталось доказать, что 
$\exists\,\varepsilon\in (0,\min\{\delta_1,\frac{c}{2}\}):$
$\forall\,B\in O_\varepsilon(A)$ верно \re{asdfasdf55543}.

Нетрудно доказать, что для каждой строки $u\in X^{\leq k}$
$$\by
\exists \, \varepsilon_u>0: 
\forall\,\varepsilon\in (0,\varepsilon_u),
\forall\,B\in O_\varepsilon(A)\quad |A^u-B^u|<\frac{\delta_1}{3}.\ey$$
(для доказательства этого факта 
можно использовать следующее
соображение: $\forall\, x\in X$
обозначим символом $M^x$ матрицу порядка $n$,
элемент в строке $i$ и столбце $j$ которой имеет вид 
$A^x_{ij}+t^x_{ij}$, где $t^x_{ij}$  -- 
различные переменные, и если $u=x_1\ldots x_s$,
то $M^u\eam M^{x_1}\ldots M^{x_s}$, 
тогда выражение $|A^u-M^u|$ определяет 
непрерывную функцию от переменных 
$t^x_{ij}$, значение которой равно 0 в том случае
когда значения всех переменных $t^x_{ij}$ равны 0, 
и т.д.)

Поскольку число строк в $X^{\leq k}$ конечно,
то в качестве искомого $\varepsilon$
можно взять $\min\{\delta_1,\frac{c}{2},\varepsilon_u \;(u\in X^{\leq k})\}$. $\blackbox$\\

Нижеследующая теорема является усилением 
теоремы \ref{th04433d2d733225t432}.\\

\refstepcounter{theorem}
{\bf Теорема \arabic{theorem}\label{th04433d2d733225t433}}.

Пусть задан ВА 
$A=(\xi^0,\{A^x\mid x\in X\},\lambda)$,
и 
$$\exists\, l: \forall\,u\in X^l\;\;A^u>0.$$
 
Тогда $\forall\,a\in I(A)\;\;
A$ устойчив относительно $a$.\\

{\bf Доказательство}.

Пусть $a\in I_\delta(A)$, и $|S_A|=n$.

Как и в доказательстве теоремы \ref{th04433d2d733225t432}, 
для доказательства
теоремы \ref{th04433d2d733225t433} 
достаточно доказать, что $\exists\,\varepsilon\in(0,\delta_1)$,
где $\delta_1\eam
\frac{\delta}{4n|\lambda|}$:
\be{1asdfasdf555334343}
\by\forall\,B
\in O_{\varepsilon}(A),\;\;
\forall\,u\in X^*\quad
|A^{u}-B^{u}|<\delta_1.\ey\ee

По предположению, $\forall\,u\in X^l\;\exists
\,c_u>0:A^u\geq c_u$.

Определим 
$c\eam \min\limits_{u\in X^l}c_u$.
Таким образом, $\forall\,u\in X^l\;A^u\geq c$.

Выберем $k$ так, чтобы было верно неравенство
$(1-c)^{[k/l]}<\frac{\delta_1}{3}.$

Аналогично доказательству теоремы \ref{th04433d2d733225t432},
 доказываются свойства
\bi\i 
  $\forall\,u\in X^k\;
\; |\!|A^u|\!|\leq (1-c)^{[k/l]}$, и 
\i $\exists \,\varepsilon>0$:
 $\forall\,B
\in O_{\varepsilon}(A)$, $\forall\,u\in X^{\leq k}\;\;
 | A^{u}-B^{u}|\leq \frac{\delta_1}{3} $,
\ei
из которых следует \re{1asdfasdf555334343}.
$\blackbox$\\

Можно доказать, что если $A=(\xi^0,\{A^x\mid x\in X\},\lambda)$ -- 
ВА, то 
$\forall\,a\in I(A)\;\;
A$ устойчив относительно $a$ 
тогда и только тогда, когда
$\exists\,\delta<1:\forall\,x\in X\;\;|\!|A^x|\!|<\delta.$

\chapter{Вероятностные языки}

\section{Понятие вероятностного языка}

Пусть $A$ --
ВА, и $a\in [0,1)$.

{\bf Вероятностным языком (ВЯ)} 
ВА $A$ с точкой сечения 
$a$ называется множество 
строк $A_a\subseteq X^*$ (где $X$ -- 
множество входных сигналов $A$), 
определяемое следующим образом:
$$A_a \eam \{u\in X^*\mid f_A(u)>a\}.$$

Понятие ВЯ может использоваться 
для решения различных задач, в частности для 
моделирования процесса обучения.
Одна из моделей обучения имеет
следующий вид. Задано конечное множество 
$S$, элементы которого называются
{\bf реакциями}  на некоторый стимул,
причём каждая реакция $s\in S$ рассматривается
либо как правильная, либо как неправильная.
Обозначим символом $\lambda$ вектор-столбец, 
компоненты которого индексированы реакциями 
$s\in S$, причем те компоненты $\lambda$, 
индексы которых являются правильными реакциями,
равны 1, а остальные компоненты равны 0.
С обучаемой системой в каждый момент времени $t=0,1,\ldots$
связано некоторое распределение 
$\xi\in S^\bigtriangleup$, в соответствии
с которым она реагирует на стимул в этот момент:
для каждого $s\in S$ вероятность того, что система 
выдаст реакцию $s$ в момент $t$, равна $s^\xi$.
Из данных определений следует, что вероятность выдачи
правильной реакции в момент $t$ имеет вид $\xi\lambda$.
Предполагается, что задано конечное множество
$\{A^x\mid x\in X\}$ {\bf обучающих операторов}, 
каждый из которых является стохастической
матрицей порядка $|S|$ и определяет изменение
реакции системы на стимул по следующему правилу: 
если\bi\i в текущий момент времени 
система реагировала на стимул
в соответствии с распределением $\xi\in S^\bigtriangleup$,
и \i в этот момент  был применён обучающий оператор
$A^x$, \ei
то в следующий момент времени
система будет реагировать на стимул
в соответствии с распределением $\xi A^x$.
Таким образом, путем применения к обучаемой системе
обучающих операторов можно добиться изменения 
вероятности правильной реакции на стимул.
Система считается {\bf обученной} в некоторой момент времени,
если вероятность того, что она выдаст в этот момент
правильную реакцию на стимул, превышает некоторое
заданное значение $a\in[0,1)$.
Одна из задач, связанных с обучением, имеет следующий
вид: задано начальное распределение
$\xi^0\in S^\bigtriangleup$,
требуется описать все  последовательности 
$(A^{x_1},\ldots,A^{x_n})$
обучающих операторов, каждая из которых обладает
следующим свойством:
после последовательного применения операторов, 
входящих в эту последовательность, 
система станет обученной.
Нетрудно видеть, что каждая такая 
последовательность
соответствует строке 
$x_1\ldots x_n\in X^*$, такой, что $\xi^0A^{x_1}\ldots
A^{x_n}\lambda >a$, т.е. строке из ВЯ $A_a$, где 
$A=(\xi^0,\{A^x\mid x\in X\}, \lambda).$

\section{Свойства вероятностных языков}

%

\refstepcounter{theorem}
{\bf Теорема \arabic{theorem}\label{th0442211315}}.

Пусть $A=(\xi^0, \{A^x\mid x\in X\},\lambda)$ -- 
ВА, и $a\in [0,1)$.

Тогда $A_a=B_a$, где ВА $B$ имеет вид
  $$B=(\vec e_1,
\left(\by 0&\xi^0 A^x\cr
{\bf 0}&A^x\cr\ey
\right),
 \left(\by \xi^0 \lambda\cr
\lambda\cr\ey
\right)),$$ 
где  $\vec e_1$ -- вектор-строка длины $|S_A|+1$, 
у которой
первая компонента равна 1, а остальные компоненты равны 0,
и {\bf 0} -- вектор-столбец длины $|S_A|$, все компоненты которого равны 0.
\\

{\bf Доказательство}. 

Нетрудно доказать, что $\forall\, u\in X^*\setminus \{\varepsilon\}\quad
B^u=\left(\by 0&\xi^0 A^u\cr
{\bf 0}&A^u\cr\ey
\right)$, откуда  следует
равенство $f_A=f_B$. $\blackbox$\\
 
\refstepcounter{theorem}
{\bf Теорема \arabic{theorem}\label{th04422113151}}.

Пусть задан
ВА $A=(\xi^0, \{A^x\mid x\in X\},\lambda)$, где 
каждая компонента $\lambda_j$ вектора $\lambda$
принадлежит отрезку
$[0,1]$,
и $a\in [0,1)$.

Тогда $A_a=B_a$,
где $B=(\xi^0_B, \{B^x\mid x\in X\},\lambda_B)$,
$|S_B|=2|S_A|$, и
\bi
\i $\xi^0_B$ получается из $\xi^0$ добавлением 0 после каждой
компоненты,
\i $\forall\,x\in X$ матрица $B^x$ получается из 
матрицы $A^x$ заменой каждого её элемента $A^x_{ij}$
на матрицу
$\left(\by \lambda_jA^x_{ij}&(1-\lambda_j)A^x_{ij}\cr
\lambda_jA^x_{ij}&(1-\lambda_j)A^x_{ij}\cr\ey
\right)$,
\i $\lambda_B=(1\,0\,\ldots\,1\,0)^\sim$. 
\ei

{\bf Доказательство}.

Нетрудно доказать, что $\forall\, u\in X^*\setminus \{\varepsilon\}$
 матрица $B^u$ получается из 
матрицы $A^u$ заменой каждого её элемента $A^u_{ij}$
на матрицу 
$\left(\by \lambda_jA^u_{ij}&(1-\lambda_j)A^u_{ij}\cr
\lambda_jA^u_{ij}&(1-\lambda_j)A^u_{ij}\cr\ey
\right)$, 
откуда  следует
равенство $f_A=f_B$. $\blackbox$\\

\refstepcounter{theorem}
{\bf Теорема \arabic{theorem}\label{th04422113152}}.

Пусть $A$ -- ВА, и $a,b\in [0,1)$.
Тогда $\exists$ ВА $B:$
\be{asffsdgsfdgsfdgd}A_a= B_b.\ee

{\bf Доказательство}.

Пусть ВА $A$ имеет вид $(\xi_A^0, \{A^x\mid x\in X\}, \lambda_A)$.


Рассмотрим отдельно случаи $b<a$ и $b>a$.
\bi
\i Если $b<a$, то $b=\alpha a$, где $\alpha\in[0,1)$.

   В этом случае $|S_B|=2|S_A|$, и 
компоненты $\xi^0_B, B^x, \lambda_B$
искомого ВА $B$ имеют следующий вид:
\bi
\i $\xi^0_B = (\xi_A^0, {\bf 0}\,)$,
где
${\bf 0}$ -- вектор-строка длины $|S_A|$, все компоненты которого равны 0,
\i $\forall\, x\in X\quad B^x\eam \left(\by 
\alpha A^x&(1-\alpha) A^x\cr
\alpha A^x&(1-\alpha) A^x\cr\ey
\right)$,
\i $\lambda_B=\left(\by 
\lambda_A\cr
{\bf 0}\cr\ey
\right)$, где
${\bf 0}$ -- вектор-столбец длины $|S_A|$, все компоненты которого равны 0.
\ei

Нетрудно доказать, что 
$$\forall\,u\in X^*\setminus\{\varepsilon\}\quad
B^u=\left(\by 
\alpha A^u&(1-\alpha) A^u\cr
\alpha A^u&(1-\alpha) A^u\cr\ey
\right),$$ откуда следует, что 
$\forall\,u\in X^*
\;\;f_B(u)=\alpha f_A(u)=\frac{b}{a}f_A(u)$.

Из последнего соотношения 
следует \re{asffsdgsfdgsfdgd}.

\i Если $b>a$, то $b=\alpha+(1-\alpha)a$, 
где $\alpha\in(0,1)$.

   В этом случае $|S_B|=1+|S_A|$, и 
компоненты $\xi^0_B, B^x, \lambda_B$
искомого ВА $B$ имеют следующий вид:
\bi\i
$\xi^0_B = (\alpha,(1-\alpha)\xi^0_A)$,
\i $\forall\,x\in X\quad B^x=\left(\by 
1&0\cr
0&A^x\cr\ey
\right)$,
\i $\lambda_B=\left(\by 
1\cr
\lambda_A\cr
\ey
\right).$
\ei

Нетрудно доказать, что $\forall\,u\in X^*$
$$\by
f_B(u)=\alpha+(1-\alpha)
f_A(u)=\frac{b-a}{1-a}+\frac{1-b}{1-a}f_A(u),
\ey$$ 
откуда следует \re{asffsdgsfdgsfdgd}.
$\blackbox$
\ei

\section{Языки, представимые вероятностными
автоматами общего вида}

Пусть $A=(X,Y,S,P,\xi^0)$ -- ВА общего вида, 
$y\in Y$, и $a\in [0,1)$.
Обозначим записью $A_{y,a}$
множество
$$A_{y,a}\eam \{ux\mid u\in X^*,x\in X,
\xi^0 A^u A^{xy}I >a\},$$ где $A^u\eam
    \sum\limits_{v\in Y^*}A^{u,v}.$

Если строка $u\in X^*$ имеет вид 
$x_0\ldots x_{k-1}$, то  значение $\xi^0 A^u A^{xy}I$
можно интерпретировать как
вероятность того, что если 
\bi\i в моменты времени $0,1,\ldots, k-1$
на вход $A$ последовательно
   поступали элементы строки $u$
   (т.е. 
      в момент $0$ поступил сигнал $x_0$,
   в момент $1$ поступил сигнал $x_1$,
   $\ldots$, 
   в момент $k-1$ поступил сигнал $x_{k-1}$),
   и \i в момент $k$ поступил сигнал $x$, \ei то
в момент времени $k$
   выходный сигнал $A$ 
   равен $y$.

Таким образом, множество $A_{y,a}$ 
можно интерпретировать как совокупность всех 
строк $u\in X^*\setminus\{\varepsilon\}$, обладающих следующим 
свойством: вероятность того, что 
\bi\i если, начиная с момента 0,
      на вход $A$ последовательно
   поступали элементы строки $u$
\i то
при подаче на вход $A$ последнего 
входного сигнала из  $u$
выходной сигнал $A$ в этот момент времени 
равен $y$,\ei
больше $a$.

Множество  $A_{y,a}$ называется {\bf языком, 
представимым ВА общего вида $A$} 
выходным сигналом $y$ и
точкой сечения $a$.\\

\refstepcounter{theorem}
{\bf Теорема \arabic{theorem}\label{th04422113153}}.

Пусть $A=(X,Y,S,P,\xi^0)$ -- ВА общего вида, 
$y\in Y$, и $a\in [0,1)$.

Тогда $A_{y,a}
 =B_a\setminus\{\varepsilon\}$, где
$B$ -- ВА вида
$$B=\Bigg((\xi^0,{\bf 0}), \{
\left(\by \sum\limits_{y'\neq y}A^{xy'}&A^{xy}\cr
\sum\limits_{y'\neq y}A^{xy'}&A^{xy}\cr\ey
\right)
\mid x\in X\},
({\bf 0}, {\bf 1})^\sim\Bigg)$$
где
${\bf 0}$ и ${\bf 1}$
-- вектор-строки длины $|S_A|$, 
все компоненты которых равны 0 и 1,
соответственно.\\

{\bf Доказательство}.

Нетрудно доказать, что 
$$\forall\,u\in X^*,\,\forall\,x\in X\quad
B^{ux}=\left(\by 
A^u\sum\limits_{y'\neq y}A^{xy'}&A^uA^{xy}\cr
A^u\sum\limits_{y'\neq y}A^{xy'}&A^uA^{xy}\cr\ey
\right),$$ 
откуда следует соотношение
$$\forall\,u\in X^*,\,\forall\,x\in X\quad
f_B(ux)=\xi^0A^uA^{xy}I,$$
которое влечёт доказываемое утверждение.
$\blackbox$

\section{Регулярность вероятностных языков}

\subsection{Понятие регулярного языка}
\label{dsfsadfsadfsfds323}

{\bf Регулярный язык} над конечным множеством
$X$ -- это подмножество 
$U\subseteq X^*$, такое, что существует
автомат Мура $M$ вида \re{asdfasdfasdf},
удовлетворяющий условиям:
\be{sdffsadfdsafasddsa}
|S|<\infty,\;\;
Y=\{0,1\}, \;\; U=\{u\in X^*\mid f_M(u)=1\}.\ee

{\bf Весом} регулярного языка $U$
называется наименьшее число $n$, такое, что 
$\exists$ автомат Мура $M$ вида \re{asdfasdfasdf},
такой, что $|S|=n$ и выполнены условия
\re{sdffsadfdsafasddsa}. Мы будем обозначать 
вес регулярного языка $U$ записью $w(U)$.

Можно доказать, что для любого языка $U\subseteq X^*$
следующие утверждения эквивалентны:
\bi
\i 
язык $U$ является
 регулярным, 
 \i число классов 
эквивалентности $\sieq{U}$ на множестве $X^*$,
определяемой следующим образом: $\forall\,u,u'\in X^*$
\be{dfasdfasdfasdf}\by
u\sieq{U} u'\;\Leftrightarrow\;
\forall\,v\in X^*\;(uv\in U \Leftrightarrow u'v\in U)
\ey\ee
является конечным,
\ei
и если $U$ регулярен, то 
$w(U)$ совпадает с числом классов 
эквивалентности $\sieq{U}$.\\

\refstepcounter{theorem}
{\bf Теорема \arabic{theorem}\label{th04422113315475}}.

Пусть $X$ -- конечное множество, и $U\subseteq X^*$ -- 
регулярный язык. Тогда $U$ -- ВЯ.

{\bf Доказательство}.

Пусть  $M=(X, \{0,1\}, S, \delta, \lambda, s^0)$ -- автомат Мура 
такой, что $U=\{u\in X^*\mid f_M(u)=1\}$.

Обозначим записью $\tilde M$
ВА $(\xi_{s^0},\{A^x\mid x\in X\}, \lambda)$,
множество состояний  и 
отображение $\lambda$ которого совпадают с соответствующими
компонентами автомата $M$, и 
$$\forall\,x\in X,\forall\,s,s'\in S\quad 
A^x_{s,s'}\eam
\left\{\by 1,&\mbox{если } \delta(s,x)=s',\\
0,&\mbox{иначе.}
\ey
\right.$$

Нетрудно доказать, что $U=\tilde M_0$. $\blackbox$

\subsection{Изолированные точки сечения}

Пусть задан ВА $A=(\xi^0,\{A^x\mid x\in X\}, \lambda)$.

Число $a\in [0,1)$ называется {\bf изолированной точкой сечения (ИТС)}
 для ВА $A$, если $\exists\,\delta>0$:
\be{dsfsadfdsafdsfdsa}
\forall\,u\in X^*\quad
|f_A(u)-a|> \delta.\ee

Для каждого ВА $A$ и каждого $\delta>0$
запись $I_\delta(A)$ обозначает совокупность всех 
ИТС $a$
для $A$, обладающих свойством \re{dsfsadfdsafdsfdsa}.
Запись 
$I(A)$ обозначает множество
$\bigcup\limits_{\delta>0}I_\delta(A)$.

Можно доказать, что проблема выяснения 
истинности утверждения $a\in I(A)$ 
(где $a$ и все 
численные значения ВА $A$ -- 
рациональные числа) 
 алгоритмически неразрешима.\\

\refstepcounter{theorem}
{\bf Теорема \arabic{theorem}\label{th0442211331547325}}.

Пусть задан ВА $A=(\xi^0,\{A^x\mid x\in X\}, \lambda)$,
где $\lambda\in [0,1]^n$, $n=|S_A|$,
и $\exists\,\delta>0:
a\in I_\delta(A)$.

Тогда $A_a$ -- регулярный язык, и 
\be{dsafsadfasdfsa44}\by
w(A_a)\leq (1+\frac{1}{\delta})^{n-1}.\ey\ee

{\bf Доказательство}.

Обозначим символом $R$ множество, содержащее по одному
представителю каждого 
класса эквивалентности $\sieq{A_a}$. 
Из определения $R$ следует, что
$\forall\,u,u'\in U,\;u\neq u' \Rightarrow\exists$ 
$v \in X^*$:
\be{asfdsafasdfr43r43}
uv\in A_a \not \Leftrightarrow u'v\in A_a.
\ee

Т.к. $a\in I_\delta(A)$, то
из \re{asfdsafasdfr43r43} и \re{dsfsadfdsafdsfdsa}
следует неравенство
$$|f_A(uv)-f_A(u'v)|> 2\delta,$$
т.е. \be{adsfsadfsadfsa}
|\xi^0(A^{u}-A^{u'})A^{v}\lambda|> 2\delta.\ee

Поскольку $A^{v} I = I$ и $\lambda \in [0,1]^n$,
то $\lambda'\eam  A^{v}\lambda \in [0,1]^n$.

Можно доказать, 
что существует частичная функция ${\it Vol}_{n-1}$ 
с неотрицательными значениями 
на подмножествах 
множества ${\bf R}^{n}$, называемая {\bf $n-1$--мерным 
объемом}, 
и обладающая следующими свойствами:
\bi
\i значение
${\it Vol}_{n-1}(\{1,\ldots, n\}^\bigtriangleup)$
 определено и положительно,\\
 (обозначим это значение символом $C$) 
\i если  $Vol_{n-1}(M)$ определено, то 
$\forall\,\vec x=(x_1,\ldots, x_n)\in {\bf R}^n$
$$
   Vol_{n-1}(\vec x+ M)=Vol_{n-1}(M),$$
  где 
$\vec x+ M\eam \{
(x_1+y_1,\ldots, x_n+y_n)\mid (y_1,\ldots, y_n)\in 
M\}$,
\i если  $Vol_{n-1}(M)$ определено, то 
$\forall\,d>0$
   $$ Vol_{n-1}(dM) = d^{n-1}Vol_{n}(M),$$
  где 
$dM\eam \{
(dx_1,\ldots, dx_n)\mid (x_1,\ldots, x_n)\in 
M\}$,
\i 
если  $Vol_{n-1}(M)$ и $Vol_{n-1}(M')$
определены, и $M\cap M'=\emptyset$,
то 
$${\it Vol}_{n-1}(M\cup M')={\it Vol}_{n-1}(M)+{\it Vol}_{n-1}(M'),$$
\i 
если  $Vol_{n-1}(M)$ и $Vol_{n-1}(M')$
определены, и $M\subseteq  M'$,
то 
$${\it Vol}_{n-1}(M)\leq {\it Vol}_{n-1}(M').$$\ei

Введём следующие обозначения:
пусть $d>0$ и $u\in R$, тогда записи 
$\Delta_d$ и $\Delta^u_d$ 
обозначают 
множество $d\{1,\ldots, n\}^\bigtriangleup$ и
$\xi^0A^u+ \Delta_d$ соответственно.
Из сказанного выше следует, что 
\be{dfsadfsadfsa2}\forall\,d>0\quad
{\it Vol}_{n-1}(\Delta_d)=
{\it Vol}_{n-1}(\Delta^u_d)=d^{n-1}C.
\ee

Нетрудно видеть, что 
\bi
\i[(A)] $\forall\,u\in R\quad
   \Delta^u_\delta\subseteq \Delta_{1+\delta}$, т.к. 
   если $\vec x\in \Delta^u_\delta=
   \xi^0A^u+ \Delta_\delta$, то 
   $$\vec x=\xi^0A^u+\delta\vec y, \mbox{ где }\vec y\in \Delta_1.$$
   Все компоненты $\vec y$ неотрицательны и $\vec y I=1$,
   поэтому все компоненты $\vec x$ неотрицательны
   и $$\vec x I = (\xi^0A^u+\delta\vec y)I = 
   \xi^0A^u I+\delta\vec y I = 1+ \delta,$$ откуда следует, что
   $\vec x \in \Delta_{1+\delta}$.

\i[(B)] $\forall\,u,u'\in R,\;u\neq u' \;\Rightarrow\;
\Delta^u_\delta \cap \Delta^{u'}_\delta=\emptyset$, т.к. 
 если
 $\exists\,\vec x\in \Delta^u_\delta \cap \Delta^{u'}_\delta$,
то $\exists\,\vec y, \vec z\in \Delta_1$:
\be{fdasdfsdfs44343}\vec x =  \xi^0A^u + \delta \vec y,\quad
   \vec x =  \xi^0A^{u'} + \delta \vec z.\ee
Из \re{fdasdfsdfs44343} следует, что 
$\xi^0(A^u-A^{u'})=\delta(\vec z-\vec y)$,
откуда на основании \re{adsfsadfsadfsa}
получаем:
$
|\delta(\vec z-\vec y) \lambda'|> 2\delta$,
или \be{dsfdsafasd44445}
|(\vec z-\vec y) \lambda'|> 2,\mbox{ где } \lambda'\in [0,1]^n.\ee
Если $\vec z = (z_1,\ldots, z_n), 
\vec y = (y_1,\ldots, y_n), 
\lambda'=(\lambda'_1,\ldots, \lambda'_n)^\sim$,
то $$\by|(\vec z-\vec y) \lambda'|= 
|\sum\limits_{i=1}^n(z_i-y_i)\lambda'_i|\leq
\sum\limits_{i=1}^n|z_i-y_i|\lambda'_i\leq\\\leq 
\sum\limits_{i=1}^n|z_i-y_i|\leq
\sum\limits_{i=1}^n z_i +
\sum\limits_{i=1}^n y_i = 2,\ey$$
что противоречит соотношению \re{dsfdsafasd44445}.
\ei

Из (A), (B) и из перечисленных выше свойств функции ${\it Vol}_{n-1}$ следует, что если $R$
содержит $k$  элементов $u_1,\ldots, u_k$,
то $$\sum\limits_{i=1}^k
{\it Vol}_{n-1}(\Delta^{u_i}_\delta)\leq
{\it Vol}_{n-1}(\Delta_{1+\delta}),$$
откуда, ввиду \re{dfsadfsadfsa2}, следует неравенство
$$k\delta^{n-1}C\leq (1+\delta)^{n-1}C,$$
которое эквивалентно неравенству 
$k\leq  (1+\frac{1}{\delta})^{n-1}$,
откуда следует регулярность $A_a$ и
неравенство \re{dsafsadfasdfsa44}.
$\blackbox$\\

Следующая теорема показывает, что 
в ВА ограниченного размера можно представлять 
сколь угодно сложные регулярные языки.\\

\refstepcounter{theorem}
{\bf Теорема \arabic{theorem}\label{th044dd73325}}.

Пусть 
$A=(\xi^0, \{A^x\mid x\in X\},\lambda)$ -- 
ВА, где 
$$\by
|S_A|=2,\;X=\{0,2\},\;
\xi^0=(1\;\;0),\\
A^0= \left(\by
1&0\cr
2/3&1/3\cr
\ey\right),\;
A^2= \left(\by
1/3&2/3\cr
0&1\cr
\ey\right),\;
\lambda = \left(\by
0\cr
1\cr
\ey\right),
\ey$$
и $\forall\,n\geq 1\;\;a_n$ -- число из $(0,1)$,
имеющее в троичной записи вид 
$0.2\ldots 211$ (количество двоек =
$n-1$).

Тогда 
$\forall\,n\geq 1\;\;a_n\in I(A)$ и 
$w(A_{a_n})\;\geq n.$\\

{\bf Доказательство}.
  
Индукцией по длине строки $u\in X^*$ 
доказывается, что 
если $u$ имеет вид 
$x_1\ldots x_k$, то 
$f_A(u)=
0.x_k\ldots x_1$ (в троичной записи).

Обозначим символом $D$ топологическое замыкание
множества $\{f_A(u)\mid u\in X^*\}$.
Множество $D$ называется {\bf канторовым дисконтинуумом}.
   Нетрудно доказать, что $[0,1]\setminus D\subseteq I(A)$.

 Из определения $a_n$  следует, что
$a_n<f_A(u)\;\Leftrightarrow
\;u=u_12\ldots 2$ (количество двоек $\geq n$),
   поэтому если $u\in A_{a_n}$, то $|u|\geq n$,
   откуда следует свойство $w(A_{a_n})\geq n$. 
   $\blackbox$\\

%
%
%
%

Нижеследующая теорема является обобщением теоремы
\ref{th0442211331547325}.\\

\refstepcounter{theorem}
{\bf Теорема \arabic{theorem}\label{th044d2d733225}}.

Пусть $M$ -- автомат Мура вида
$(X, \{0,1\}, S, \delta, \lambda, s^0)$,
причем 
$S$  -- компактное 
метрическое пространство 
 с метрикой $\rho$, 
такой, что  для любой пары 
$s_1,s_2$ достижимых состояний автомата $M$
выполнены  условия:
\be{ffasdfasdfasdgadsgfd}\forall\,x\in X\quad
\rho(s_1x,s_2x)\leq \rho(s_1,s_2),\ee
\be{fgfdgfsdgfd}\exists\,\delta>0:\;
\lambda(s_1)\neq\lambda(s_2)
\;\Rightarrow\;
\rho(s_1,s_2)\geq \delta.\ee

Тогда язык $U\eam\{u\in X^*\mid f_M(u)=1\}$ регулярен, и
$w(U)$ не превосходит 
числа элементов
в минимальном (по числу элементов)
покрытии множества $S$ открытыми шарами радиуса $\delta/2$.\\

{\bf Доказательство}.

Обозначим символом $R$ множество, содержащее
 по одному представителю  каждого класса эквивалентности
 $\sieq{U}$.
 Из \re{fgfdgfsdgfd}
и \re{ffasdfasdfasdgadsgfd}
  следует, что $\forall\,u,u'\in R$: 
$u\neq u'$ $\exists \,v\in X^*:$
\be{dfadfssafdsafdsa}\delta\leq
\rho(s^0 u v, s^0  u' v)\leq
\rho(s^0 u,s^0  u' ).\ee

Пусть ${\cal C}$ --  минимальное (по числу элементов)
конечное покрытие  $S$ открытыми шарами радиуса $\delta/2$. 
Если $u,u'$ -- различные элементы $R$, 
то из \re{dfadfssafdsafdsa} следует, что $s^0u$
и $s^0u'$ не могут попасть в один и тот же шар из ${\cal C}$,
поэтому $|R|\leq |{\cal C}|$.
$\blackbox$\\

Теорема 
\ref{th0442211331547325} является следствием
теоремы \ref{th044d2d733225}, т.к. 
если заданы ВА $A=(\xi^0,\{A^x\mid x\in X\},\lambda)$
и число $a\in [0,1)$, удовлетворяющие условиям
теоремы \ref{th0442211331547325},
то, полагая 
$$M\eam (X, \{0,1\}, S_A^\bigtriangleup, 
\delta, \lambda_M, \xi^0),$$
где $\delta(\xi,x)\eam \xi A^x,\;\; 
\lambda_M(\xi)\eam \left\{\by
1,&\mbox{если } \xi\lambda>a,\\
0,&\mbox{иначе}, \ey
\right.$
получаем: 
$$\forall\,u\in X^*\quad 
u\in A_a\;\Leftrightarrow\;
f_M(u)=1.$$ Множество
 $S_A^\bigtriangleup$ является компактным метрическим пространством
относительно метрики \be{asdfasdfasfasfa33s}\rho(\vec x,\vec y\,)=\max\limits_{i=1\ldots n}|x_i-y_i|,\ee
т.к. оно является ограниченным подмножеством 
${\bf R}^n$.
Нетрудно доказать, что 
для любой пары $s_1,s_2$ достижимых состояний автомата $M$
условия
\re{ffasdfasdfasdgadsgfd}
и \re{fgfdgfsdgfd} выполнены.\\

Следующая теорема также является следствием
теоремы \ref{th044d2d733225}.\\

\refstepcounter{theorem}
{\bf Теорема \arabic{theorem}\label{th04433d2d733225}}.

Пусть $M$ -- автомат Мура вида 
$$(X,\{0,1\}, \{1,\ldots, n\}^\bigtriangleup,
\delta,\lambda, \xi^0),$$
где $|X|<\infty$, и 
 для любой пары $s_1,s_2$ достижимых состояний 
автомата $M$ выполнены условия 
\re{ffasdfasdfasdgadsgfd} и
\re{fgfdgfsdgfd},  
где метрика  $\rho$ определяется соотношением
 \re{asdfasdfasfasfa33s}.

Тогда язык $U\eam\{u\in X^*\mid f_M(u)=1\}$ регулярен, и
\be{sdfsdafsadfdsfsadfdsa}
w(U)\leq C^m_{n+m-1},\quad\mbox{где $m=]\frac{2}{\delta}[$.}\ee

{\bf Доказательство}.

Регулярность $U$ непосредственно следует из
теоремы \ref{th044d2d733225}.

Для доказательства неравенства
\re{sdfsdafsadfdsfsadfdsa}
определим  покрытие
множества $\{1,\ldots, n\}^\bigtriangleup$ 
открытыми шарами радиуса $\frac{1}{m}$
(поскольку $\frac{1}{m} \leq \frac{\delta}{2}$, то, 
согласно теореме \ref{th044d2d733225},
$w(U)$ не превосходит числа элементов в этом покрытии).

Обозначим записью $F_m^n$ совокупность точек 
  $\{1,\ldots, n\}^\bigtriangleup$, имеющих вид
 $(\frac{k_1}{m},\ldots, \frac{k_n}{m})$,
 где $k_1,\ldots, k_n$ -- неотрицательные целые числа, 
 сумма которых равна $m$.
 Нетрудно видеть, что $$\by\forall\,\vec x\in \{1,\ldots, n\}^\bigtriangleup\;
 \exists \,\vec y\in F_m^n: \rho(\vec  x,\vec y)<\frac{1}{m}\ey$$
 (если $\vec x=(x_1,x_2,\ldots)$,
 то $\vec y=(\,]x_1m[, ](x_1+x_2)m[-]x_1m[, \ldots)$),
 поэтому множество открытых шаров радиуса 
 $\frac{1}{m}$ с центрами в точках из $F_m^n$
 является покрытием множества $\{1,\ldots, n\}^\bigtriangleup$.
Число элементов в этом покрытии равно 
$|F_m^n| =C^m_{n+m-1}$.
$\blackbox$

\section{Дефинитные языки}

Пусть задано конечное множество $X$. 

Мы будем использовать
следующие обозначения.

\bi\i $\forall\,S\subseteq X^*,\;\forall\,
k\geq 0$ записи $S_k$,
$S_{\geq k}$, 
и т.д.
  обозначают множества $S\cap X^k$, $S\cap X^{\geq k}$, 
и т.д., соответственно.
\i $\forall\,S_1,S_2\subseteq X^*$
записи $S_1+ S_2$ и $S_1\cdot S_2$ (точка в этой записи
обычно опускается) обозначают множества
$$S_1\cup S_2\quad\mbox{и}\quad\{uv\mid u\in S_1,v\in S_2\}$$
соответственно.
 \ei
$S\subseteq X^*$ называется
{\bf дефинитным языком (ДЯ)}, если
 $$\exists\,k\geq 0: \;\;S_{\geq k}=X^*\cdot S_k.$$

\refstepcounter{theorem}
{\bf Теорема \arabic{theorem}\label{th04433d2d733225t431}}.

Пусть задан ВА 
$A=(\xi^0,\{A^x\mid x\in X\},\lambda)$,
и 
$\forall\,x\in X\;\;A^x>0$.
 
Тогда $\forall\,a\in I(A)\;\;
A_a$ -- ДЯ.\\

{\bf Доказательство}.


По предположению, 
$\exists\,c>0$:
$\forall\,x\in X\;\; A^x\geq c$, и $\exists\,\delta>0$: 
\be{safdds334afsadga}
\forall\,u\in X^*\quad
|f_A(u)-a|>\delta.\ee

Выберем  $k$: $(1-2c)^{k-1}<\frac{2\delta}{n |\lambda|}$, 
где $n=|S_A|$.

По теореме \ref{dasfsadfasdfasd3334fsaf}, 
\be{adfsadfdsafsaf334s}\by\forall\,u\in X^k\;\;
|\!|A^u|\!|\leq(1-2 c)^{k-1}<
\frac{2\delta}{n |\lambda|}.\ey\ee

$\forall\,u\in X^k$, $\forall\,v\in X^*$
$$\by |f_A(vu)-f_A(u)|=
|\xi^0 A^{vu}\lambda-\xi^0 A^u\lambda|=
|\xi^0 (A^{vu}-A^u)\lambda|\leq\\\leq
|A^{vu}-A^u|\cdot n \cdot |\lambda|=
|A^{v}A^{u}-A^u|\cdot n \cdot |\lambda|\leq
|\!|A^u|\!|\cdot n \cdot |\lambda|<2\delta.\ey$$
(предпоследнее неравенство верно 
по теореме \ref{da2asdfa554sd3334fsaf},
а последнее -- согласно \re{adfsadfdsafsaf334s}).
Таким образом,
\be{sdfdsafasdfsaee3}\forall\,u\in X^k, \,\forall\,v\in X^*\quad
|f_A(vu)-f_A(u)|<2\delta.\ee

Докажем, что $(A_a)_{\geq k}=X^*\cdot(A_a)_k$,
т.е. $\forall\,u\in X^k$, $\forall\,v\in X^*$
\be{sfdgfd54w56}
vu\in A_a\;\;\Leftrightarrow\;\; u\in A_a.\ee

Если для некоторых $u\in X^k, v\in X^*$
соотношение \re{sfdgfd54w56} неверно, то, 
согласно определению ВЯ $A_a$ и 
 соотношению \re{safdds334afsadga},
$$|f_A(vu)-f_A(u)|> 2\delta$$
что противоречит соотношению \re{sdfdsafasdfsaee3}.
$\blackbox$\\

\refstepcounter{theorem}
{\bf Теорема \arabic{theorem}.\label{th04433d2d733225c431}}

Пусть ВА $A=(\xi^0,\{A^x\mid x\in X\},\lambda)$ --
эргодичный.

Тогда $\forall\,a\in I(A)\quad A_a$ -- ДЯ.\\

{\bf Доказательство}.

Пусть $a\in I_\delta(A)$, и $|S_A|=n$.

Из эргодичности $A$ следует, что 
$|\!|A^u|\!|\to 0$ при $|u|\to \infty$, поэтому
$\exists\,k$:  
$$\by\forall\,u\in X^k\quad
|\!|A^u|\!|<
\frac{2\delta}{n |\lambda|}.\ey$$

Оставшаяся часть доказательства совпадает
с частью доказательства
теоремы \ref{th04433d2d733225t431},
идущей после соотношения \re{adfsadfdsafsaf334s}.
$\blackbox$

\section{Языки, представимые линейными автоматами}


Пусть задан ЛА $L$.
Для каждого $a\in {\bf R}$ запись $L_a$
обозначает подмножество множества $X^*$,
определяемое следующим образом:
\be{asfasdfsdags555}L_a\eam \{u\in X^*\mid
f_L(u)>a\}.\ee
Множество \re{asfasdfsdags555}
называется
{\bf языком},
представимым ЛА $L$ с точкой сечения $a$.\\

\refstepcounter{theorem}
{\bf Теорема \arabic{theorem}.\label{th04422113154}}

Пусть $L=(\xi^0,\{L^x\mid x\in X\}, \lambda)$ -- 
ЛА, 
и $a\in {\bf R}$.

Тогда существует
ВА $A$ вида \be{dvvdfvdfvxczx}(\vec e_1,\{A^x\mid 
x\in X\}, e_1^\downarrow),\ee
такой, что 
$$|S_A|=n+4, \quad
L_a=A_{\frac{1}{n+4}},\quad\mbox{где $n=\dim L$}.$$

{\bf Доказательство}.

Определим ЛА $L'$ следующим образом:
$$L'\;\eam\;\Bigg((\xi^0,1),\{\left(\begin{array}{ccc}
L^x&{\bf 0}\cr
{\bf 0}&1\cr\ey
\right)\mid x\in X\}, \left(\begin{array}{c}
\lambda\cr
-a\cr\ey
\right)\Bigg),$$
где символы {\bf 0} обозначают строку и столбец 
соответствующего размера
с нулевыми компонентами.
Отметим, что
\bi\i $\dim L'=\dim L+1$, и
\i $f_{L'}=f_L-a,\;\;\Rightarrow\;\; L_a=L'_0$,
\ei
поэтому для 
доказательства теоремы \ref{th04422113154}
достаточно доказать следующее
утверждение: 
для каждого ЛА $L$
существует ВА $A$ вида \re{dvvdfvdfvxczx},
такой, что 
$|S_A|=n+3$ и
$L_0=A_{\frac{1}{n+3}}$, где $n=\dim L$.

Для доказательства этого утверждения рассмотрим отдельно случаи $f_L(\varepsilon)>0$
и $f_L(\varepsilon)\leq 0$.
Пусть $L=(\xi^0,\{L^x\mid x\in X\},\lambda)$.

\bi
\i Если $f_L(\varepsilon)>0$, то $\varepsilon\in L_0$.
   Определим функцию $f\in {\bf R}^{X^*}$ следующим образом:
   $$\forall\,u\in X^*\quad
   f(u)\;\eam \left\{\by
   1,&\mbox{если } u=\varepsilon\\
   f_L(u),&\mbox{иначе.}
   \ey\right.$$

Обозначим записью $L_1$ ЛА $(\xi^0_1,\{L_1^x\mid x\in X\},\lambda_1)$, где 
$$\xi^0_1 =(\xi^0,1),\quad
 L_1^x=\left(\begin{array}{ccc}
L^x&{\bf 0}\cr
{\bf 0}&0\cr\ey
\right),\quad
\lambda_1= \left(\begin{array}{c}
\lambda\cr
1-f_L(\varepsilon)\cr\ey
\right),$$
где символы {\bf 0} обозначают строку и столбец 
соответствующего размера
с нулевыми компонентами.

Пусть $P$ -- невырожденная матрица порядка 
$\dim L_1$, первый столбец которой 
равен $\lambda_1$, а остальные столбцы
ортогональны $\xi^0_1$.

Нетрудно видеть, что
$Pe_1^\downarrow=\lambda_1$, $\xi^0_1P=\vec e_1$.

Обозначим записью $L_2$ ЛА 
$(\vec e_1,\{P^{-1}L_1^xP\mid x\in X\},e_1^\downarrow)$.

Нетрудно видеть, что $f_{L_2}=f_{L_1}$.

Затем $L_2$ преобразуется в искомый ВА $A$
(в соответствии с построением, изложенным в доказательстве
теоремы \ref{th0131335}), 
такой, что $\exists\,b>0$:
$$\forall\,u\in X^*
\quad
f_A(u) =  \left\{\by 1,\;\;\mbox{если }
u=\varepsilon,\\
b^{|u|+1}f_{L_2}(u)+\frac{1}{n+3},\;\;\mbox{иначе}.
\ey\right.
$$

\i Если $f_L(\varepsilon)\leq 0$, то $\varepsilon\not\in L_0$.
   Определим функцию $f\in {\bf R}^{X^*}$ следующим образом:
   $$\forall\,u\in X^*\quad
   f(u)\;\eam \left\{\by
   0,&\mbox{если } u=\varepsilon\\
   f_L(u),&\mbox{иначе.}
   \ey\right.$$
   
   Обозначим записью $L_1$ ЛА $(\xi^0_1,\{L_1^x\mid x\in X\},\lambda_1)$, где 
$$\xi^0_1 =(\xi^0,1),\quad
 L_1^x=\left(\begin{array}{ccc}
L^x&{\bf 0}\cr
{\bf 0}&0\cr\ey
\right),\quad
\lambda_1= \left(\begin{array}{c}
\lambda\cr
-f_L(\varepsilon)\cr\ey
\right),$$
   
Пусть $P$ -- невырожденная матрица порядка 
$\dim L_1$, первый столбец которой 
равен $\lambda_1$, все столбцы кроме последнего
ортогональны $\xi^0_1$, и $\xi^0_1 P^\downarrow_{n+1}=1$,
где $n=\dim L$ и
$P^\downarrow_{n+1}$ -- последний столбец $P$.

Нетрудно видеть, что
$Pe_1^\downarrow=\lambda_1$, $\xi^0_1P=\vec e_{n+1}$.

Обозначим записью $L_2$ ЛА 
$(\vec e_{n+1},\{P^{-1}L_1^xP\mid x\in X\},e_1^\downarrow)$.

Нетрудно видеть, что $f_{L_2}=f_{L_1}$.

Затем $L_2$ преобразуется в искомый ВА $A$
(в соответствии с построением, изложенным в доказательстве
теоремы \ref{th0131335}), 
такой, что $\exists\,b>0$:
$$\by \forall\,u\in X^*
\quad
f_A(u) = \left\{\by 0,\;\;\mbox{если }
u=\varepsilon,\\
b^{|u|+1}f_{L_2}(u)+\frac{1}{n+3},\;\;\mbox{иначе}.
\;\;\blackbox\ey\right.
\ey$$
\ei

\refstepcounter{theorem}
{\bf Теорема \arabic{theorem}.\label{th044221133155}}

Пусть $X$ -- конечное множество, и
$S\subseteq  X^*$.

Следующие условия эквивалентны:
\bi
\i $S$ -- ВЯ,
\i $\exists$ ЛАФ $f\in {\bf R}^{ X^*}$, $\exists\,a\in {\bf R}$:
   $S=f_a\eam \{u\in X^*\mid f(u)>a\}$,
\i $\exists$ ЛАФ $f\in {\bf R}^{ X^*}: S=f_0$.
\ei
В доказательстве этой теоремы используется
утверждение из главы \ref{dfdsafasfdsagfa43}, 
что если $f$ -- ЛАФ и $a\in {\bf R}$, то  $f-a$   
 -- ЛАФ. $\blackbox$




 \chapter{Алгебраические вопросы теории линейных автоматов}
 \label{dfdsafasfdsagfa43}

  В этой главе мы рассматриваем 
  некоторые алгебраические вопросы, относящиеся к линейным
  автомамам и связанным с ними функциям на строках.
  Материал этой главы будет использоваться в последующих 
  частях.
  
  \section{Алгебраические свойства множества функций
  на строках}
\label{1sdvsdfdsfvzdxvfrr}

Пусть задано конечное множество $X$.

В пункте \ref{sdvsdfdsfvzdxvfrr} было введено понятие 
функции на строках из $X^*$, и на множестве ${\bf R}^{X^*}$
таких функций были определены алгебраические 
операции суммы, 
разности,  
и умножения на числа из ${\bf R}$
(соотношениями    \re{sdfdsafdsaf55566} 
 и    \re{sdfdsafdsaf55566111}).

Определим другие алгебраические операции на  
${\bf R}^{X^*}$

\bi
\i Для функций
   $f_1,f_2\in {\bf R}^{X^*}$ их 
   {\bf произведение} $f_1f_2$
и {\bf свёртка}    $f_1\circ f_2$ 
определяются следующим образом:
   \be{1sdfdsafdsaf55566}
   \forall\,u\in X^*
   \left\{
   \by 
   (f_1f_2)(u)\eam
   f_1(u) f_2(u),\\
(f_1\circ f_2)(u)\eam
   \sum\limits_{u_1u_2=u}
   f_1(u_1)f_2(u_2).   \ey
   \right.\ee
\i Для каждой функции
     $f\in {\bf R}^{X^*}$
     {\bf инвертирование}    $\tilde f$ 
     определяется следующим образом:
   $\forall\,u\in X^*\quad
   \tilde f(u)\eam
   f(\tilde u).
   $

\ei

  Для каждого подмножества $S\subseteq X^*$
мы будем обозначать записью $\chi_S$ характеристическую
функцию этого подмножества, т.е. 
функцию из ${\bf R}^{X^*}$,
которая отображает каждый элемент из $S$ в 1, и все остальные 
элементы из $S$ -- в 0. Если множество $S$ 
одноэлементно и имеет вид $\{s\}$, то мы будем обозначать
функцию $\chi_{\{s\}}$ более короткой записью $\chi_s$.\\

\refstepcounter{theorem}
{\bf Теорема \arabic{theorem}\label{th44433015ee3s}}.

Множество ${\bf R}^{X^*}$ является
кольцом, в котором сумма элементов определяется
первым соотношением в \re{sdfdsafdsaf55566}
а в качестве умножения выступает операция свёртки. 
Функция $\chi_\varepsilon$
является единицей этого кольца.\\

{\bf Доказательство}.

Ассоциативность свёртки 
следует из того, что
$\forall\,f_1, f_2,f_3\in {\bf R}^{X^*}$ и
$\forall\,u\in X^*$ значения 
$((f_1\circ f_2)\circ f_3)(u)$
и $(f_1\circ (f_2\circ f_3))(u)$
равны значению выражения
$$
\sum\limits_{u_1u_2u_3=u}
f_1(u_1)f_2(u_2)f_3(u_3).
$$
и, следовательно, функции 
$(f_1\circ f_2)\circ f_3$
и $f_1\circ (f_2\circ f_3)$
совпадают. 

Остальные свойства кольца для ${\bf R}^{X^*}$
устанавливаются непосредственной проверкой.
$\blackbox$\\


\refstepcounter{theorem}
{\bf Теорема \arabic{theorem}\label{th44433015}}.

Если $f\in {\bf R}^{X^*}$ и
$f(\varepsilon) \neq 0$, то
существует единственная функция
$g\in {\bf R}^{X^*}$, удовлетворяющая условию
\be{gfdgsdfgsrrr}
f\circ g=g\circ f=\chi_\varepsilon.\ee

{\bf Доказательство}.

 \re{gfdgsdfgsrrr}
эквивалентно конъюнкции следующих соотношений:
\be{sadfsadf55566}\by f(\varepsilon)g(\varepsilon)=1,\\
\forall\,x\in X\quad 
f(\varepsilon)g(x)+
f(x)g(\varepsilon)=0,\\
\forall\,x_1,x_2\in X\quad 
f(\varepsilon)g(x_1x_2)+
f(x_1)g(x_2)+\\\hspace{50mm}+
f(x_1x_2)g(\varepsilon)=0,\\\mbox{и т.д.}
\ey\ee
Определим функцию 
$g\in {\bf R}^{X^*}$
индуктивно следующим образом:
$$\by g(\varepsilon)\eam (f(\varepsilon))^{-1},\\
\forall\,x\in X\quad 
g(x)\eam-(f(\varepsilon))^{-1}
f(x)g(\varepsilon),\\
\forall\,x_1,x_2\in X\quad 
g(x_1x_2)
\eam -(f(\varepsilon))^{-1}
(f(x_1)g(x_2)+
f(x_1x_2)g(\varepsilon)),\\
\mbox{и т.д.}\ey$$
Нетрудно видеть, что определённая таким образом
функция $g$ является единственной функцией,
удовлетворяющей  соотношениям
\re{sadfsadf55566}.
$\blackbox$\\

Мы будем использовать следующие обозначения.
\bi\i
Функцию $g$, удовлетворяющую 
условию  \re{gfdgsdfgsrrr}, мы будем обозначать записью $f^{-1}$.
\i Если $f\in {\bf R}^{X^*}$ и $k\geq 1$, то запись $f^{\circ k}$ обозначает 
\bi
\i функцию  $f$, если $k=1$, и
\i функцию  $f^{\circ (k-1)}\circ f$, если $k>1$.
\ei
\ei

{\bf Лемма}.

 Если $f(\varepsilon) =0$ и 
  $u\in X^{<k}$, то $f^{\circ k}(u)=0$.\\

{\bf Доказательство}.

      Нетрудно видеть, что 
  $f^{\circ k}(u)$
равно сумме
\be{rrrr444444}
 \sum\limits_{u_1\ldots u_k=u}f(u_1)\ldots f(u_k).\ee
Каждое слагаемое в \re{rrrr444444}
 равно 0, т.к.
если 
$u_1\ldots u_k=u$ и 
$|u|<k$, то $\exists\,i\in\{1,\ldots,k\}:
u_i=\varepsilon$, откуда по предположению
следует, что $f(u_i) =f(\varepsilon) =0.\;\;\blackbox$\\

Если функция $f\in {\bf R}^{X^*}$ такова, что 
$f(\varepsilon)=0$, то запись 
$\sum\limits_{k\geq 1}f^{\circ k}$ обозначает 
функцию из ${\bf R}^{X^*}$, 
значение которой на каждой строке $u\in X^*$
равно сумме всех чисел вида $f^{\circ k}(u)$,
где $k\geq 1$.
Из вышесказанного следует, что лишь конечное число
чисел такого вида будет отлично от 0.

Пусть $f\in {\bf R}^{X^*}$ и 
$f(\varepsilon) \neq 1$. Поскольку функция
$\chi_\varepsilon-f$ принимает на $\varepsilon$
значение, не равное 0, то, согласно теореме \ref{th44433015},
определена функция $(\chi_\varepsilon-f)^{-1}$.
Обозначим записью $f^+$ функцию
\be{sadfsdf4657}f^+\eam (\chi_\varepsilon-f)^{-1}-\chi_\varepsilon.\ee
Мы будем называть функцию $f^+$ {\bf итерацией}
функции $f$.\\

\refstepcounter{theorem}
{\bf Теорема \arabic{theorem}\label{th4443301335}}.

Если функция $f\in {\bf R}^{X^*}$ такова, что 
$f(\varepsilon)=0$, то 
$f^+=\sum\limits_{k\geq 1}f^{\circ k}$.

{\bf Доказательство.}

Доказываемое равенство эквивалентно равенству
\be{fsvgsfsdgdgff}(\chi_\varepsilon-f)\circ
(\chi_\varepsilon+\sum\limits_{k\geq 1}f^{\circ k})=\chi_\varepsilon,\ee
которое, в силу дистрибутивности свёртки относительно сложения
и вычитания, эквивалентно равенству
\be{dsfsadfasdfdsafsa}\chi_\varepsilon- f+\sum\limits_{k\geq 1}f^{\circ k}-
f\circ\sum\limits_{k\geq 1}f^{\circ k}
=\chi_\varepsilon.\ee
\re{dsfsadfasdfdsafsa} можно переписать в виде
$$f+f\circ\sum\limits_{k\geq 1}f^{\circ k}= \sum\limits_{k\geq 1}f^{\circ k}.
$$
Нетрудно доказать, что последнее равенство  верно.
$\blackbox$

\section{Операции на линейных автоматах}

Пусть задано конечное множество $X$.

В пункте \ref{linearautomaton} было введено
понятие линейного автомата (ЛА) над $X$.
На множестве всех ЛА над $X$ можно определить 
операции, аналогичные тем, которые были определены 
в пунктах \ref{sdvsdfdsfvzdxvfrr}
и            \ref{1sdvsdfdsfvzdxvfrr}
на множестве
${\bf R}^{X^*}$.

Для определения этих операций мы 
будем использовать \bi\i обозначения, введенные 
в пункте \ref{fgdsfgdsfgsdgsfdghsfd},\i
а также следующее обозначение:
если $A$ и $B$ -- матрицы, и $A$ имеет вид 
   $\left(\begin{array}{ccccc} a_{11}&\ldots&a_{1n}\cr\ldots&\ldots&\ldots\cr
   a_{m1}&\ldots&a_{mn}\cr\ey\right)$,
то запись $A\otimes B$
   обозначает матрицу 
   $$\left(\begin{array}{ccccc} a_{11}B&\ldots&a_{1n}B\cr\ldots&\ldots&\ldots\cr
   a_{m1}B&\ldots&a_{mn}B\cr\ey\right).$$\ei

Пусть $L_i=(\xi_i^0, \{L_i^x\mid x\in X\},
\lambda_i)\;\;(i=1,2)$ -- ЛА над $X$.
Определим их {\bf сумму} $L_1+L_2$, 
{\bf произведение} $L_1 L_2$ и {\bf свёртку}
$L_1\circ L_2$ следующим образом.

\bi
\i $L_1+ L_2\eam (({\xi_1}^0, {\xi_2}^0), 
\{\left(\by L_1^x&0\cr 0& L_2^x\ey\right)\mid x\in X\},\left(\by \lambda_1\\
\lambda_2\ey\right))$.

\i $L_1 L_2\eam
({\xi_1}^0\otimes {\xi_2}^0,
\{L_1^x\otimes L_2^x\mid x\in X\},  \lambda_1
\otimes \lambda_2)$.
\i $L_1\circ L_2\eam\\\eam (({\xi_1}^0, {\xi_1}^0 M),
\{\left(\by L_1^x&L_1^x M\cr
0&L_2^x\cr\ey\right)\mid x\in X\}, 
\left(\by 0^\downarrow\\ \lambda_2\ey\right))$,\\
где $M=\lambda_1 {\xi_2}^0$, и
$0^\downarrow$ -- вектор-столбец 
с нулевыми компонентами, размерность которого 
равна размерности $\lambda_1$.
\ei

Пусть $L=(\xi^0, \{L^x\mid x\in X\},
\lambda)$ -- ЛА над $X$, и $a\in {\bf R}$.
Определим ЛА
$aL$, 
$\tilde L$ и 
$L^+$ 
(называемые соответственно {\bf произведением}
 $a$ на $L$, {\bf инвертированием} $L$ и {\bf итерацией} $L$)
следующим образом:

\bi
\i $aL\eam(\xi^0, \{L^x\mid x\in X\}, a\lambda)$,
\i $\tilde L\eam (\tilde\lambda, \{(L^x)^\sim\mid x\in X\}, \tilde \xi^0)$,
\i $L^+\eam(\xi^0, \{L^x (E+\lambda\xi^0)\mid x\in X\},
\lambda)$, где $E$ -- единичная матрица.
\ei

\refstepcounter{theorem}
{\bf Теорема \arabic{theorem}\label{th444330133435}}.

\bn\i Пусть $X$ -- конечное множество, и 
$L_1,L_2$ -- ЛА над $X$. Тогда
\be{sdfsdfds55553}f_{L_1+ L_2}=f_{L_1} + f_{L_2},\ee
\be{s4dfsdfds55553}f_{L_1 L_2}=f_{L_1}  f_{L_2},\ee
\be{sd3fsdfds55553}f_{L_1\circ L_2}=f_{L_1} \circ f_{L_2}.\ee
\i Для любого ЛА $L$  
\be{sd3fsdfds555533} 
\forall\,a\in {\bf R}\quad f_{a L}=a f_L,\ee
\be{sd3fsdfds5555333}f_{\tilde L}=(f_L)^\sim,\ee
\be{sd3fsdfds55553343}
\mbox{если $f_L(\varepsilon)=0$,
то $f_{L^+}=(f_L)^+$.}\ee
\en

{\bf Доказательство}.

\bn\i  Пусть $L_i\;(i=1,2)$ имеют вид
$(\xi^0_i, \{L_i^x\mid x\in X\}, \lambda_i)$.
\bn\i
\re{sdfsdfds55553} доказывается непосредственной проверкой,
с использованием того, что 
$$\forall\,u\in X^*\quad
(L_1+L_2)^u=\left(\by L_1^u&0\cr 0& L_2^u\cr\ey\right).$$

\i \re{s4dfsdfds55553}  непосредствено вытекает из следующего
утверждения: 
если матрицы $A,B,C,D$ таковы, что определены
произведения $AC$ и $BD$, то верно 
равенство
$$(A\otimes B)(C\otimes D)=(AC)\otimes(BD).$$

\i Докажем \re{sd3fsdfds55553}.

Нетрудно доказать (индукцией по $|u|$), что $\forall\,u\in X^*$
\be{sadfasdfsadfdsaaf}(L_1\circ L_2)^u=\left(\by L_1^u&
\sum\limits_{u_1u_2=u}L_1^{u_1} M
L_2^{u_2}-ML_2^u\cr
0&L_2^u\cr\ey\right)\ee
где $M=\lambda_1\xi^2_0$.
Поэтому $\forall\,u\in X^*$
$$\by f_{L_1\circ L_2}(u)=(\xi^0_1,\;\xi_1^0 M)
(L_1\circ L_2)^u  \left(\by 0^\downarrow\cr
\lambda_2\ey\right)=
\\=
(\xi^0_1,\;\xi_1^0 M) 
\left(\by \Big(\sum\limits_{u_1u_2=u}L_1^{u_1} M
L_2^{u_2}-ML_2^u\Big)\lambda_2\cr
L_2^u \lambda_2\ey\right)=\\=
\xi^0_1 \Big(\sum\limits_{u_1u_2=u}L_1^{u_1} M
L_2^{u_2}-ML_2^u\Big)\lambda_2+
\xi_1^0 M L_2^u \lambda_2=\\=
\xi^0_1 \Big(\sum\limits_{u_1u_2=u}L_1^{u_1} M
L_2^{u_2}\Big)\lambda_2=
\sum\limits_{u_1u_2=u}\xi^0_1 L_1^{u_1} \lambda_1\xi_0^2
L_2^{u_2}\lambda_2=\\=
\sum\limits_{u_1u_2=u}f_{L_1}(u_1)f_{L_2}(u_2)=
(f_{L_1}\circ f_{L_2})(u).\ey$$
\en
\i Пусть $L$ имеет вид
$(\xi^0, \{L^x\mid x\in X\}, \lambda)$.

\bn\i \re{sd3fsdfds555533} 
доказывается непосредственной проверкой:
$$\by\forall\,u\in X^*&
f_{a L}(u)=\xi^0 L^u (a\lambda)=
a \xi^0 L^u\lambda=\\&=a f_L(u)
=(a f_L)(u).\ey$$

\i Для доказательства \re{sd3fsdfds5555333} мы используем
утверждение о том, 
что если $A,B$ -- матрицы, для которых определено
произведение $AB$, то $(AB)^\sim=\tilde B\tilde A$:
$$\by \forall\,u\in X^*& f_{\tilde L}(u)=
\tilde \lambda 
\tilde L^u\tilde \xi^0=
\tilde \lambda 
(L^{\tilde u})^\sim\tilde \xi^0=\\&=
(\xi^0 L^{\tilde u} \lambda)^\sim=
(f_L(\tilde u))^\sim=f_L(\tilde u).\ey$$

\i
Докажем соотношение \re{sd3fsdfds55553343}, 
т.е. 
$\forall\,u\in X^*$
\be{ds443fsdfsdeee55e}
f_{L^+}(u)=(f_L)^+(u).\ee

Если $u=\varepsilon$, то левая часть \re{ds443fsdfsdeee55e}
равна $$f_{L^+}(\varepsilon) =\xi^0\lambda=f_L(\varepsilon)=0.$$
Нетрудно доказать, что правая часть 
\re{ds443fsdfsdeee55e} в случае $u=\varepsilon$ тоже 
равна 0.

Пусть $u\neq \varepsilon$. Рассмотрим обе части 
\re{ds443fsdfsdeee55e} в этом случае,
 и докажем, что они  совпадают.
\bn
\i Согласно теореме \ref{th4443301335},
из 
$f_L(\varepsilon)=0$ следует, что правая часть 
\re{ds443fsdfsdeee55e} равна сумме 
$\sum\limits_{k\geq 1}f_L^{\circ k}(u)$, которая 
равна сумме  
\be{dsfsdfsdeeeee4e}\sum\limits_{\scriptsize \begin{array}{c} 
k\geq 1,\;
u_1\ldots u_k=u\\
\forall\,i\in\{1,\ldots, k\}\;\;
u_i\neq \varepsilon \ey}
f_L(u_1)\ldots f_L(u_k).\ee

\i Левая часть 
\re{ds443fsdfsdeee55e} имеет вид 
\be{sdfsadf44445}
\xi^0L^{x_1}(E+\lambda\xi^0)\ldots
L^{x_k}(E+\lambda\xi^0)\lambda\ee
где $u=x_1\ldots x_k$, $\forall\,i=1,\ldots, k\;\;x_i\in X$.

Т.к. $\xi^0\lambda=0$, то $(E+\lambda\xi^0)\lambda= \lambda$,
поэтому можно переписать \re{sdfsadf44445} в виде
\be{sdfsadf4444445}
\xi^0 L^{x_1}(E+\lambda\xi^0)
\ldots L^{x_k}\lambda.\ee
Раскроем все скобки
\re{sdfsadf4444445}. Нетрудно видеть, что каждое слагаемое
в получившейся сумме имеет вид
\be{sadfsadfsdfdsa445f}
\xi^0 L^{u_1}\lambda \ldots \xi^0 L^{u_k}\lambda \ee
где $u_1\ldots u_k=u$ и $\forall\,i\in\{1,\ldots, k\}\;\;
u_i\neq \varepsilon$.
Согласно определению функции $f_L$,
 \re{sadfsadfsdfdsa445f} совпадает с 
 соответствующим слагаемым (определяемым 
 тем же 
представлением $u$ 
 в виде конкатенации непустых подстрок $u_1\ldots u_k$)
 в сумме
\re{dsfsdfsdeeeee4e}.

\en
Таким образом, каждому слагаемому в разложении выражения
\re{sdfsadf4444445} 
взаимно однозначно соответствует
равное ему слагаемое  в сумме \re{dsfsdfsdeeeee4e}. 
Следовательно, значения выражений
\re{sdfsadf4444445} и
\re{dsfsdfsdeeeee4e} совпадают.
$\blackbox$
\en
\en

\section{Алгебраические свойства мно\-же\-с\-т\-ва
ли\-ней\-но-\-ав\-то\-мат\-ных
функций}\label{dfgdsfgsfdgsfdfs}

Пусть задано конечное множество $X$.

Обозначим символом 
${\cal L}^{X}$ множество всех ЛАФ из ${\bf R}^{X^*}$.

Отметим, что $\chi_\varepsilon\in
{\cal L}^{X}$, т.к.
$\chi_\varepsilon=f_L$, где 
 $L$ -- ЛА вида \re{vdsvczvzxcvzx} размерности 1,
у которого $\xi^0=(\rm{1})$, 
$\forall\,x\in X\;\;L^x=(0)$, 
$\lambda=(\rm{1})$.

Согласно вышесказанному и
теореме \ref{th44433015ee3s},
${\cal L}^{X}$ является
кольцом относительно 
определённых в пункте \ref{sdvsdfdsfvzdxvfrr}
операций сложения и свёртки (рассматриваемой 
в данном случае как умножение).
Единицей этого 
кольца является
ЛАФ $\chi_\varepsilon$.
Ниже мы будем опускать символ свёртки 
$\circ$  в обозначении
произведения элементов кольца ${\cal L}^{X}$
(т.е. для любых элементов $a,b$ кольца ${\cal L}^{X}$
 будем обозначать их произведение
$a \circ b$ записью $ab$).

Обозначим записью 
${\cal L}^{X}_n$ кольцо  квадратных матриц порядка $n$
над  ${\cal L}^{X}$. Единицей этого кольца является диагональная
матрица $E_\varepsilon$, все компоненты диагонали 
которой равны
$\chi_\varepsilon$. 

$\forall\, M\in {\cal L}^{X}_n$ и $\forall\,u\in X^*$
мы будем обозначать записью $M(u)$ матрицу 
над ${\bf R}$, каждая компонента $M(u)_{ij}$
которой равна значению соответствующей функции $M_{ij}$
на строке $u$.\\

\refstepcounter{theorem}
{\bf Теорема \arabic{theorem}\label{th4443301334345}}.

Пусть $X$ -- конечное множество, и 
матрица $A\in {\cal L}^{X}_n$ такова, что 
матрица $A(\varepsilon)$ -- нулевая. Тогда
существует матрица из ${\cal L}^{X}_n$, 
обозначаемая записью $\sum\limits_{k\geq 1}A^k$,
для которой верно равенство
\be{fgdffgdgsdff5555}\Big(E_\varepsilon-A\Big)\Big(E_\varepsilon+\sum\limits_{k\geq 1}A^k\Big)=
E_\varepsilon.\ee


{\bf Доказательство}. 


$\forall\,u\in X^*$,
$\forall\,k> |u|$ матрица $A^k(u)$ является нулевой, 
т.к. каждая компонента матрицы $A^k$ является
суммой произведений вида $a_{i_1j_1}\ldots a_{i_kj_k}$,
где $a_{i_sj_s}\;(s=1,\ldots, k)$ -- компоненты матрицы $A$.
Нетрудно видеть, что
\be{rrrr44444443}
(a_{i_1j_1}\ldots a_{i_kj_k})(u)=
 \sum\limits_{u_1\ldots u_k=u}a_{i_1j_1}(u_1)\ldots 
 a_{i_kj_k}(u_k).\ee
Каждое слагаемое в сумме в правой части
\re{rrrr44444443}
 равно 0, т.к.
если 
$u_1\ldots u_k=u$ и 
$|u|<k$, то $\exists\,s\in\{1,\ldots,k\}:
u_s=\varepsilon$, откуда по предположению
следует, что $a_{i_sj_s}(u_s) =a_{i_sj_s}(\varepsilon) =0.$
  
Искомая матрица $\sum\limits_{k\geq 1}A^k$
определяется как
матрица из ${\cal L}^{X}_n$,
значение которой на каждой строке $u\in X^*$
равно сумме всех матриц  вида $A^{k}(u)$,
где $k\geq 1$.
Из вышесказанного следует, что лишь конечное число
матриц такого вида  отлично от нулевой.

Равенство \re{fgdffgdgsdff5555}
доказывается так же, как доказывается
аналогичное ему равенство \re{fsvgsfsdgdgff},
и следует из того, что $\forall\,u\in X^*$
\be{dsfsdfsdfdsfsa}\Big(A+A\,\Big(\,\sum\limits_{k\geq 1}A^k\,\Big)\Big)(u)=
\Big(\sum\limits_{k\geq 1}A^k\Big)(u).\ee

Обоснуем \re{dsfsdfsdfdsfsa}. 

Пусть  $B=
\sum\limits_{k= 1}^{|u|}A^k$ и 
$C=
\sum\limits_{k>|u|}A^k$.
Нетрудно видеть, что 
\bi\i матрица
$(AC)(u)$ -- нулевая, т.к.
каждая компонента матрицы $AC$ является суммой произведений
вида $a_{ij}c_{jk}$, где $a_{ij}$ и $c_{jk}$  --
компоненты $A$
и $C$ соответственно, и поскольку 
\be{sdfsdafs4455}
(a_{ij}c_{jk})(u)=\sum\limits_{u_1u_2=u}a_{ij}(u_1)c_{jk}(u_2)\ee
то из $|u_2|\leq |u|$ следует,
что  матрица $C(u_2)$ является нулевой, в частности
$c_{jk}(u_2) = 0$, поэтому правая (а значит и левая) часть
\re{sdfsdafs4455} равна 0,
\i   $\sum\limits_{k\geq 1}A^k =B+C$, поэтому,
используя утверждение из предыдущего пункта, левую часть
\re{dsfsdfsdfdsfsa} можно переписать в виде
\be{sdfsadfsdgfdgdsfgt}\!\!\!\!\!
\by (A+A(B+C))(u)=A(u)+(AB)(u)+(AC)(u)=\\= A(u)+(AB)(u)+0=
A(u)+\sum\limits_{k=2}^{|u|+1}A^k(u).
\ey\ee
Значение последнего выражения в цепочке 
равенств \re{sdfsadfsdgfdgdsfgt} 
совпадает с правой частью \re{dsfsdfsdfdsfsa}.
    $\blackbox$
\ei      

Пусть $X$ -- конечное множество.
Мы будем использовать следующие обозначения.
\bi \i Символ
$\Omega$ обозначает совокупность операций
на ${\bf R}^{X^*}$, состоящую из 
определённых в пункте \ref{sdvsdfdsfvzdxvfrr}
операций сложения, свёртки, 
итерации (для тех функций, которые отображают $\varepsilon$ в 0)
и умножения на числа из ${\bf R}$.
\i 
Для произвольного подмножества $F\subseteq {\bf R}^{X^*}$
запись $\Omega F$ обозначает
множество всех функций из ${\bf R}^{X^*}$, которые 
могут быть получены из 
функций из $F$
при помощи применения операций из $\Omega$.
\ei

\refstepcounter{theorem}
{\bf Теорема \arabic{theorem}\label{th44433013343425}}.

Пусть $X$ -- конечное множество, $n$ -- натуральное число, 
$A$ -- матрица из ${\cal L}^{X}_n$, 
и $B,C$ -- вектор-столбцы размерности $n$
над ${\cal L}^{X}$, 
имеющие вид 
$$A=\left(\by 
a_{11}&\ldots&a_{1n}\cr
\ldots&\ldots&\ldots\cr
a_{n1}&\ldots&a_{nn}\cr
\ey\right),\quad
B=\left(\by b_1\cr\ldots\cr b_n\ey\right),\quad
C=\left(\by c_1\cr\ldots\cr c_n\ey\right)$$
причём матрица $A(\varepsilon)$ -- нулевая,
и верно равенство
\be{dsfgdfgt44t6y}(E_\varepsilon-A)B=C.\ee

Тогда 
\be{dsffgsdgsd5554}\forall\,i=1,\ldots, n\quad
b_i\in \Omega\{\chi_\varepsilon,
a_{ij},
c_i\mid i,j=1,\ldots, n\}.\ee

{\bf Доказательство}. 

Докажем теорему индукцией по $n$.

Если $n=1$, то \re{dsfgdfgt44t6y} имеет вид
$(\chi_\varepsilon-a_{11}) b_1=c_1$.
Поскольку по предположению $a_{11}(\varepsilon)=0$,
то к $a_{11}$ можно применить операцию
итерации, и, согласно определению 
\re{sadfsdf4657}, верны равенства
$$b_1=
(\chi_\varepsilon+a^+_{11})
(\chi_\varepsilon-a_{11}) b_1=(\chi_\varepsilon+a^+_{11})c_1$$
т.е. в данном случае  \re{dsffgsdgsd5554} верно.

Пусть $n>1$. 
Перепишем \re{dsfgdfgt44t6y} в виде
системы равенств
\be{dsffdsgsdfeeew3}\left\{\begin{array}{llllll} 
(\chi_\varepsilon-a_{11}) b_1-\ldots
-a_{1,n-1} b_{n-1}-a_{1n} b_n&=c_1\cr
\ldots\cr
-a_{n1} b_1-\ldots-a_{n,n-1}
 b_{n-1}+(\chi_\varepsilon-a_{nn}) b_n&=c_n\cr
\ey\right.\ee

Используя рассуждения, аналогичные вышеизложенным, 
получаем, что последнее равенство в \re{dsffdsgsdfeeew3}
равносильно равенству
\be{sdfgdfgfsdgsfdsr}
b_n=(\chi_\varepsilon+a_{nn}^+)(c_n+a_{n1}b_1+
\ldots+a_{n,n-1}b_{n-1}).\ee
Заменим в \re{dsffdsgsdfeeew3}
во всех равенствах, кроме последнего, 
выражение $b_n$ на правую часть 
равенства \re{sdfgdfgfsdgsfdsr}.
После раскрытия скобок и приведения подобных членов
совокупность этих $n-1$ равенств будет
представлять собой систему, 
которая в матричной записи имеет вид
\be{dsfgdf44gt44t6y}(E_\varepsilon-A')B'=C',\ee
где компоненты $a'_{ij}, b'_{i}, c'_{i}$
$(i,j=1,\ldots, n-1)$ матриц $A',B',C'$ имеют следующий вид:
\be{afgsdgdsfs55655376}\by
a'_{ij}=
a_{ij}+a_{in} (\chi_\varepsilon+a_{nn}^+)a_{nj},\cr
b'_i=b_i,\cr
c'_i=c_i+a_{in} (\chi_\varepsilon+a_{nn}^+)c_n.
\ey\ee

Поскольку $\forall\,f,g\in {\bf R}^{X^*}$ верно равенство
$(fg)(\varepsilon)=f(\varepsilon)g(\varepsilon)$, то
из \re{afgsdgdsfs55655376} и из того, что матрица $A(\varepsilon)$
нулевая, следует, что матрица $A'(\varepsilon)$
нулевая. Таким образом, к матрицам $A',B',C'$
можно применить индуктивное предположение,
согласно которому будет верно утверждение теоремы
\ref{th44433013343425}, в котором матрицы $A,B,C$
заменены на $A',B',C'$ соответственно, т.е. 
\be{dsffgsd3gsd5554}\forall\,i=1,\ldots, n-1\quad
b_i\in \Omega\{\chi_\varepsilon,
a'_{ij},
c'_i\mid i,j=1,\ldots, n-1\}.\ee
Из \re{afgsdgdsfs55655376} следует, что 
\be{afg44sdgdsfs55655376}
\forall\,i,j=1,\ldots, n-1\quad
a'_{ij}, c'_i\in \Omega\{\chi_\varepsilon,
a_{ij},
c_i\mid i,j=1,\ldots, n\}.\ee
Из \re{dsffgsd3gsd5554} и \re{afg44sdgdsfs55655376}
следует, что 
\be{dsf55fgsd3gsd5554}\forall\,i=1,\ldots, n-1\quad
b_i\in \Omega\{\chi_\varepsilon,
a_{ij},
c_i\mid i,j=1,\ldots, n\}.\ee
Из \re{sdfgdfgfsdgsfdsr} и \re{dsf55fgsd3gsd5554}
следует, что 
\be{dsf4455fgsd3gsd5554}
b_n\in \Omega\{\chi_\varepsilon,
a_{ij},
c_i\mid i,j=1,\ldots, n\}.\ee
Из \re{dsf55fgsd3gsd5554} и
\re{dsf4455fgsd3gsd5554}
следует утверждение теоремы \ref{th44433013343425}.
$\blackbox$\\

\refstepcounter{theorem}
{\bf Теорема \arabic{theorem}\label{th444330133433245}}.

Пусть $X$ -- конечное множество.
Тогда 
\be{asfdsd4465}{\cal L}^{X}=\Omega \{\chi_\varepsilon, \chi_x\mid
x\in X\}.\ee

{\bf Доказательство}.

Как было отмечено в начале пункта \ref{dfgdsfgsfdgsfdfs}, 
$\chi_\varepsilon\in {\cal L}^{X}$. Кроме того, 
$\forall\,x\in X\;\;\chi_x\in {\cal L}^{X}$, 
т.к. $\chi_x=f_L$, где 
$L$ -- ЛА  вида \re{vdsvczvzxcvzx} размерности 2, 
компоненты которого имеют следующий вид:
$$\mbox{$\xi^0=(1,0),\;L^x=\left(\begin{array}{ccccc} 0&1\cr
0&0\cr\ey\right)$, $L^{y}=\left(\begin{array}{ccccc} 0&0\cr
0&0\cr\ey\right)$ для $y\neq x$,
$\lambda=\left(\begin{array}{ccccc} 0\cr
1\cr\ey\right)$.}$$
Согласно теореме 
\ref{th444330133435}, 
${\cal L}^{X}$ замкнуто относительно операций из $\Omega$.
Следовательно, правая часть соотношения
\re{asfdsd4465} содержится в левой его части.

Докажем, что верно и обратное включение.

Пусть $f=f_L$, где $L$ -- ЛА вида
$(\xi^0,\{L^x\mid x\in X\}, \lambda)$.

$\forall\,u\in X^*$ обозначим записью $L_u$
матрицу из  ${\cal L}^{X}_n$ 
(где $n$ -- размерность ЛА $L$), 
имеющую вид $L^u \chi_u$
(т.е. каждая компонента $L_u$ представляет собой
произведение соответствующей компоненты матрицы
$L^u$ и ЛАФ $\chi_u$).

Обозначим символом $A$ матрицу $\sum\limits_{x\in X}L_x$.
Нетрудно видеть, что 
\be{dffdsdrrtt677}\forall\,k\geq 1\;\;
A^k=\sum\limits_{x_1,\ldots, x_k\in X}L_{x_1}\ldots L_{x_k}=
\sum\limits_{u\in X^k}L_u.\ee
Последнее равенство в \re{dffdsdrrtt677}
верно потому, что
\be{dffdsdrrtt6773}\forall\,x_1,\ldots, x_k\in X\quad
L_{x_1}\ldots L_{x_k}=
L_{x_1\ldots x_k}.\ee \re{dffdsdrrtt6773} следует из равенства
$\chi_{x_1}\ldots \chi_{x_k}=\chi_{x_1\ldots x_k}$.

Поскольку матрица $A(\varepsilon)$ -- нулевая, 
то, согласно теореме \re{th4443301334345},
существует матрица
$\sum\limits_{k\geq 1}A^k$, для которой 
верно равенство
\re{fgdffgdgsdff5555}.

Компоненты матрицы $A$ представляют собой 
суммы ЛАФ вида $a\chi_x$, где $a\in {\bf R}$
и $x\in X$. Следовательно, 
используя равенство
\re{fgdffgdgsdff5555} и 
теорему \ref{th44433013343425}, применяемую к столбцам 
матрицы $E_\varepsilon+\Big(\sum\limits_{k\geq 1}A^k\Big)$,
можно заключить, что 
компоненты этой матрицы 
принадлежат множеству 
$\Omega \{\chi_\varepsilon, \chi_x\mid x\in X\}$.

Из \re{dffdsdrrtt677} следует, что 
\be{asdfdsgsfdgfsdgeee}\forall\,u\in X^*\quad
\Big(E_\varepsilon+\Big(\sum\limits_{k\geq 1}A^k\Big)\Big)(u)=L^u.\ee

Напомним, что $\forall\,u\in X^*$
\be{sfgsfdgfsdgsd443332}
f_L(u)=\xi^0L^u\lambda.\ee
Обозначим компоненты векторов $\xi^0$, $\lambda$ и матриц
$L^u$,
$E_\varepsilon+\Big(\sum\limits_{k\geq 1}A^k\Big)$
записями $\xi^0_i$, $\lambda_j$, $L^u_{ij}$ и
$a_{ij}$ соответственно (где $i,j=1,\ldots, n$). 
В этих обозначениях
равенство \re{sfgsfdgfsdgsd443332}
можно переписать в виде
\be{sfgsfdgfsdgsd43344567}
f_L(u)=\sum\limits_{i=1}^{n}\sum\limits_{j=1}^{n}
\xi^0_iL^u_{ij}\lambda_j\ee

Определим функцию
$f$ как линейную комбинацию функций $a_{ij}$
вида
\be{dwfsdsffsgsdf4443333}
f\;\eam \;\sum\limits_{i=1}^{n}\sum\limits_{j=1}^{n}
\xi^0_i\lambda_ja_{ij}.\ee

Докажем, что $f$ совпадает с $f_L$.
По определению $f$,
$\forall\,u\in X^*$
\be{sdfdsfasdfsafdasgsdf}
f(u)=\sum\limits_{i=1}^{n}\sum\limits_{j=1}^{n}
\xi^0_i\lambda_ja_{ij}(u).\ee
Из \re{asdfdsgsfdgfsdgeee} следует, что
$\forall\,i,j=1,\ldots,n\;\;
a_{ij}(u)=L^u_{ij}$, поэтому  можно переписать 
\re{sdfdsfasdfsafdasgsdf}
в виде
\be{sd32fdsfasdfsafdasgsdf}
f(u)=\sum\limits_{i=1}^{n}\sum\limits_{j=1}^{n}
\xi^0_i\lambda_jL^u_{ij}.\ee
Поскольку правая часть \re{sd32fdsfasdfsafdasgsdf}
совпадает с правой частью
\re{sfgsfdgfsdgsd43344567}, то, следовательно,
и левые части этих  равенств совпадают, т.е.
верно равенство $f(u)=f_L(u)$, что и требовалось доказать.

Как было отмечено выше, 
все функции $a_{ij}$  принадлежат  множеству
$\Omega \{\chi_\varepsilon, \chi_x\mid x\in X\}$, поэтому,
согласно определению 
\re{dwfsdsffsgsdf4443333},
функция $f$ (т.е. $f_L$) тоже принадлежит этому  множеству.
$\blackbox$

\section{Линейные пространства, связанные с ли\-ней\-но-\-ав\-то\-мат\-ны\-ми фун\-к\-ци\-я\-ми}

Пусть $X$ -- конечное множество, и $f\in {\bf R}^{X^*}$.

Мы будем использовать следующие обозначения:
$\forall\,u\in X^*\;\;f_u$ обозначает
функцию из ${\bf R}^{X^*}$, определяемую следующим образом:
$$\forall\,u_1\in X^*\quad f_u(u_1)\eam f(uu_1),$$
и $E_f$ 
обозначает подпространство
линейного пространства ${\bf R}^{X^*}$, 
порождённое
множеством
$\{f_u\mid u\in X^*\}$.\\


\refstepcounter{theorem}
{\bf Теорема \arabic{theorem}\label{th4443301334333245}}.

Пусть $X$ -- конечное множество, $f\in {\bf R}^{X^*}$,
и $n$ -- натуральное число.
Следующие условия 
эквивалентны:
\bn
\i существует ЛА $L$ размерности не больше $n$ 
над $X$, такой, что $f_L=f$,
\i $\dim E_f\leq n$.
\en

{\bf Доказательство}.

Докажем, что из условия 1 следует условие 2. 

Пусть ЛА $L=(\xi^0, \{L^x\mid x\in  X\}, \lambda)$ 
имеет размерность $k\leq n$ и удовлетворяет условию
$f_L=f$, т.е. 
$$\forall\,u\in X^*\quad f(u)=\xi^0L^u\lambda,$$
откуда следует, что $\forall\,u\in X^*$ функция $f_u$
удовлетворяет  условию
$$\forall\,u_1\in X^*\quad 
f_u(u_1)=f(uu_1)=\xi^0L^{uu_1}\lambda=
(\xi^0L^{u})L^{u_1}\lambda.$$
 $\forall\,i=1,\ldots,k$ обозначим 
\bi\i записью $e_i$
вектор--строку из ${\bf R}^k$, $i$--я компонента которой равна 1, 
а все остальные компоненты равны 0, и
\i записью $f_i$ -- функцию из ${\bf R}^{X^*}$,
определяемую следующим образом:
$$\forall\,u\in X^*\quad f_i(u)=e_iL^u\lambda.$$
\ei

Поскольку $\{e_1,\ldots, e_k\}$ -- базис в ${\bf R}^k$,
то существуют числа $a_1,\ldots, a_k\in {\bf R}$,
такие, что $\xi^0L^{u} = \sum\limits_{i=1}^k a_ie_i$.
Следовательно, 
$$\forall\,u_1\in X^*\quad 
f_u(u_1)=\sum\limits_{i=1}^k a_ie_i L^{u_1}\lambda=
\sum\limits_{i=1}^k a_if_i(u_1),$$
т.е. $f_u=\sum\limits_{i=1}^k a_if_i$.

Таким образом, $\forall\,u\in X^*$ функция $f_u$
принадлежит линейному подпространству в ${\bf R}^{X^*}$,
порождённому функциями $f_1,\ldots, f_k$.
Отсюда непосредственно следует условие 2.

Теперь докажем, что из условия 2 следует условие 1.

Определим компоненты $\xi^0, L^x\;(x\in X)$, $\lambda$
искомого ЛА $L$ следующим образом.
Выберем базис  в пространстве $E_f$, имеющий вид 
$\{f_{u_1}, \ldots, f_{u_k}\}$, где $u_1,\ldots, u_k\in X^*$.
\bi
\i Т.к. $f=f_\varepsilon\in E_f$, 
то $\exists\,\xi^0_1,\ldots, \xi^0_k\in {\bf R}:
f = \sum\limits_{i=1}^k \xi^0_if_{u_i}.$

Определим $\xi^0\eam (\xi^0_1,\ldots, \xi^0_k)$.
\i $\forall\,x\in X\;\;L^x\eam 
\left(\by 
a_{11}&\ldots&a_{1k}\cr
\ldots&\ldots&\ldots\cr
a_{k1}&\ldots&a_{kk}\cr
\ey\right)$, где $\forall \,i=1,\ldots,k$ 
числа $a_{i1}$, $\ldots$, $a_{ik}$
являются коэффициентами разложения функции 
$f_{u_ix}$ по базису $\{f_{u_1}, \ldots, f_{u_k}\}$:
$f_{u_ix}=\sum\limits_{j=1}^k a_{ij}f_{u_j}.$
\i $\lambda\eam 
\left(\by f_{u_1}(\varepsilon)\cr\ldots\cr 
f_{u_k}(\varepsilon)\ey\right).$
\ei

Докажем (индукцией по $|u|$), что $\forall\,u,v\in X^*$ верно равенство
\be{sdfdsfdsfsdf}
L^u \left(\by f_{u_1}(v)\cr\ldots\cr 
f_{u_k}(v)\ey\right)=\left(\by f_{u_1}(uv)\cr\ldots\cr 
f_{u_k}(uv)\ey\right).\ee 
\bi
\i Если $u=\varepsilon$, то \re{sdfdsfdsfsdf}
верно по определению $\lambda$.
\i Пусть \re{sdfdsfdsfsdf} верно для некоторого $u$
и произвольного $v$.
Тогда 
$\forall\,x\in X$
$$\by
L^{ux} \left(\by f_{u_1}(v)\cr\ldots\cr 
f_{u_k}(v)\ey\right)=
L^{u}  L^{x} \left(\by f_{u_1}(v)\cr\ldots\cr 
f_{u_k}(v)\ey\right)=
L^{u}   \left(\by \sum\limits_{j=1}^k a_{1j}f_{u_j}(v)\cr\ldots\cr 
\sum\limits_{j=1}^k a_{kj}f_{u_j}(v)\ey\right)=\\=
L^{u}   \left(\by f_{u_1x}(v)\cr\ldots\cr 
f_{u_kx}(v)\ey\right)=
L^{u}   \left(\by f_{u_1}(xv)\cr\ldots\cr 
f_{u_k}(xv)\ey\right)=
\left(\by f_{u_1}(uxv)\cr\ldots\cr 
f_{u_k}(uxv)\ey\right).
\ey$$
\ei
Таким образом, \re{sdfdsfdsfsdf} верно для произвольных $u,v\in 
X^*$.

Используя \re{sdfdsfdsfsdf},
докажем, что $f_L=f$.
$$\by \forall\,u\in X^*\quad 
f_L(u)=\xi^0L^u\lambda=\xi^0L^u\left(\by f_{u_1}(\varepsilon)\cr\ldots\cr 
f_{u_k}(\varepsilon)\ey\right)=\\=\xi^0
\left(\by f_{u_1}(u)\cr\ldots\cr 
f_{u_k}(u)\ey\right)=\sum\limits_{i=1}^k\xi^0_{i}f_{u_i}(u)
=f(u).\;\blackbox
\ey$$

Из теоремы \ref{th4443301334333245} следует, что 
если для функции $f\in {\bf R}^{X^*}$ существует ЛА
$L$ над $X$, такой, что $f_L=f$, то $\dim E_f$
является наименьшей возможной размерностью такого ЛА.\\



Пусть $X$ -- конечное множество.
Мы будем обозначать записью 
${\bf R}\langle X\rangle$ кольцо многочленов
с коэффициентами из ${\bf R}$
от некоммутирующих переменных из множества $X$.
Каждый элемент $p\in {\bf R}\langle X\rangle$ можно рассматривать как
формальную сумму вида 
$\sum\limits_{i=1}^n a_iu_i$, где $\forall\,i=1,\ldots, n\;\;
a_i\in {\bf R}, 
u_i\in X^*$.

Для каждой функции $f\in {\bf R}^{X^*}$
мы будем обозначать тем же символом $f$
линейное продолжение этой функции на ${\bf R}\langle X\rangle$,
т.е. функцию вида ${\bf R}\langle X\rangle\to {\bf R}$
  определяемую следующим образом:
$$\forall\,p\in {\bf R}\langle X\rangle,\;\;
\mbox{если }p=\sum\limits_{i=1}^n a_iu_i,
\mbox{ то }f(p)\eam \sum\limits_{i=1}^n a_if(u_i).$$

  $\forall\,f\in {\bf R}^{X^*}$ мы будем обозначать
  записью $\sim_f$
 отношение эквивалентности на ${\bf R}\langle X\rangle$,
  определяемое следующим образом:
$$\forall\,p_1,p_2\in {\bf R}\langle X\rangle\;\;
p_1\sim_f p_2\;\Leftrightarrow\;
\forall\,u\in X^*\;\;f(p_1u)=f(p_2u).$$

Для каждого $p\in {\bf R}\langle X\rangle$ мы будем 
обозначать записью $[p]$
класс эквивалентности $\sim_f$, содержащий $p$.
Нетрудно видеть, что $\sim_f$ сохраняет операции
сложения и умножения на числа из ${\bf R}$,
поэтому определено факторпространство 
${\bf R}\langle X\rangle/_{\sim_f}$.\\

\refstepcounter{theorem}
{\bf Теорема \arabic{theorem}\label{th44433013343332245}}.

Пусть $X$ -- конечное множество, $f\in {\bf R}^{X^*}$. Тогда 
$$\dim E_f=
\dim\Big({\bf R}\langle X\rangle/_{\sim_f}\Big).$$

{\bf Доказательство}.

Докажем, что для каждого натурального числа $n$
верно соотношение
\be{sdfsdfsdfgsadgsfd}
\dim E_f\leq n\;\;\Leftrightarrow\;\;
\dim {\bf R}\langle X\rangle/_{\sim_f} \leq n.\ee

Левая часть \re{sdfsdfsdfgsadgsfd}
 равносильна условию
\be{fdsdfsdgsdfgdttt55}
\by
\exists\,u_1,\ldots,u_n\in X^*:
\forall\,u\in X^*\;\exists \,a_1,\ldots, a_n\in {\bf R}:\\
\forall\,v\in X^*\quad f(uv)=\sum\limits_{i=1}^na_if({u_i}v).\ey\ee

Правая часть \re{sdfsdfsdfgsadgsfd}
равносильна условию
\be{fdsdfsdgsdf2gdttt553}
\by
\exists\,p_1,\ldots,p_n\in {\bf R}\langle X\rangle:
\forall\,p\in {\bf R}\langle X\rangle
\;\exists \,a_1,\ldots, a_n\in {\bf R}:\\
\mbox{$[p]$}
=\sum\limits_{i=1}^na_i[p_i].
\ey\ee

Нетрудно доказать, что 
\re{fdsdfsdgsdf2gdttt553} эквивалентно 
условию
\be{fds44dfsdgsdf2gdttt553}
\by
\exists\,u_1,\ldots,u_n\in {X^*}:
\forall\,u\in {X^*}
\;\exists \,a_1,\ldots, a_n\in {\bf R}:\\
\mbox{$[u]$}
=\sum\limits_{i=1}^na_i[u_i]=[\sum\limits_{i=1}^na_iu_i].
\ey\ee
Равенство $[u]=[\sum\limits_{i=1}^na_iu_i]$
равносильно условию $u\sim_f \sum\limits_{i=1}^na_iu_i$, т.е.
соотношению
$$\forall\,v\in X^*\quad
 f(uv)=f(\Big(\sum\limits_{i=1}^na_iu_i\Big)v)=
\sum\limits_{i=1}^na_if(u_iv).$$

Используя последнее замечание, заключаем, что 
условия 
\re{fdsdfsdgsdfgdttt55}
и
\re{fds44dfsdgsdf2gdttt553} эквивалентны,
откуда следует доказываемое соотношение 
\re{sdfsdfsdfgsadgsfd}.
$\blackbox$\\

Пусть $X$ -- конечное множество, и $f\in {\bf R}^{X^*}$.

Мы будем использовать следующие обозначения.
\bi\i
$\forall\,u\in X^*$ запись $\hat f_u$ обозначает функцию из 
${\bf R}^{X^*\times X^*}$,
т.е. 
$$\hat f_u: X^*\times X^*\to {\bf R},$$ определяемую следующим образом:
$$\forall\,v_1,v_2\in X^*\quad \hat f_u(v_1,v_2)\eam f(v_1uv_2).$$
\i Запись $E_{\hat f}$  
обозначает подпространство 
линейного пространства 
${\bf R}^{X^*\times X^*}$, порождённое
множеством
$\{\hat f_u\mid u\in X^*\}$. 
\i Запись
$I_f$ обозначает идеал кольца ${\bf R}\langle X\rangle$,
определяемый следующим образом:
$$I_f\eam \{p\in {\bf R}\langle X\rangle\mid \forall\,p_1,p_2
\in {\bf R}\langle X\rangle\;
f(p_1pp_2)=0\}
$$
Нетрудно доказать, что $I_f$ действительно 
является идеалом, и, следовательно,
определено фактор-кольцо
${\bf R}\langle X\rangle/{I_f}$, которое также является
линейным пространством над ${\bf R}$.
$\forall\,p\in {\bf R}\langle X\rangle$ мы будем обозначать
соответствующий элемент  фактор-кольца 
${\bf R}\langle X\rangle/{I_f}$
записью $[p]$.
\ei

\refstepcounter{theorem}
{\bf Теорема \arabic{theorem}\label{th4443301334333224335}}.

Пусть $X$ -- конечное множество, $f\in {\bf R}^{X^*}$. Тогда 
$$\dim E_{\hat f}=
\dim {\bf R}\langle X\rangle/{I_f}.$$

{\bf Доказательство}.

Докажем, что для каждого натурального числа $n$
верно соотношение
\be{sdfsdfsdfgsadgs33fd}
\dim E_{\hat f} \leq n\;\;\Leftrightarrow\;\;
\dim {\bf R}\langle X\rangle/{I_f} \leq n.\ee

Левая часть \re{sdfsdfsdfgsadgs33fd}
 равносильна условию
\be{fdsdfsdgsdfgdttt5335}
\by
\exists\,u_1,\ldots,u_n\in X^*:
\forall\,u\in X^*\;\exists \,a_1,\ldots, a_n\in {\bf R}:\\
\forall\,v_1,v_2\in X^*\quad 
f(v_1uv_2)=\sum\limits_{i=1}^na_i f({v_1u_iv_2}).\ey\ee

Правая часть \re{sdfsdfsdfgsadgs33fd}
равносильна условию
\be{1fdsdfsdgsdf2gdttt553}
\by
\exists\,p_1,\ldots,p_n\in {\bf R}\langle X\rangle:
\forall\,p\in {\bf R}\langle X\rangle
\;\exists \,a_1,\ldots, a_n\in {\bf R}:\\
\mbox{$[p]$}
=\sum\limits_{i=1}^na_i[p_i].
\ey\ee

Нетрудно доказать, что 
\re{1fdsdfsdgsdf2gdttt553} эквивалентно 
условию
\be{1fds44dfsdgsdf2gdttt553}
\by
\exists\,u_1,\ldots,u_n\in {X^*}:
\forall\,u\in {X^*}
\;\exists \,a_1,\ldots, a_n\in {\bf R}:\\
\mbox{$[u]$}
=\sum\limits_{i=1}^na_i[u_i]=[\sum\limits_{i=1}^na_iu_i].
\ey\ee
Равенство $[u]=[\sum\limits_{i=1}^na_iu_i]$
равносильно условию $u-\sum\limits_{i=1}^na_iu_i\in I_f$, 
т.е.
соотношению
$$\forall\,v_1,v_2\in X^*\quad
f(v_1(u-\sum\limits_{i=1}^na_iu_i)v_2)=0$$
которое можно переписать в виде
$$\forall\,v_1,v_2\in X^*\quad
f(v_1uv_2)=\sum\limits_{i=1}^na_i
f(v_1u_iv_2).$$
 
Используя последнее замечание, заключаем, что 
условия 
\re{fdsdfsdgsdfgdttt5335}
и
\re{1fds44dfsdgsdf2gdttt553} эквивалентны,
откуда следует доказываемое соотношение 
\re{sdfsdfsdfgsadgs33fd}.
$\blackbox$\\

\refstepcounter{theorem}
{\bf Теорема \arabic{theorem}\label{th44433013343332243335}}.

Пусть $X$ -- конечное множество, $f\in {\bf R}^{X^*}$. Тогда 
для каждого натурального числа $n$
верны импликации
\be{1sdfsdfsdfgsa32dg3s33fd}
\dim E_{\hat f}\leq n\;\;\Rightarrow\;\;
\dim E_{f}\leq n,\ee
\be{2sdfsdfsdfgsa32dg3s33fd}
\dim E_{f}\leq n\;\;\Rightarrow\;\;
\dim E_{\hat f}\leq n^2.\ee

{\bf Доказательство.}

Докажем импликацию \re{1sdfsdfsdfgsa32dg3s33fd}.

Как было отмечено в доказательстве теоремы 
\ref{th4443301334333224335}, 
левая часть \re{1sdfsdfsdfgsa32dg3s33fd}
 равносильна условию
\re{fdsdfsdgsdfgdttt5335}, из которого
(полагая $v_1\eam\varepsilon$) получаем условие
\re{fdsdfsdgsdfgdttt55}, которое, как было отмечено в 
доказательстве теоремы \ref{th44433013343332245},
равносильно 
правой части \re{1sdfsdfsdfgsa32dg3s33fd}.

Теперь докажем импликацию 
\re{2sdfsdfsdfgsa32dg3s33fd}.

Согласно теореме 
\ref{th4443301334333245}, 
из левой части \re{2sdfsdfsdfgsa32dg3s33fd} следует, что
существует ЛА $L$ размерности $k\leq n$ 
над $X$, такой, что $f_L=f$. 

Пусть $L$ имеет вид
$(\xi^0, \{L^x\mid x\in  X\}, \lambda)$, и 
$u$ -- произвольная строка из $X^*$,
тогда 
\be{sdafdsafdsaf4433}
\by\forall\,v_1,v_2\in X^*\\
\hat f_u(v_1,v_2)=f(v_1uv_2)=
\xi^0L^{v_1uv_2}\lambda=
\xi^0L^{v_1}L^{u}L^{v_2}\lambda.\ey\ee

Обозначим записью $E_{ij}$ (где $i,j=1,\ldots, k$)
квадратную матрицу порядка $k$, 
в которой компонента в строке $i$ столбце $j$ 
равна 1, а все остальные компоненты равны 0.
Используя эти обозначения, 
можно представить матрицу $L^u$ в виде суммы
\be{1sdafdsafdsaf4433}
L^u=\sum\limits_{i,j=1,\ldots, k}a_{ij}E_{ij},\ee
где $a_{ij}$ -- компонента матрицы $L^u$ 
в строке $i$ столбце $j$.

Из \re{sdafdsafdsaf4433} и \re{1sdafdsafdsaf4433}
следует, что
\be{sda33fdsafdsaf4433}
\forall\,v_1,v_2\in X^*\quad
\hat f_u(v_1,v_2)=\sum\limits_{i,j=1,\ldots, k}a_{ij}
\xi^0L^{v_1}E_{ij}L^{v_2}\lambda.\ee

Обозначим записью $f_{ij}$ (где $i,j=1,\ldots, k$)
функцию из ${\bf R}^{X^*\times X^*}$,
определяемую следующим образом:
\be{s32da33fdsafdsaf4433}
\forall\,v_1,v_2\in X^*\quad
f_{ij}(v_1,v_2)=
\xi^0L^{v_1}E_{ij}L^{v_2}\lambda.\ee

Из \re{sda33fdsafdsaf4433} и
\re{s32da33fdsafdsaf4433} следует, что
\be{sd21a33fdsafdsaf4433}
\forall\,v_1,v_2\in X^*\quad
\hat f_u(v_1,v_2)=\sum\limits_{i,j=1,\ldots, k}a_{ij}
f_{ij}(v_1,v_2).\ee

Из \re{sd21a33fdsafdsaf4433}
следует, что функция $\hat f_u$
принадлежит подпространству линейного пространства 
${\bf R}^{X^*\times X^*}$, порождённому функциями
$f_{ij}$ (где $i,j=1,\ldots, k$). 
Это подпространство одинаково для всех $u\in X^*$.
Поскольку размерность 
этого подпространства не превосходит $k^2$, и $k\leq n$,
то, следовательно, верна правая часть 
импликации
\re{2sdfsdfsdfgsa32dg3s33fd}.
$\blackbox$\\
 
Можно доказать, что $\forall\,f\in {\bf R}^{X^*}\;\;
\dim E_{\hat f} = (\dim E_f)^2$.

\section{Счетномерные линейные автоматы и их языки}

\subsection{Вспомогательные понятия}

В этом пункте
мы будем использовать следующие обозначения.
\bi
\i Символ ${\bf N}$ обозначает множество 
 положительных натуральных 
чисел $(1$, $2,$ $\ldots)$.
\i Запись 
${\bf R}^{\bf N}$
обозначает множество последовательностей 
действительных чисел,
$\forall\,f\in {\bf R}^{\bf N},\;\forall
\,i\in {\bf N}$ $i$--й член последовательности
$f$ обозначается записью $f_i$.
\i Символ 
${\Xi}$ обозначает множество всех 
последовательностей 
$\xi\in {\bf R}^{\bf N}$,
удовлетворяющих условию
$\sum\limits_{i\geq 1}|\xi_i|<\infty$.

Мы будем рассматривать элементы ${\Xi}$
как счетномерные вектор-строки.
\i Символ ${\Lambda}$ обозначает множество 
всех последовательностей
$\lambda\in 
{\bf R}^{\bf N}$, удовлетворяющих условию
$\sup\limits_{i\geq 1}|\lambda_i|<\infty$.

Мы будем рассматривать элементы ${\Lambda}$
как счетномерные вектор-столбцы.

\i Символ 
${\cal H}$ обозначает множество всех 
бесконечных матриц $H$, компоненты
которых индексированы парами натуральных
чисел, и 
удовлетворяют условию
$\sup\limits_{i\geq 1}\sum\limits_{j\geq 1}|H_{ij}|<\infty$.

$\forall\,H,H'\in {\cal H}$ запись $HH'$
обозначает матрицу из ${\cal H}$, компоненты которой 
определяются следующим образом:
$$\forall\,i,j\in {\bf N}\quad
(HH')_{ij}=\sum\limits_{k\geq 1}H_{ik}H_{kj}.$$

Нетрудно видеть, что \bi\i операция умножения матриц 
из ${\cal H}$ ассоциативна, и \i нейтральным элементом
относительно этой операции 
является матрица $E$, определяемая следующим образом:
$\forall\,i,j\in {\bf N}\;\;
E_{ij}\eam 1$, если $i=j$, и $0$, иначе.\ei

\i $\forall\,\xi\in \Xi,\;
\forall\,H\in {\cal H}$ запись $\xi H$
обозначает вектор-строку из $\Xi$, компоненты которой 
определяются следующим образом:
$$\forall\,i\in {\bf N}\quad
(\xi H)_{i}=\sum\limits_{k\geq 1}\xi_k H_{ki}.$$

\i $\forall\,\lambda\in \Lambda,\;
\forall\,H\in {\cal H}$ запись $H\lambda$
обозначает вектор-столбец из $\Lambda$, 
компоненты которого
определяются следующим образом:
$$\forall\,i\in {\bf N}\quad
(H\lambda)_{i}=\sum\limits_{k\geq 1}H_{ik}\lambda_k.$$
\ei

\subsection{Понятие счетномерного линейного автомата
и связанные с ним понятия}
\label{dsfsadfsadfasdfdsa}

Пусть задано конечное множество $X$.

{\bf Счетномерным линейным автоматом (СЛА)} 
мы будем называть тройку $L$ вида
\be{vdsvczvzxcv3zx}L=(\xi^0, \{L^x\mid x\in  X\}, \lambda)\ee
где $\xi^0\in \Xi,\;\forall\,x\in X\;\;L^x\in {\cal H},
\lambda\in \Lambda$.

Так же, как и для обычных ЛА, для каждого 
СЛА $L$ можно определить понятие
реакции $f_L\in {\bf R}^{X^*}$ и языка $L_a\subseteq X^*$, представимого СЛА с заданной точкой сечения $a\in {\bf R}$:
\bi
\i $\forall\,u\in X^*\;\;f_L(u)\eam \xi^0L^u\lambda$,
где $$L^u\eam \left\{\by E,& \mbox{если } u=\varepsilon,\\ 
L^{x_1}\ldots L^{x_n},& \mbox{если } u=x_1\ldots x_n, 
\mbox{где }
x_1,\ldots, x_n\in X,\ey\right.$$
\i язык $L_a$ определяется соотношением 
\re{asfasdfsdags555}.
\ei

Пусть $L=(\xi^0, \{L^x\mid x\in  X\}, \lambda)$ -- СЛА.

{\bf Базисные матрицы} СЛА $L$ -- 
это бесконечные матрицы 
$M_L$ и $N_L$, обладающие следующими свойствами:
\bi
\i строки матрицы $M_L$ образуют базис 
линейного подпространства 
пространства $\Xi$, порожденного строками вида $\xi^0 L^u$,
где $u\in X^*$,
\i столбцы матрицы $N_L$ образуют базис 
линейного подпространства 
пространства $\Lambda$, 
порожденного столбцами вида $L^u\lambda$,
где $u\in X^*$.
\ei

Базисные матрицы $M_L$ и $N_L$
определяют конгруэнции 
на линейных пространствах $\Xi$ и $\Lambda$
соответственно, 
обозначаемые символом $\sim_L$ и определяемые
следующим образом:
$$\by
\forall\,\xi,\xi'\in \Xi&\xi \sim_L \xi'&\Leftrightarrow&
\xi N_L = \xi' N_L,\\
\forall\,\lambda,\lambda'\in \Lambda& \lambda 
\sim_L \lambda'&\Leftrightarrow&
M_L\lambda = M_L\lambda'.
\ey$$

Нетрудно доказать, что если матрица $M_L$ ($N_L$)
имеет конечное число строк (столбцов), то конгруэнция
$\sim_L$ на $\Xi$ ($\Lambda$)
имеет конечный индекс.

\subsection{Свойства счетномерных линейных автоматов}

\refstepcounter{theorem}
{\bf Теорема \arabic{theorem}\label{th044221133156}}.

Пусть $L$ -- СЛА вида \re{vdsvczvzxcv3zx}, 
 $a\in {\bf R}$, и $E$ = $\Xi$ или $\Lambda$.

Тогда $L_a$ -- ВЯ 
$\Leftrightarrow$
$\exists$ линейная конгруэнция $\sim$ конечного ранга на $E$,
обладающая следующими свойствами:
\bi
\i 
\be{dfasdfsadfgsfdgfsdgfsd}
\forall\,x\in X\;\exists\,
L^x_\sim:\quad
\diagrw{{E}&\pright{L^x}&{E}\cr
\pdownl{}&&\pdownr{}\cr
{E}_\sim&\pdashright{L^x_\sim}&{E}_\sim}\ee
где ${E}_\sim$ -- фактор-пространство, и 
$E\to {E}_\sim$ -- каноническая проекция,
\i $\exists$ линейное отображение
$\lambda_\sim: E_\sim\to {\bf R}$,
 $\exists$ $\xi^0_\sim\in E_\sim$,
 $\exists\, b\in {\bf R}$:
\ei
\be{redfdsfasdgfsdgfsdfgsd}
u\in L_a\quad\Leftrightarrow\quad \lambda_\sim
(\xi^0_\sim A_\sim^u)>b.
\ee

{\bf Доказательство}.

Обоснуем лишь импликацию ``$\Rightarrow$''.

Пусть $\exists$ ВА $B$,
$\exists \,b\in {\bf R}: L_a=B_b.$

Обозначим символом $\pi$
эпиморфизм $E\to {\bf R}^{|S_B|}$.

Искомая конгруэнция $\sim$ определяется следующим образом:
\bi\i если $E=\Xi$, то
$\xi'\sim \xi''\;\Leftrightarrow\;
\pi(\xi')B^u =\pi(\xi'')B^u$, 
\i если $E=\Lambda$, то 
$\lambda'\sim \lambda''\;\Leftrightarrow\;
 B^u \pi(\lambda')= B^u\pi(\lambda'')$.
$\blackbox$\ei

Отметим, что для каждого СЛА $L$
\bi
\i конгруэнция $\sim_L$ (определённая в конце
пункта \ref{dsfsadfsadfasdfdsa})
обладает свойством
\re{dfasdfsadfgsfdgfsdgfsd}, и
\i если конгруэнция $\sim_L$ имеет конечный индекс, то 
   для языка $(f_L)_0$
 выполнены условия теоремы \ref{th044221133156}.\\
 (напомним, что  $\forall\, f\in {\bf R}^{X^*}\;
 f_0\eam \{u\in X^*\mid f(u)>0\}$)
\ei

Пусть $G$ -- некоторая полугруппа.
Мы будем использовать следующие обозначения.
\bi
\i Запись ${\bf R}^G$ обозначает множество функций
вида $f:G\to {\bf R}$.  

${\bf R}^G$ является линейным
пространством, в котором операции сложения 
и умножения на действительные числа определяются
стандартным образом.
\i $\forall\,f\in {\bf R}^G$, 
$\forall \,u\in G$ запись $f_u$ обозначает
функцию из ${\bf R}^G$, которая сопоставляет
каждому $u'\in G$ число $f(uu')$.
\i $\forall\,f\in {\bf R}^G$ запись $T_f$ обозначает множество 
$\{f_u\mid u\in G\}$.
\i $\forall\,f\in {\bf R}^G$ запись $\langle T_f\rangle$ обозначает
подпространство линейного пространства ${\bf R}^G$,
порожденное функциями из $T_f$.
\ei

\refstepcounter{theorem}
{\bf Теорема \arabic{theorem}\label{th0442211331538}}.

Пусть $G$ -- полугруппа, 
$f\in {\bf R}^G$,
и
 $|Im(f)|<\infty$.
Тогда $$\dim\,\langle T_f\rangle<\infty\;\Leftrightarrow\;
|T_f|<\infty.$$

{\bf Доказательство}.  

Импликация ``$\Leftarrow$'' является очевидной.

Докажем импликацию
``$\Rightarrow$''. 

$\langle T_f\rangle$ можно рассматривать как
 нормированное
конечномерное линейное пространство, где $|\!|f|\!|=
\max \limits_{u\in G}|f(u)|$.

 $ T_f$  -- ограниченное подмножество 
  $\langle T_f\rangle$, поэтому $T_f$ -- компакт.
  
  Из $|Im(f)|<\infty$ следует, что $\exists\,c>0$:
  \be{asdfsfgsfdgsd}\forall\,f_{u_1}\neq f_{u_2}\in T_f\quad
  |\!|f_{u_1}-f_{u_2}|\!|\geq c
  \ee 
  Из компактности
$T_f$ следует, что  множество открытых шаров
  с центрами в точках из $T_{f}$ радиуса $\frac{c}{2}$ 
  содержит конечное подмножество, покрывающее $T_f$.
  Согласно \re{asdfsfgsfdgsd}, отсюда следует, что
  $|T_{f}|<\infty$. $\blackbox$\\
  

\refstepcounter{theorem}
{\bf Теорема \arabic{theorem}\label{th044221133157}}.

Пусть $L$ -- СЛА вида \re{vdsvczvzxcv3zx}, 
 $a\in {\bf R}$, и $E$ = $\Xi$ или $\Lambda$.

Тогда язык $L_a$ регулярен 
$\Leftrightarrow$
$\exists$ линейная конгруэнция $\sim$ конечного ранга на $E$,
такая, что 
\bi\i $\sim$ обладает свойствами, изложенными в 
теореме \ref{th044221133156},
\i $|\{\lambda_\sim(\xi^0_\sim A_\sim^u)\mid u\in X^*\}|<\infty$.
\ei

{\bf Доказательство}.

Импликация ``$\Rightarrow$''
следует из теоремы \ref{th044221133156}, в данном случае
$B$ -- КДА, и
$$\forall\,u\in X^*\quad \lambda_\sim(\xi^0_\sim A_\sim^u)\in\{0,1\}.$$

Докажем импликацию
``$\Leftarrow$''.

Обозначим символом $f$ функцию из ${\bf R}^{X^*}$, 
которая сопоставляет каждому $u\in X^*$ число
$\lambda_\sim(\xi^0_\sim A_\sim^u)$.
Поскольку
$f$ -- ЛАФ,  то
$\dim\langle L_f\rangle<\infty$,
поэтому по теореме
\ref{th0442211331538}
$|L_f|<\infty$.

Нетрудно видеть, что язык $L_a$ совпадает с множеством
строк, которые принимаются конечным детерминированным
автоматом $A=(X, S, s^0, \delta, F)$, где 
\bi
\i $S\eam \{f_u\mid u\in X^*\}$, 
\i $s^0 \eam f_\varepsilon=f$,
\i   $\delta: S\times X\to S,\;\;\delta(f_u,x)\eam f_{ux}$,
\i $F\eam \{f_u\mid f(u)>a\}.\;\;\blackbox$
\ei

\refstepcounter{theorem}
{\bf Теорема \arabic{theorem}\label{th044221133157432}}.

Пусть
$f\in {\bf R}^{X^*},|Im(f)|<\infty$. 

Тогда
$f$ -- ЛАФ $\Leftrightarrow\;|\{f_u\mid u\in X^*\}|<\infty.\quad \blackbox$\\

\refstepcounter{theorem}
{\bf Теорема \arabic{theorem}\label{th0442211331574}}.

Пусть $S\subseteq X^*$.
Тогда $\chi_S$ (характеристическая функция
подмножества $S$) -- ЛАФ $\Leftrightarrow\;S$ регулярен. $\blackbox$\\

\refstepcounter{theorem}
{\bf Теорема \arabic{theorem}\label{th0442211331575}}.

Пусть  $f\in {\bf R}^{X^*}$, $|Im(f)|<\infty$.

Тогда $f$ -- ЛАФ $\Leftrightarrow$ $\forall\,a\in {\bf R}\;\;
\{u\in X^*\mid f(u)=a\}$ регулярен.\\

{\bf Доказательство}.

Импликация ``$\Rightarrow$'' следует из теоремы 
\ref{th044221133157}.

Докажем импликацию 
``$\Leftarrow$''.

Пусть
 $Im(f)=\{a_1,\ldots, a_m\}$, и
$\forall\,i=1,\ldots,m$ существует ДА $A_i$, который
представляет язык 
$\{u\in X^*\mid f(u)=a_i\}$.
Представим этот ДА в виде ЛА
$(\xi^0_i,\{B^x_i\mid x\in X\}, \lambda_i)$, т.е. 
\bi
\i  $\xi^0_i$ содержит единицу в позиции, соответствующей
начальному состоянию $A_i$, остальные 
 компоненты  $\xi^0_i$  равны 0, \i $(B^x_i)_{s,s'}=1$, 
если $s'=\delta_i(s,x)$, и 0 -- иначе, 
\i компоненты $\lambda_i$, соответствующие терминальным
состояниям $A_i$, равны 1, остальные компоненты $\lambda_i$ равны 0.\ei
Тогда $f(u)= (\xi^0_1\ldots\xi^0_m)
\left(\by
B_1^u\cr &\ldots&\cr
&&B_m^u\cr
\ey\right)
\left(\by
a_1\lambda_1\cr \ldots\cr
a_m\lambda_m\cr
\ey\right).\;\;\blackbox
$

 \section{Достижимость  и различимость в линейных автоматах}

\subsection{Достижимость}

Пусть задан ЛА $L=(\xi^0,\{L^x\mid x\in X\},\lambda)$, 
и $n=\dim L$.

Вектор $\xi\in {\bf R}^n$ называется {\bf достижимым} в $L$, если 
$$\exists\,u\in X^*:\;\xi=\xi^0 L^u.$$ 

{\bf Степенью достижимости} ЛА $L$ называется 
число $$
\delta(L)\eam \min\,k:\langle \xi^0 L^u\mid u\in X^*\rangle=
\langle \xi^0 L^u\mid u\in X^{\leq k}\rangle,$$
где $\forall\,V\subseteq {\bf R}^n$ запись
$\langle V\rangle$ обозначает подпространство, порожденное
векторами из множества $V$.

Обозначим записью $M_L$ бесконечную матрицу, 
строками которой являются вектора вида $\xi^0L^u\;(u\in X^*)$,
и записью $r(M_L)$  -- ранг этой матрицы.
Нетрудно видеть, что $r(M_L)\leq n$.\\

\refstepcounter{theorem}
{\bf Теорема \arabic{theorem}.\label{th0442211333231255}}

Если $L$ -- ЛА конечной размерности, то $\delta(L)\leq r(M_L)-1$.\\

{\bf Доказательство}.

Цепочка подпространств
$$V_0\subseteq V_1\subseteq 
V_2\subseteq \ldots \subseteq{\bf R}^n,$$
где $V_0\eam \langle \xi^0\rangle$ и $\forall\,k\geq 0\;\;
V_{k+1}\eam V_k+\langle\{\xi L^x\mid \xi\in V_k,x\in X\}\rangle$,
не может неограниченно возрастать.
Нетрудно видеть, что 
$$\forall\,k\geq 0 \quad V_k=\langle 
\{\xi^0 L^u\mid u\in X^{\leq k}
\}\rangle,$$
и минимальное  $k$, такое, что $V_k=V_{k+1}$, совпадает с $\delta(L)$, и удовлетворяет неравенству $$k+1\leq \dim 
V_k =
r(M_L),$$
откуда следует утверждение теоремы.
$\blackbox$

\subsection{Различимость}

Пусть задан ЛА $L=(\xi^0,\{L^x\mid x\in X\},\lambda)$, 
и $n=\dim L$.

Векторы $\xi_1,\xi_2\in {\bf R}^n$ 
называются {\bf различимыми}  в $L$, если 
\be{adsfsafasdfaaasd}
\exists\,u\in X^*:\;\xi_1 L^u\lambda\neq 
\xi_2 L^u\lambda.\ee
Ниже запись $\xi_1\not\sim_L \xi_2$ обозначает соотношение 
\re{adsfsafasdfaaasd}.

{\bf Степенью различимости} ЛА $L$ называется 
число $$
\rho(L)\eam \min \,k:\xi_1\not\sim_L \xi_2\Rightarrow\;\exists\,
u\in X^{\leq k}:\xi_1 L^u\lambda\neq 
\xi_2 L^u\lambda.$$
Из определения $\rho(L)$ следует, что 
$$\langle L^u\lambda\mid u\in X^*\rangle=
\langle L^u\lambda\mid u\in X^{\leq \rho(L)}\rangle.$$

Обозначим записью $N_L$ бесконечную матрицу, 
столбцами которой являются вектора вида $L^u\lambda\;(u\in X^*)$,
и записью $r(N_L)$  -- ранг этой матрицы.
Нетрудно видеть, что 
$r(N_L)\leq n$.\\

\refstepcounter{theorem}
{\bf Теорема \arabic{theorem}.\label{t3h04232155}}

Если $L$ -- ЛА конечной размерности, то $\rho(L)\leq r(N_L)-1$.\\ 

{\bf Доказательство}.

Теорема доказывается аналогично теореме 
\ref{th0442211333231255}:
определяется цепочка вложенных подпространств
$\{V_k\mid k
\geq 0\}$, где 
$$\forall\,i\geq 0\quad
V_k= \langle \{L^u\lambda\mid u\in X^{\leq k}\}\rangle,$$
и нетрудно доказать, что 
тот номер $k$, на котором эта цепочка стабилизируется
(т.е. минимальное $k$, такое, что $V_k=V_{k+1}$),
совпадает с $\rho(L)$.
$\blackbox$

\section{Реализация функций на строках линейными автоматами}

Пусть $X$ -- конечное множество.

{\bf Ганкелева матрица} функции $f\in {\bf R}^{X^*}$ --
это бесконечная матрица $H^f$, строки и столбцы которой 
индексированы элементами множества $X^*$
и $$\forall\,u,v\in X^*\quad H^f_{u,v}\eam f(uv),$$
где $H^f_{u,v}$ -- элемент в строке
$u$ и столбце $v$ матрицы $H^f$. 

Для каждого $u\in X^*$ 
записи $\vec H^f_u$ и ${H^f_u}^\downarrow$ 
обозначают 
строку $u$ и столбец $u$ соответственно матрицы $H^f$
(данные понятия определяются так же, как 
аналогичные понятия
из пункта \ref{fdgjhglrereerrer}).

Мы будем обозначать записью $r(f)$ ранг матрицы $H^f$,
т.е. размерность линейного пространства, порожденного
множеством её строк (или столбцов).

Подмножество $B$ строк (или столбцов) матрицы $H^f$
называется {\bf базисным}, если порождаемое
ими линейное пространство $\langle B\rangle$
совпадает с пространством,
порождаемым всеми строками (или столбцами) $H^f$, и 
ни один элемент $B$ не  является линейной комбинацией других
элементов $B$.
\\



\refstepcounter{theorem}
{\bf Теорема \arabic{theorem}.\label{th05442215}}

Если $f\in {\bf R}^{X^*}$, и 
$r(f)<\infty$, то в матрице $H^f$ можно выбрать 
такое базисное множество $B$
строк (или столбцов), что  
 индекс $u$ каждой строки (или столбца) из $B$
удовлетворяет условию 
$|u|\leq r(f)-1$. $\blackbox$\\

Будем использовать следующие обозначения.
Пусть 
$$r\eam {r(f)}<~\infty,$$ и
$U$ и $V$  -- 
последовательности строк из $X^*$ вида 
\be{dsasfsadfdsafee}
U= (u_1,\ldots, u_{r}),\quad
V=(v_1,\ldots, v_{r})\ee
соответственно, обладающие следующими свойствами:
\bi
\i $u_1=v_1=\varepsilon$,
\i $\{\vec H^f_{u}\mid u\in U\}$ и 
    $\{{H^f_{v}}^\downarrow\mid v\in V\}$ -- 
базисные множества строк и столбцов матрицы $H^f$ 
соответственно.    \ei
Тогда
\bi
\i запись $(U,V)^f$ обозначает квадратную матрицу 
порядка ${r}$, строки и столбцы которой индексированы элементами
$U$ и $V$ соответственно (мы будем называть эту матрицу
{\bf базисной матрицей} фунции $f$), и 
$$\forall\,u\in U,\;\forall\,v\in V\quad
(U,V)^f_{u,v}\eam f(uv),$$
т.е. $(U,V)^f$ -- подматрица порядка $r$  матрицы $H^f$\\
(нетрудно доказать, что $(U,V)^f$ -- невырожденная подматрица максимального порядка   матрицы $H^f$),
\i $\forall\,w\in X^*$ запись $(U,V)^{f,w}$ 
 обозначает квадратную матрицу 
порядка ${r}$, строки и столбцы которой индексированы элементами
$U$ и $V$ соответственно, и 
$$\forall\,u\in U,\;\forall\,v\in V\quad
(U,V)^{f,w}_{u,v}\eam f(uwv).$$
\ei

\refstepcounter{theorem}
{\bf Теорема \arabic{theorem}.\label{th0544225561555}}

Пусть $f\in {\bf R}^{X^*}$, $r=r(f)<\infty$, и
$U,V$  -- последовательности 
строк из $X^*$, 
такие, что  
$(U,V)^f$ базисная матрица функции $f$.

Тогда 
$$(U,V)^{f,uv}=(U,V)^{f,u}\Big((U,V)^{f}\Big)^{-1}(U,V)^{f,v}.$$

{\bf Доказательство}.

Нетрудно видеть, что 
$$\by\left(\by 
(U,V)^{f}&(U,V)^{f,v}\cr
(U,V)^{f,u}&(U,V)^{f,uv}\cr\ey
\right)=\\=((u_1,\ldots, u_r,u_1u,\ldots,u_ru),
(v_1,\ldots, v_r,vv_1,\ldots,vv_r))^f.\ey$$

Доказываемое равенство является  следствием 
того, что ранг этой матрицы равен $r$. $\blackbox$\\

Из теоремы \ref{th0544225561555} вытекает нижеследующая 
теорема.\\

\refstepcounter{theorem}
{\bf Теорема \arabic{theorem}.\label{th05442233}}

Пусть $f\in {\bf R}^{X^*}$, $r=r(f)<\infty$, и
$U,V$  -- последовательности 
строк из $X^*$, 
такие, что  
$(U,V)^f$ базисная матрица функции $f$.

Тогда $\forall\,u\in X^*$, если $u=x_1\ldots x_s$,
где $x_1,\ldots ,x_s\in X$, то 
$$(U,V)^{f,u}=(U,V)^{f,x_1}\Big((U,V)^{f}\Big)^{-1}\ldots
\Big((U,V)^{f}\Big)^{-1}(U,V)^{f,x_s}.\;\;\blackbox$$

\refstepcounter{theorem}
{\bf Теорема \arabic{theorem}.\label{th054422331}}

Пусть $f\in {\bf R}^{X^*}$, $r=r(f)<\infty$, и
$U,V$  -- последовательности 
строк из $X^*$, 
такие, что  
$(U,V)^f$ -- базисная матрица функции $f$.

Тогда $f=f_L$, где $L$ -- ЛА, определяемый следующим образом:
$$L\;\eam\;
\Big(\vec e_1,\;\;\Big\{(U,V)^{f,x}\Big((U,V)^{f}\Big)^{-1}\mid x\in X\Big\},\;\;(U,V)^{f}e_1^\downarrow\Big).\;\;
\blackbox$$

\end{document}